# Solar system science with the Wide-Field InfraRed Survey Telescope (WFIRST)

B.J. Holler[1*], S.N. Milam[2*], J.M. Bauer[3,4], C. Alcock[5], M.T. Bannister[6], G.L. Bjoraker[2], D. Bodewits[3], A.S. Bosh[7], M.W. Buie[8], T.L. Farnham[3], N. Haghighipour[9], P.S. Hardersen[10], A.W. Harris[11], C.M. Hirata[12], H.H. Hsieh[13,14], M.S.P. Kelley[3], M.M. Knight[3], E.A. Kramer[4], A. Longobardo[15], C.A. Nixon[2], E. Palomba[15], S. Protopapa[3,8], L.C. Quick[16], D. Ragozzine[17], V. Reddy[18], J.D. Rhodes[4], A.S. Rivkin[19], G. Sarid[20], A.A. Sickafoose[21,7], A.A. Simon[2], C.A. Thomas[13], D.E. Trilling[22], R.A. West[4]

[1]Space Telescope Science Institute, 3700 San Martin Drive, Baltimore, MD 21218, USA
[2]NASA Goddard Space Flight Center, 8800 Greenbelt Road, Greenbelt, MD 20771, USA
[3]Department of Astronomy, University of Maryland, College Park, MD 20742, USA
[4]Jet Propulsion Laboratory, California Institute of Technology, 4800 Oak Grove Drive, Pasadena, CA 91109, USA
[5]Harvard-Smithsonian Center for Astrophysics, 60 Garden Street, Cambridge, MA 02138, USA
[6]Astrophysics Research Centre, Queen's University Belfast, Belfast BT7 1NN, UK
[7]Department of Earth, Atmospheric, and Planetary Sciences, MIT, 77 Massachusetts Avenue, Cambridge, MA 02139, USA
[8]Southwest Research Institute, 1050 Walnut Street, Suite 300, Boulder, CO 80302, USA
[9]Institute for Astronomy, University of Hawaii-Manoa, Honolulu, HI 96822, USA
[10]University of North Dakota, Department of Space Studies, 4149 University Avenue, Stop 9008, 530 Clifford Hall, Grand Forks, ND 58202, USA
[11]MoreData! Inc., 4603 Orange Knoll Avenue, La Cañada, CA 91011, USA
[12]Center for Cosmology and Astro Particle Physics (CCAPP), The Ohio State University, 191 West Woodruff Lane, Columbus, OH 43210, USA
[13]Planetary Science Institute, 1700 East Fort Lowell Road, Suite 106, Tucson, AZ 85719, USA
[14]Institute of Astronomy and Astrophysics, Academia Sinica, P.O. Box 23-141, Taipei 10617, Taiwan
[15]INAF Istituto di Astrofisica e Planetologia Spaziali, via Fosso del Cavaliere 100, Rome, Italy
[16]Smithsonian Institution, National Air and Space Museum, Center for Earth and Planetary Studies, Independence Avenue at 6th Street, SW, Washington, DC, 20560, USA
[17]Department of Physics and Astronomy, Brigham Young University, N283 ESC, Provo, UT 84602, USA
[18]Lunar and Planetary Laboratory, University of Arizona, 1629 E University Boulevard, Tucson, AZ 85721, USA
[19]Johns Hopkins University Applied Physics Laboratory, 11101 Johns Hopkins Road, Laurel, MD 20723, USA
[20]Florida Space Institute, University of Central Florida, 12354 Research Parkway, Orlando, FL 32826, USA
[21]South African Astronomical Observatory, P.O. Box 9, 7935 Observatory, Cape Town, South Africa
[22]Department of Physics and Astronomy, Northern Arizona University, P.O. Box 6010, Flagstaff, AZ 86011, USA

## Abstract

We present a community-led assessment of the solar system investigations achievable with NASA's next-generation space telescope, the Wide Field InfraRed Survey Telescope (WFIRST). WFIRST will provide imaging, spectroscopic, and coronagraphic capabilities from 0.43-2.0 μm and will be a potential contemporary and eventual successor to JWST. Surveys of irregular satellites and minor bodies are where WFIRST will excel with its 0.28 deg$^2$ field of view Wide Field Instrument (WFI). Potential ground-breaking discoveries from WFIRST could include detection of the first minor bodies orbiting in the Inner Oort Cloud, identification of additional Earth Trojan asteroids, and the discovery and characterization of asteroid binary systems similar to Ida/Dactyl. Additional investigations into asteroids, giant planet satellites, Trojan asteroids,

[*] Corresponding authors: B.J. Holler (bholler@stsci.edu) and S.N. Milam (stefanie.n.milam@nasa.gov).

Centaurs, Kuiper Belt Objects, and comets are presented. Previous use of astrophysics assets for solar system science and synergies between WFIRST, LSST, JWST, and the proposed NEOCam mission are discussed. We also present the case for implementation of moving target tracking, a feature that will benefit from the heritage of JWST and enable a broader range of solar system observations.

## 1. Introduction

The Wide Field InfraRed Survey Telescope (WFIRST) is NASA's next flagship space observatory after the James Webb Space Telescope (JWST). WFIRST will have a 2.4-meter primary mirror, equivalent to the Hubble Space Telescope (HST), and is on track to launch in the mid-2020s, with a 6-year nominal mission. This telescope (Fig. 1) will have two instruments onboard: The Wide Field Instrument (WFI) with a 0.28 $deg^2$ FOV (Fig. 2), which includes an Integral Field Channel (IFC) that will obtain spectral information over the entirety of its FOV (3.0″×3.15″), and the Coronagraphic Instrument (CGI), which is designed to take images and spectra of super-Earths. These instruments will be capable of imaging, grism spectroscopy, and R~100 imaging spectroscopy over the near-infrared wavelength range (0.6-2.0 μm), and coronagraphic imaging and spectroscopy from 0.43-0.98 μm. For solar system observations, the primary modes will be imaging, R~100 spectroscopy, and coronagraphic imaging; grism spectroscopy will not be discussed further.

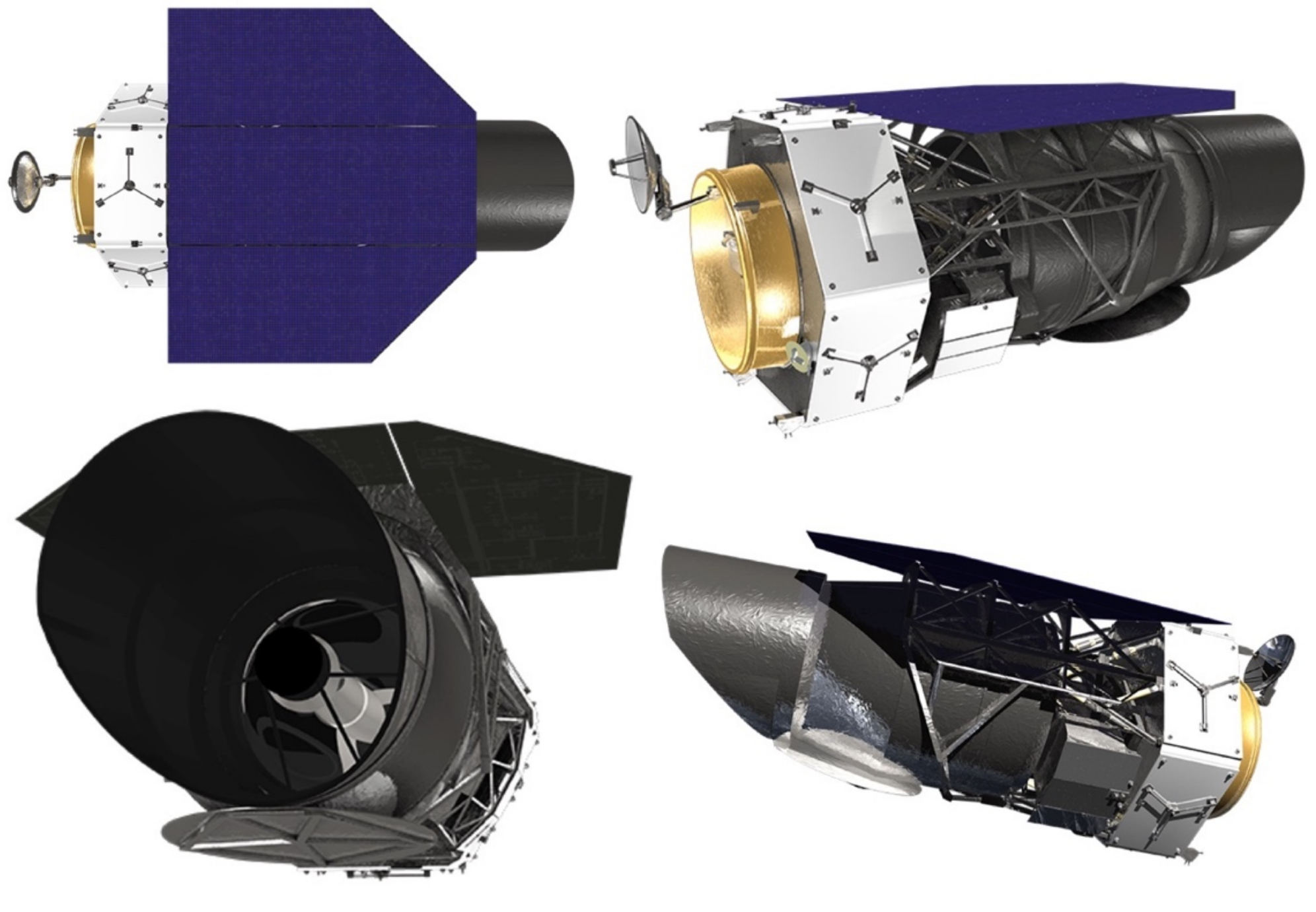

**Figure 1**: Four computer-generated models of the WFIRST spacecraft from different angles. The primary mirror is 2.4 meters in diameter, equal to the primary mirror on the Hubble Space Telescope. WFIRST will orbit around the Earth-Sun L2 point located ~1.5 million km from

Earth, similar to the James Webb Space Telescope [2]. (Spacecraft images from wfirst.gsfc.nasa.gov)

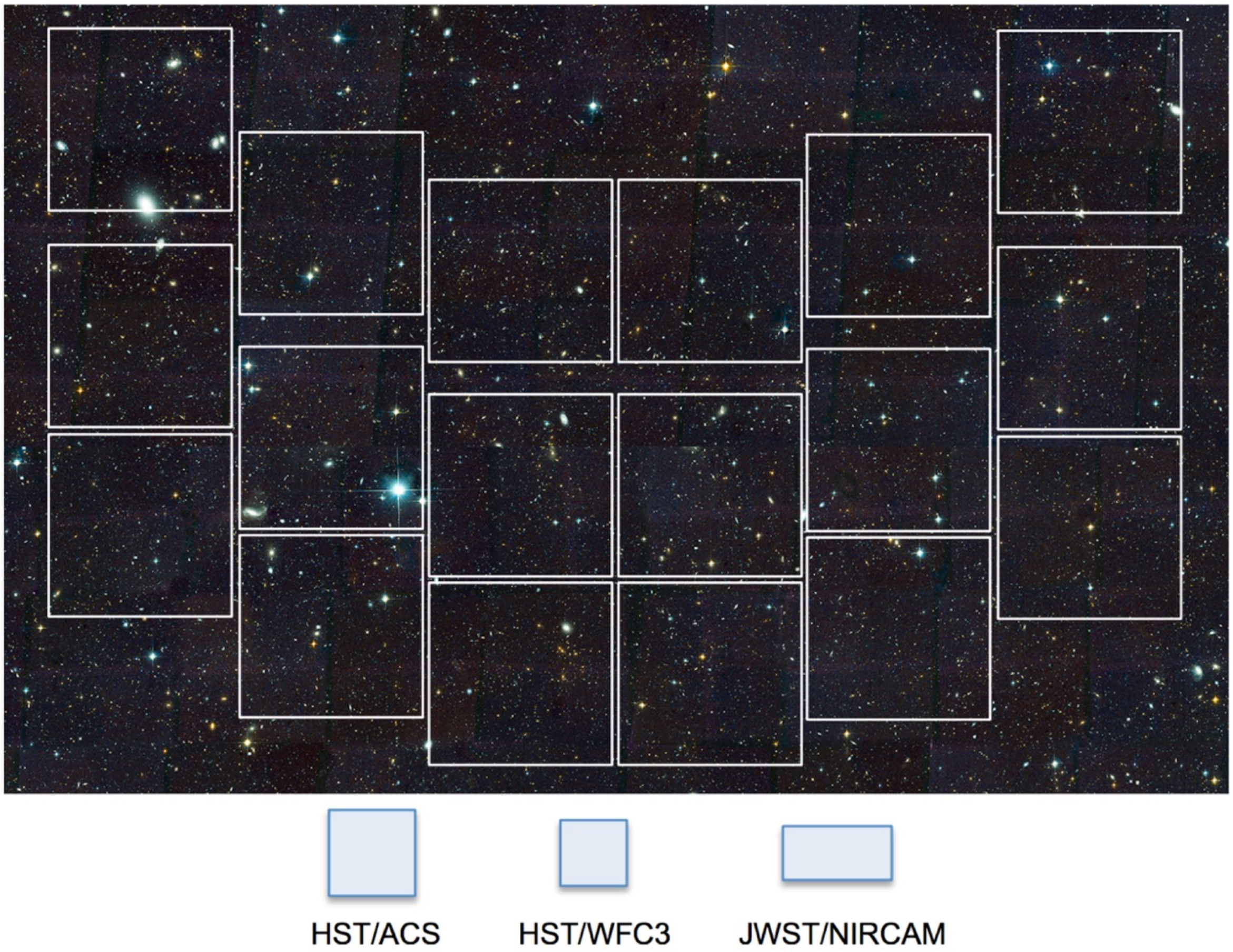

**Figure 2**: The WFI field of view covers 0.28 $deg^2$ and is composed of 18 separate 4k×4k HgCdTe detectors. The fields of view for the Advanced Camera for Surveys (ACS) and Wide Field Camera 3 (WFC3) instruments on HST and the Near-Infrared Camera (NIRCam) instrument on JWST are shown for comparison [2].

WFIRST will operate at the Earth-Sun L2 point and its primary goal will be to carry out the following science programs: (1) an exoplanet microlensing survey near the galactic bulge, (2) CGI observations of exoplanets, (3) imaging and grism spectroscopy of galaxies at high galactic latitude, (4) a supernova survey with 3 different imaging depths and a spectroscopy component, and (5) Guest Observer (GO) science. Funding for the analysis of survey data outside the defined science teams will be part of the Guest Investigator (GI) program. GO programs are expected to comprise ~1.5 years, or approximately 25%, of the nominal mission duration.

The imaging and spectroscopic capabilities of WFIRST [1,2] are well-suited for studies of solar system objects. In particular, the wavelength ranges of the IFC (0.6-2.0 μm) and the WFI (0.43-2.0 μm) cover diagnostic absorption features due to atmospheric gases, as well as minerals and ices on the surfaces of terrestrial bodies. The IFC provides spectral information over its entire

FOV (3.0″×3.15″, 0.05″ or 0.10″/pixel plate scale), enabling analysis of cloud compositions, atmospheres, terrestrial surfaces, and comet comae. The WFI will make observations through broadband filters (Table 1) with a pixel scale of 0.11″/pixel. Figure 3 shows diameter vs. observer distance for an object at the imaging depth magnitude for each WFI filter in a 1000-second exposure.

**Table 1. WFI filter specifications [2]**

| Filter | Bandpass (μm) | Center (μm) | Width (μm) | Imaging depth (5-σ) in 1000s |
|---|---|---|---|---|
| R062 | 0.480-0.760 | 0.620 | 0.280 | N/A |
| Z087 | 0.760-0.927 | 0.869 | 0.217 | 27.15 |
| Y106 | 0.927-1.192 | 1.060 | 0.265 | 27.13 |
| J129 | 1.131-1.454 | 1.293 | 0.323 | 27.14 |
| H158 | 1.380-1.774 | 1.577 | 0.394 | 27.12 |
| F184 | 1.683-2.000 | 1.842 | 0.317 | 26.15 |
| W149 | 0.927-2.000 | 1.485 | 1.030 | 27.67 |

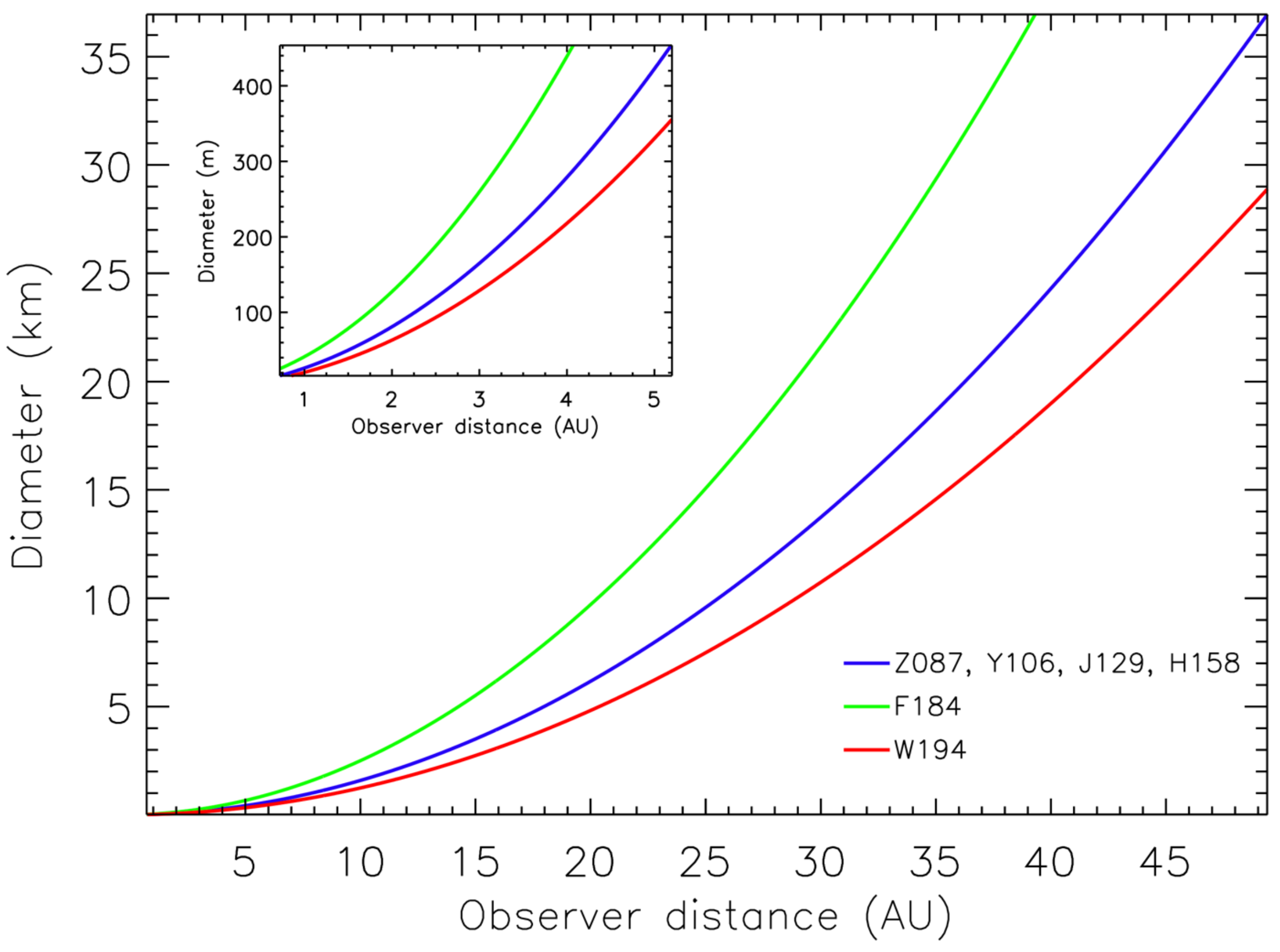

**Figure 3**: Diameters of minor bodies and irregular satellites that can be detected at the imaging depth for the WFI filters over a range of observer distances. Table 1 provides the values for the imaging depth of each filter [2]. The R062 filter is not included because imaging depth information is not yet available. The imaging depths for the Z087, Y106, J129, and H158 filters are comparable, so we used the average imaging depth for calculating the diameter of bodies detectable through these filters. For these calculations, we assumed hypothetical minor bodies with visible geometric albedos of 0.08 and observations at the maximum solar elongation angle of 126°. We then estimated radii by scaling to Charon, a spherical, neutral-colored object in the visible, using heliocentric distance, observer distance, V magnitude, visible geometric albedo, and radius. Charon's heliocentric distance, observer distance, and V magnitude were determined on an arbitrary date using JPL Horizons and the albedo and radius were taken from [3]. No additional spectral features were considered. We also assumed no loss of SNR due to streaking (though this can be a non-negligible effect for NEAs, main belt asteroids, and other inner solar system objects). The inset plot shows a zoom-in on observer distances of ~0.7-5.2 AU, which includes NEAs, main belt asteroids, Jupiter Trojans, and some comets.

The primary WFIRST instruments for solar system science are the WFI and the IFC, but observations should also be possible with the CGI. The CGI is a coronagraph designed for imaging and spectroscopy of exoplanets with a contrast of $10^{-9}$ ($\Delta$m=22.5). The CGI covers 0.43 to 0.98 μm and has an inner working angle (IWA) of ~0.15″. The field of view is 10″x10″ and the pixel scale is 0.0208″/pixel at 0.48 μm. By blocking the bright central object, the CGI could be used for the study of satellites around main belt asteroids (§5.2) and detection of activity around active asteroids and Centaurs. Due to the small inner working angle, observations of the ring systems of Uranus and Neptune (angular diameters of 3.7″ and 2.3″, respectively) would likely not be useful, since only a small fraction of the disks of these planets would be blocked.

The wide FOV of the WFI (0.28 $deg^2$) will revolutionize targeted surveys. Large regions of the sky that typically required many pointings will be observable in one pointing with WFIRST. Exploration of the entirety of the Hill spheres surrounding the giant planets could add to the population of irregular satellites already known around these planets. The efficiency of surveys aimed at identifying new minor bodies would increase. Serendipitous occultations utilizing the rapid cadence of guide star imaging may yield more occultation candidates because each of the 18 detectors of the WFI has the ability to obtain its own guide star. A significant amount of deep imaging data will be obtained during the 75% (~4.5 years) of the mission dedicated to astrophysics surveys, leading to the potential for serendipitous detection of new solar system objects, such as asteroids, comets, Centaurs, and KBOs.

However, as with any ground- or space-based telescope, there are limitations to WFIRST's ability to observe solar system targets. A major limitation is the field of regard (FOR). WFIRST's solar panels act both to power the spacecraft and shield the telescope and instruments from solar radiation. Thus, certain orientations with respect to the Sun are not allowed and only regions of the sky between solar elongation (Sun-WFIRST-target) angles of 54° and 126° will be observable at any given time (Fig. 4). This means that objects at opposition, when they are often brightest, will be unobservable. Additionally, inner solar system objects such as the Sun, Mercury, Venus, Earth, the Moon, and some Near-Earth Asteroids (NEAs) and comets will be unobservable. A further limitation is the assumed 30 milliarcsecond/second (mas/s) speed limit on non-sidereal tracking; this is the non-sidereal tracking limit for JWST [4], corresponding to the approximate maximum rate of motion of Mars, and was adopted as the nominal tracking limit for WFIRST. This affects the number of NEAs and long-period comets that can be observed (untrailed) by WFIRST (§3.1), with ~10% of NEAs and >50% of long-period comets trailing across images when they are in the FOR. A further discussion of non-sidereal tracking, and the effects on solar system science with WFIRST without it, can be found in §3.1.

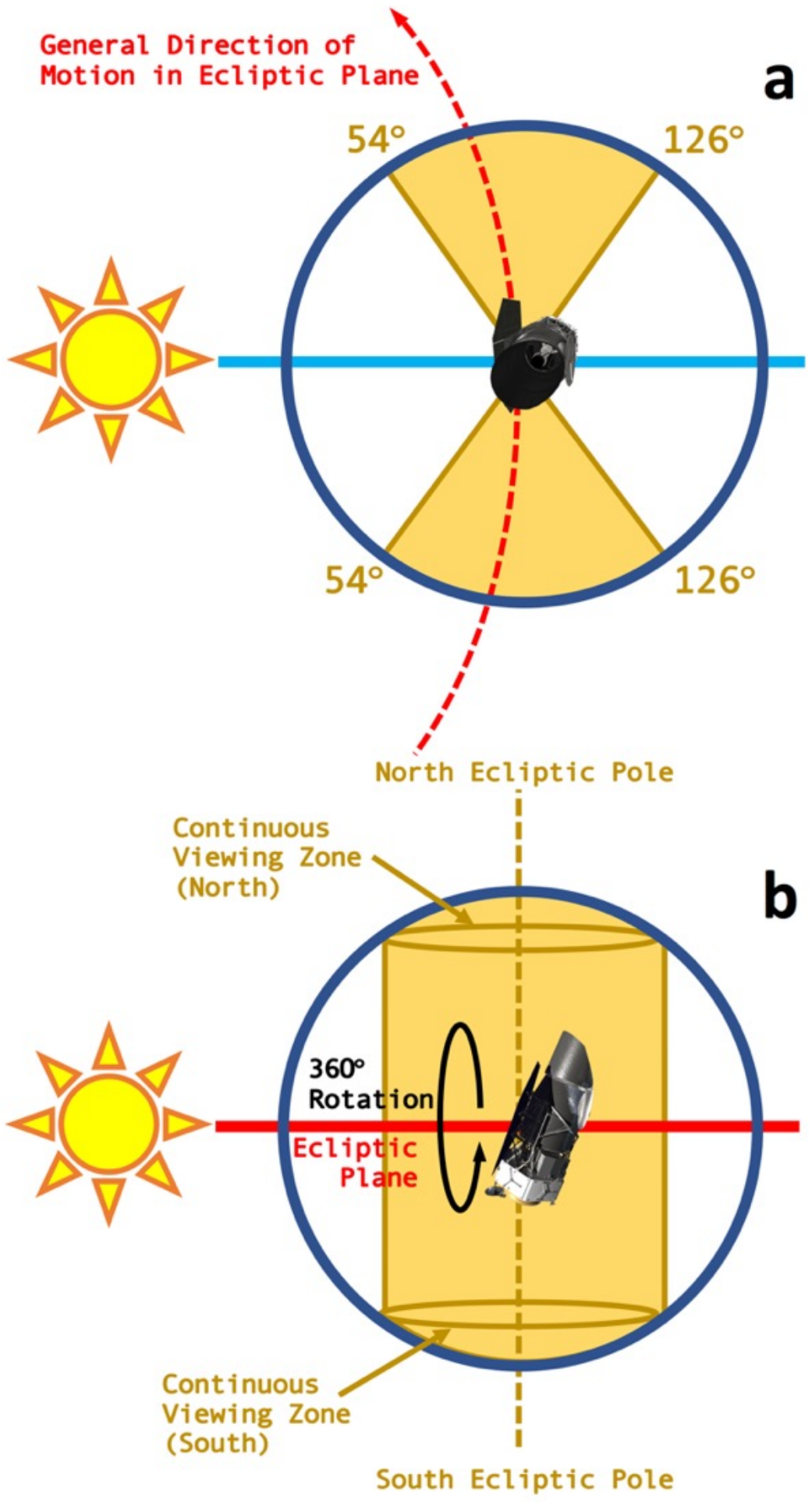

**Figure 4**: The WFIRST field of regard (FOR) is the region of the sky observable by the telescope at any given time. Observations of objects at opposition are not allowed, nor are observations of objects in the inner solar system (e.g., the Sun, Mercury, Venus, Earth, and the Moon). (a) The FOR is bounded by solar elongation angles of 54° and 126°. Viewing the FOR from above the ecliptic plane shows that there are two separate regions in opposite sides of the sky visible to WFIRST. A (relatively slow-moving) solar system object in the leading field will be observable for ~72 days, then not observable for ~108 days, then enters the trailing region for ~72 days, and following that is unobservable for another ~108 days. The exact durations vary depending on the rate of motion and inclination of the object (objects further from the ecliptic will be observable for longer periods of time). (b) View of the WFIRST FOR in the plane of the ecliptic. The continuous viewing zones are the regions within 36° of the North and South ecliptic poles. These regions are observable over the course of the entire year. (Spacecraft images from wfirst.gsfc.nasa.gov)

## 2. Scope of paper

The WFIRST Solar System Working Group (SSWG) was formed in December 2016. Membership was advertised to the solar system science community through an open call for volunteers. Presently, its members span the range of solar system expertise, including the many minor body populations, giant planet atmospheres, satellite studies, and occultations. The SSWG members divided themselves into 7 sub-groups covering these topics: Asteroids/NEOs/PHAs, Giant Planets, Satellites, Titan, KBOs/TNOs/Centaurs/Binaries, Comets, and Occultations. Each sub-group was tasked with identifying the science capabilities of WFIRST, as well as key augmentations to facilitate an active planetary science program. In response, the sub-groups produced science cases in their respective fields and are presented in more detail in later sections, with a summary provided in Table 2. Tables 3 and 4 list brightness estimates for a variety of solar system targets, some of which are explicitly discussed in this paper.

In the following sections, we discuss the desired instrumental capabilities and proposal categories for solar system observations (§3), the successful use of past and present astrophysics assets for solar system science (§4), the high-impact solar system investigations possible with WFIRST (§5), the usefulness of the astrophysics surveys for serendipitous detection of minor bodies (§6), and additional science investigations achievable with WFIRST (§7).

The science cases discussed in §5, §6, and §7 were based on the baseline WFIRST mission, which includes the WFI, IFC, and CGI instruments, wavelength coverage from 0.6-2.0 μm, and moving target tracking. A large fraction of our proposed science investigations would be negatively impacted if the IFC or moving target tracking were removed from the mission. Figure 5 provides an impact assessment for each of the science cases when considering the removal of moving target tracking, the IFC, or both from the baseline mission. Only two investigations, detection of Inner Oort Cloud objects and occultations (both targeted and serendipitous), would be unaffected by the proposed changes; all other investigations would be negatively impacted, with most becoming impossible to carry out. We therefore stress the importance of both moving target tracking and the IFC for observations of solar system objects.

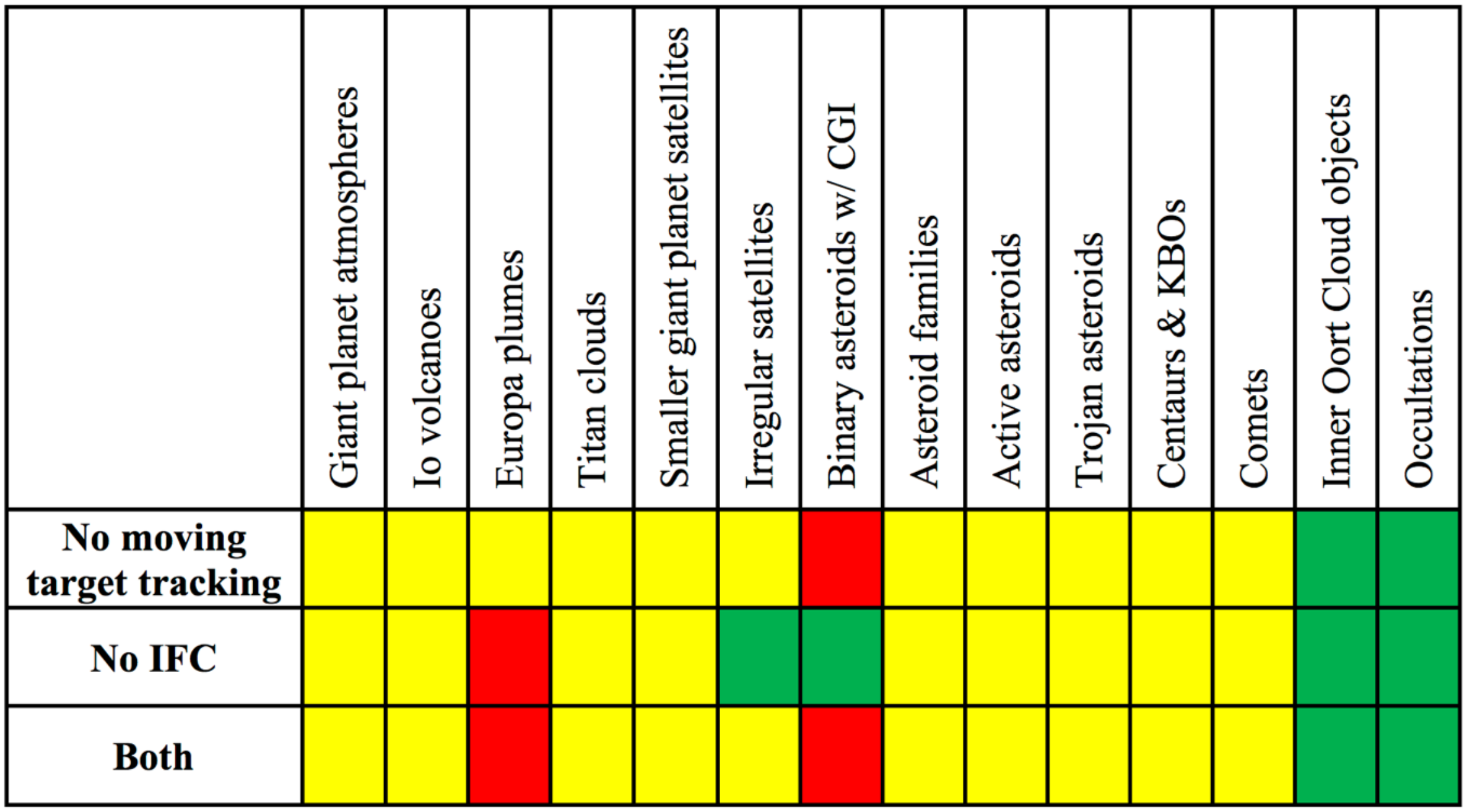

**Figure 5**: Red, yellow, and green boxes correspond to the three different levels of impact for the two potential cases of no moving target tracking and removal of the IFC. Red means that the entire science investigation, as outlined in §5, §6, and §7, would no longer be feasible, yellow means that a large fraction of the science investigation would be significantly impacted, and green means that there would be no impact.

**Table 2. WFIRST Solar system science cases**

| Science case | Rate (mas/s) | Instrument(s) | GO/GI |
|---|---|---|---|
| §5.1: Earth Trojan asteroids | ≤90 | WFI/IFC | GO |
| §5.2: Binary asteroids with the CGI | ≤40 | CGI | GO |
| §5.3: Irregular satellites | 1.5-10 | WFI | GO/GI |
| §5.4: Inner Oort Cloud objects | ≤0.5 | WFI | GO/GI |
| §5.5.1: Targeted occultations | N/A | WFI | GO |
| §5.5.2: Serendipitous occultations | N/A | WFI | GO/GI |
| §6.1: High-Latitude Survey (HLS) | N/A | WFI | GO/GI |
| §6.2: Microlensing survey | N/A | WFI | GO/GI |
| §6.3: Supernova survey | N/A | WFI | GO/GI |
| §6.4: CGI survey | N/A | WFI | GO/GI |
| §7.1: Giant planets | 1.5-10 | WFI/IFC | GO |
| §7.2.1: Tracking volcanoes on Io | 10 | WFI/IFC | GO |
| §7.2.2: Europa plumes | 10 | IFC | GO |
| §7.2.3: Clouds and surface features on Titan | 5.5 | IFC | GO |
| §7.2.4: Spectra of smaller satellites | 1.5-10 | IFC | GO |
| §7.3: Comets | ≤60 | WFI/IFC | GO/GI |
| §7.4.1: Asteroid families | ≤40 | WFI/IFC | GO/GI |
| §7.4.2: Active asteroids | ≤40 | WFI/IFC/CGI | GO/GI |
| §7.4.3: Jupiter Trojan asteroids | 10 | WFI/IFC | GO/GI |
| §7.5.1: Centaurs | 1.5-10 | WFI/IFC/CGI | GO/GI |
| §7.5.2: Kuiper Belt Objects | 1.5-10 | WFI/IFC | GO/GI |

**Table 3. Surface brightness estimates of extended Solar system targets**

| Target | $p_V$ | D (km) | r (AU) | Δ (AU) | 1.5 µm (Jy/arcsec$^2$) |
|---|---|---|---|---|---|
| *Giant Planets* | | | | | |
| Jupiter | 0.538 | 142984 | 5.09 | 4.45 | 29 |
| Saturn | 0.499 | 120536 | 9.53 | 8.92 | 7.7 |
| Uranus | 0.488 | 51118 | 19.51 | 18.90 | 1.8 |
| Neptune | 0.442 | 49528 | 29.89 | 29.29 | 0.69 |
| *Large Satellites* | | | | | |
| Io | 0.62 | 3643 | 5.09 | 4.45 | 34 |
| Europa | 0.68 | 3122 | 5.09 | 4.45 | 37 |
| Ganymede | 0.44 | 5262 | 5.09 | 4.45 | 24 |
| Callisto | 0.19 | 4821 | 5.09 | 4.45 | 10 |
| Rhea | 0.7 | 1527 | 9.53 | 8.92 | 11 |
| Titan | 0.22 | 5150 | 9.53 | 8.92 | 3.4 |
| Iapetus (dark) | 0.05 | 1470 | 9.53 | 8.92 | 0.77 |
| Iapetus (bright) | 0.5 | 1470 | 9.53 | 8.92 | 7.7 |
| *Comet Example* | | | | | |
| Coma | N/A | N/A | 3.0 | 2.5 | 0.19 |
| *Main Belt Asteroids* | | | | | |
| 1 Ceres | 0.090 | 939 | 2.91 | 2.32 | 15 |
| 2 Pallas | 0.101 | 545 | 3.39 | 2.70 | 12 |
| 4 Vesta | 0.423 | 525 | 2.16 | 1.40 | 127 |
| 10 Hygiea | 0.072 | 407 | 3.44 | 2.76 | 8.5 |

**Note.** Objects considered in this table are at least 2 WFI pixels in diameter and are therefore extended objects. For the "Comet Example," we take the coma/nucleus brightness ratio be 250, with information on the "Nucleus" found in Table 4. The heliocentric (r) and observer (Δ) distances were obtained using JPL Horizons, with the observer at the Earth-Sun L2 point in 2025; the observer distance corresponds to the smallest value in 2025 when the object is in WFIRST's field of regard. D is the diameter of the object and $p_V$ is the visible geometric albedo. The flux density at 1.5 µm ($\lambda_0$ of the W149 filter), in Janskys, was calculated by considering the reflected solar flux and the emitted thermal flux from each object. The solid angle, in units of square arcseconds, was calculated as $\pi(D/2\Delta)^2$. We assume the solid angle of an extended comet coma is 0.20 $arcsec^2$.

**Table 4. Brightness estimates of compact solar system targets**

| Target | $p_V$ | D (km) | r (AU) | Δ (AU) | 1.5 µm (mJy) |
|---|---|---|---|---|---|
| *Comet Example* | | | | | |
| Nucleus | 0.04 | 10 | 3.0 | 2.5 | 0.15 |
| *Asteroid Families* | | | | | |
| 153 Hilda | 0.062 | 171 | 4.17 | 3.50 | 18 |
| 434 Hungaria | 0.726 | 8.9 | 1.84 | 1.04 | 33 |
| 8 Flora | 0.226 | 147 | 2.36 | 1.63 | 692 |
| 298 Baptistina | 0.131 | 21 | 2.09 | 1.35 | 15 |
| 44 Nysa | 0.482 | 71 | 2.06 | 1.32 | 689 |
| 15 Eunomia | 0.248 | 232 | 2.36 | 1.63 | 1892 |
| *Jupiter Trojans* | | | | | |
| 624 Hektor | 0.025 | 225 | 5.15 | 4.51 | 4.9 |
| 617 Patroclus | 0.047 | 180 | 4.67 | 4.01 | 9.1 |
| *Saturnian Satellites* | | | | | |
| Mimas | 0.6 | 396 | 9.53 | 8.92 | 27 |
| Enceladus | 1.0 | 504 | 9.53 | 8.92 | 74 |
| Tethys | 0.8 | 1062 | 9.53 | 8.92 | 262 |
| Dione | 0.7 | 1123 | 9.53 | 8.92 | 256 |
| Hyperion | 0.3 | 270 | 9.53 | 8.92 | 6.3 |
| Phoebe | 0.08 | 213 | 9.53 | 8.92 | 1.1 |
| *Uranian Satellites* | | | | | |
| Miranda | 0.32 | 472 | 19.51 | 18.90 | 1.1 |
| Ariel | 0.39 | 1158 | 19.51 | 18.90 | 8.1 |
| Umbriel | 0.21 | 1170 | 19.51 | 18.90 | 4.4 |
| Titania | 0.27 | 1578 | 19.51 | 18.90 | 10 |
| Oberon | 0.23 | 1522 | 19.51 | 18.90 | 8.2 |
| Sycorax | 0.04 | 150 | 19.51 | 18.90 | 0.014 |
| *Neptunian Satellites* | | | | | |
| Triton | 0.72 | 2707 | 29.89 | 29.29 | 14 |
| Nereid | 0.16 | 340 | 29.89 | 29.29 | 0.051 |
| *Neptune Trojans* | | | | | |
| 2013 $KY_{18}$ | 0.08 | 200 | 29.04 | 28.43 | 0.0098 |
| (316179) 2010 $EN_{65}$ | 0.08 | 200 | 29.14 | 28.55 | 0.0097 |
| *Centaurs* | | | | | |
| (10199) Chariklo | 0.045 | 302 | 17.53 | 16.91 | 0.098 |
| (5145) Pholus | 0.044 | 190 | 30.30 | 29.69 | 0.0041 |
| (2060) Chiron | 0.15 | 166 | 18.42 | 17.82 | 0.080 |
| (60558) Echeclus | 0.077 | 59 | 13.41 | 12.81 | 0.019 |
| *Kuiper Belt Objects (KBOs)* | | | | | |
| (134340) Pluto | 0.68 | 2374 | 35.27 | 34.67 | 5.4 |
| Charon | 0.31 | 1212 | 35.27 | 34.67 | 0.64 |
| (136199) Eris | 0.96 | 2274 | 95.53 | 94.94 | 0.13 |
| (225088) 2007 $OR_{10}$ | 0.089 | 1964 | 89.60 | 89.00 | 0.011 |
| (136472) Makemake | 0.81 | 1713 | 52.71 | 52.11 | 0.66 |
| (136108) Haumea | 0.804 | 1366 | 49.88 | 49.27 | 0.52 |
| (50000) Quaoar | 0.109 | 1347 | 42.64 | 42.03 | 0.13 |
| (90377) Sedna | 0.32 | 1190 | 83.09 | 82.50 | 0.020 |
| (90482) Orcus | 0.231 | 1014 | 48.01 | 47.41 | 0.098 |

**Note.** The objects in this table are considered point sources because their apparent angular diameters are smaller than one WFI pixel. Values for the "Nucleus" in the "Comet Example" are assumed. The diameter of 617 Patroclus is the effective diameter of the binary system, and the geometric albedo is the system albedo. The heliocentric (r) and observer (Δ) distances were obtained using JPL Horizons, with the observer at the Earth-Sun L2 point in 2025; the observer distance corresponds to the smallest value in 2025 when the object is in WFIRST's field of regard. D is the diameter of the object and $p_V$ is the visible geometric albedo. The flux density at 1.5 μm ($\lambda_0$ of the W149 filter), in millijanskys, was calculated by considering the reflected solar flux and the emitted thermal flux from each object.

## 3. Desired mission enhancements

### 3.1. Non-sidereal tracking

WFIRST will excel in the serendipitous identification of new irregular satellites and minor bodies and will not require moving target tracking for this purpose. However, many of the other imaging and spectral investigations discussed in this paper require moving target tracking. Solar system objects moving less than 30 mas/s, the nominal tracking rate based on the value for JWST, can be tracked exactly without blurring the object on the detector. Objects moving faster than 30 mas/s, including near-Earth asteroids (NEAs) and long-period comets, can be tracked at the maximum rate, but will smear across the detector, even in the 2.7-second readout time for each individual WFI detector. As shown in Table 2, a 30 mas/s track rate (the maximum apparent rate of Mars) is adequate for most investigations discussed in this paper; about 10% of near-Earth asteroids (NEAs) and >50% of long-period comets (§7.3) would be unobservable at this rate. Increasing this rate to 60 mas/s would enable observations of >99% of NEAs and would significantly increase the observability of long-period comets, primitive remnants of the formation of the solar system. We understand that such a high track rate may not be achievable, but any increase over the nominal 30 mas/s rate would nonetheless help increase the number of NEAs and long-period comets trackable with WFIRST.

Without moving target tracking, targeted observations of all solar system objects, even the slowest moving objects in the Kuiper Belt, will be affected by blurring. Distribution of an object's flux over a larger number of pixels is not ideal, as this means more pixels must be included in the photometric extraction aperture, reducing the signal-to-noise ratio through an increase in read noise and the contribution of background sources (thermal, zodiacal, etc.). In Figure 6 we present the blurring of flux in individual pixels with and without moving target tracking of 30 mas/s. The WFI point spread function (PSF) is slightly larger than one pixel, but the results in Figure 6 can be applied to each individual pixel in the PSF, as well as each pixel in an extended object. Thus, this effect is even more significant for realistic point sources and extended objects. It would be important to commission moving target tracking at 30 mas/s, and possibly 60 mas/s, at the beginning of the mission so that WFIRST can carry out the full suite of solar system observations described in this paper, as well as complement JWST observations. Based on mission lifetimes, WFIRST may one day be the only operational space-based near-infrared facility, so reducing the constraints on observable moving targets will be necessary to enable a wide range of solar system studies into the 2030s.

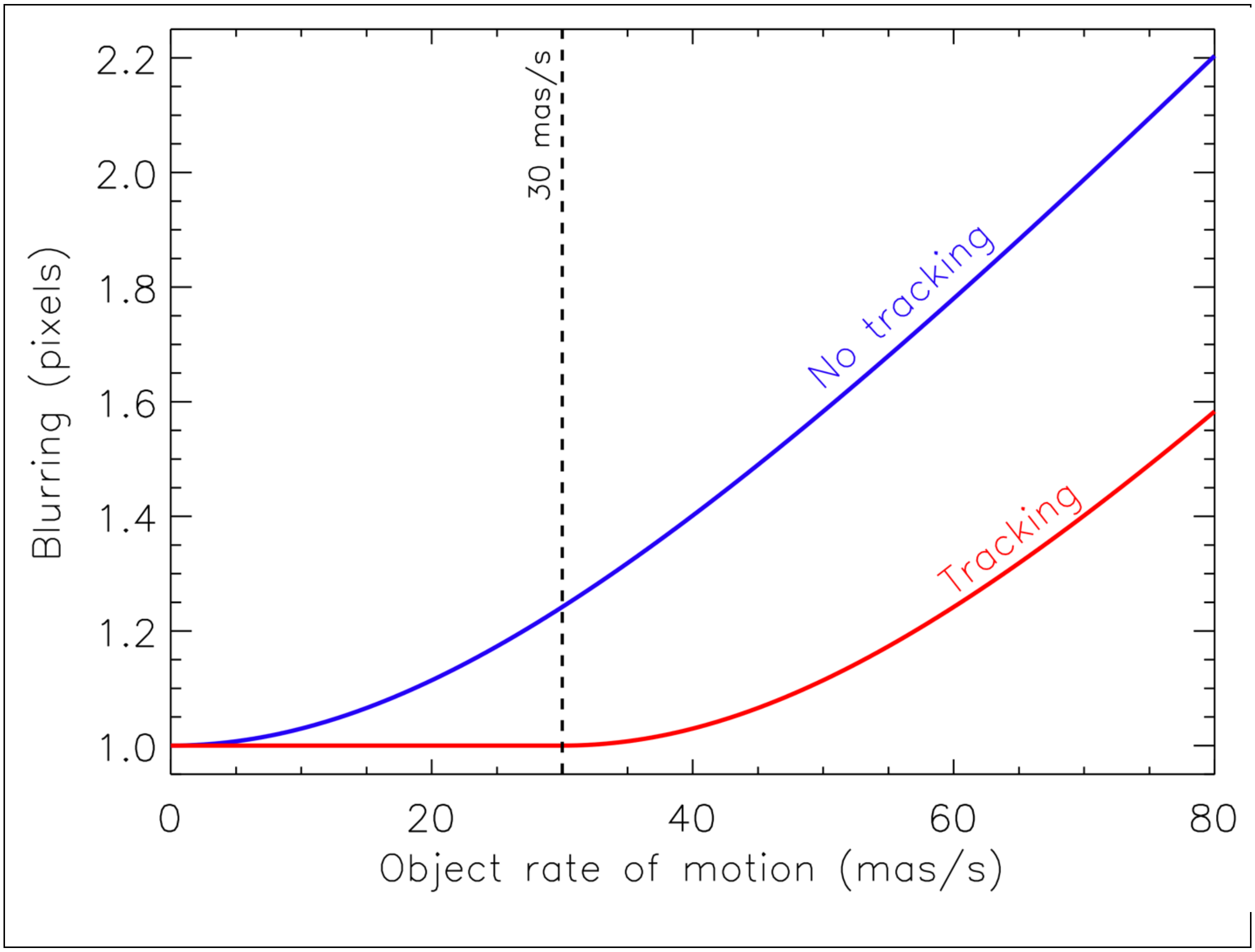

**Figure 6**: Blurring of the flux contained within one WFI pixel (0.11″/pixel) in one detector readout time (2.7 seconds) for the cases of no moving target tracking and moving target tracking of up to 30 mas/s. For example, a value of 2.0 pixels means that the flux that would be contained within any pixel of a stationary target's PSF is now evenly spread over 2 adjacent pixels, resulting in a reduction in the SNR when performing photometry. This of course assumes the target moves exactly in the *x*- or *y*-direction; the flux would be spread over additional pixels if the target moved at any intermediate angle, further reducing the SNR. The WFI PSF is larger than 1 pixel in all filters, but the blurring quantified here applies to every pixel in the PSF. Objects moving up to 30 mas/s can be tracked exactly, resulting in no blurring; objects moving faster than 30 mas/s, including near-Earth asteroids (NEAs) and long-period comets would experience significantly less blurring by making use of the maximum track rate. Observations of all objects, even those moving at the slowest rates, would be affected if moving target tracking is not supported for WFIRST.

### 3.2. K-band filter

The inclusion of a K-band (~2.0-2.4 μm) filter in the WFI filter wheel would result in a wider variety, as well as more comprehensive, solar system investigations. For instance, imaging in the K-band would provide higher contrast on Io between emission from the surface and emission from active volcanic eruptions (§7.2.1). A K-band filter would also support imaging to different depths within the atmospheres of the giant planets (§7.1) and Titan (§7.2.3) compared to the currently commissioned filters and provide a useful test of the color dichotomy among Jupiter Trojans identified in ground-based data (§7.4.3).

Two different groups of Jupiter Trojans have been identified based on their near-infrared colors, suggesting two different origins (§7.4.3). Only a few dozen Trojans have measured near-infrared colors, but further characterization of these groups and an evaluation of their origins with WFIRST would not be possible without a K-band filter. The groups were identified in J-K vs. 0.85 μm-J color space [5]; WFIRST can approximate these measurements with the Z087, J129, and a K-band filter to further evaluate the existence of the two-color classifications. Extending these measurements to smaller members of the Trojan population, an investigation not currently achievable with ground-based telescopes, is important for testing collisional theories and the origin of this population of minor bodies [6].

In general, observations through a K-band filter would support more accurate compositional determinations of fainter minor bodies for which spectroscopy is not feasible [7]. The strongest water ice absorption band spans ~1.9-2.1 μm [8] and presents the best opportunity to detect this ice species, even in smaller quantities on very faint minor bodies, and K-band photometry would provide a more definitive identification than photometry with the F184 filter (Fig. 7). For detections of water ice, the H158 filter covers a region that includes the 1.5 μm amorphous water ice and 1.65 μm crystalline water ice features, but these features are weaker than the 2.0 μm feature and will be harder to detect photometrically for fainter objects. The detection would be made more difficult if crystalline water ice is not present on the surface, as this feature would no longer be present in the spectrum. Together, the H158 filter and a proposed K-band filter could be used to make definitive determinations of water ice on outer solar system minor bodies.

Extending photometric measurements to these longer wavelengths would in turn extend the range of near-infrared spectral photometry. Comparison of spectral slopes provides useful information for comparing compositional properties between bodies. The wider the overall wavelength range, the more useful the comparison, since an ice or mineral present on one body but not another will be evident in the spectral slope. The K-band is an especially useful indicator due to strong absorption features of many potential surface components, including water ice, methane, pyroxene, and methanol. K-band photometry could therefore be the decisive measurement for determining whether or not two bodies have similar surface compositions.

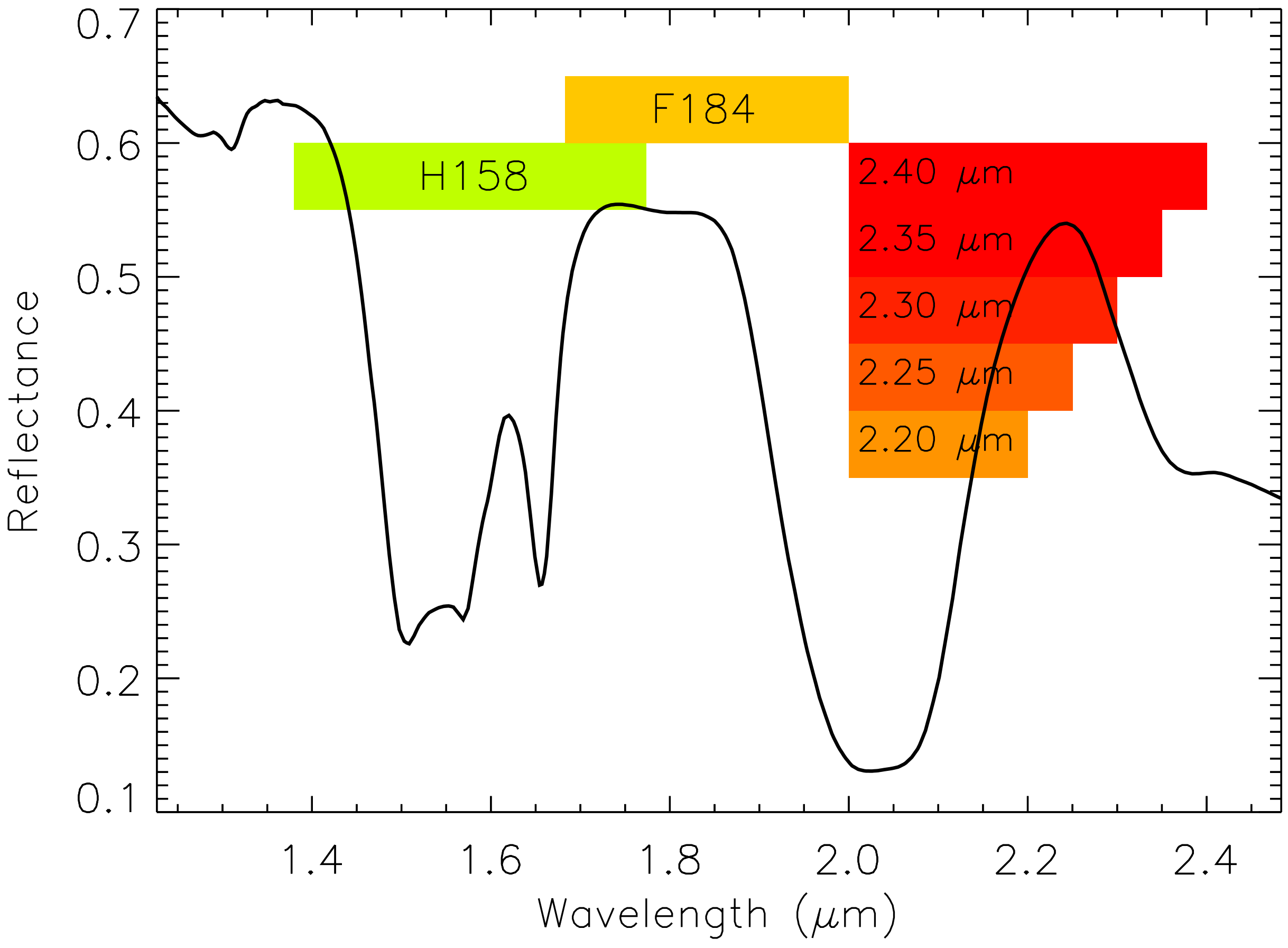

**Figure 7**: Near-infrared spectrum of crystalline water ice measured in a laboratory setting at 40 K [8]. Only a handful of spectra of minor bodies and satellites in the outer solar system exhibit strong water ice absorption features. It is more common for the spectra of these objects, if water ice is present, to show a broad, shallow absorption feature at 2.0 μm, and very little detectable absorption at shorter wavelengths [9,10]. Without a K-band filter, constraining the presence of water ice on small outer solar system bodies with WFIRST will not be possible. The various options for WFIRST K-band filters (Table 5) are identified by their maximum wavelengths and plotted against the H158 and F184 filters for comparison.

Table 5 presents the limiting AB magnitudes (roughly equivalent to V magnitude) and sensitivities for 5 different K-band filter options. These numbers were calculated for a telescope operating temperature of 260 K. The aperture for the extended sources is circular with a half-light radius of 0.3″. Comparison of the numbers in the third column of Table 5 to Table 3 and the fifth column of Table 5 to Table 4 shows that any of the K-band filter options could be useful for observations of objects throughout the solar system. Any of these filters would be especially useful for extended sources, since extended sources in the solar system tend to be very bright compared to point sources; very high SNRs will be achievable on the giant planets, their large satellites, and bright main belt asteroids.

The solar system community will be able to make use of a visible-band filter if that is chosen instead of a K-band filter. (The most important WFIRST capability for solar system observations is moving target tracking, discussed in §3.1.) Such observations with a visible band filter would be comparable to those with HST and include searching for and characterizing satellite orbits around minor bodies due to higher flux and tighter PSF FWHM in the visible [11-13], atmospheric mapping on the giant planets [14], and the study of activity around asteroids, Centaurs, and comets, to name a few potential investigations.

**Table 5. Estimated K-filter sensitivities**

| Wavelength range (μm) | AB mag (extend. source, 3600s, 5σ) | Surface brightness (Jy/arcsec$^2$) | AB mag (pt. source, 3600s, 5σ) | Flux density (mJy) |
|---|---|---|---|---|
| 2.00-2.20 | 25.086 | $1.19 \times 10^{-6}$ | 26.082 | 0.00013 |
| 2.00-2.25 | 25.027 | $1.25 \times 10^{-6}$ | 26.019 | 0.00014 |
| 2.00-2.30 | 24.935 | $1.36 \times 10^{-6}$ | 25.922 | 0.00016 |
| 2.00-2.35 | 24.829 | $1.50 \times 10^{-6}$ | 25.812 | 0.00017 |
| 2.00-2.40 | 24.710 | $1.68 \times 10^{-6}$ | 25.689 | 0.00019 |

### 3.3. Parallel observations

Assuming there will be no data volume issues, the ability to propose for parallel observations through the GO program would benefit efforts for serendipitous detection of minor bodies. Specifically, use of the WFI during fixed target observations with the IFC or CGI would provide coverage over the adjacent 0.28 deg$^2$ WFIRST FOV for analysis of the presence of moving targets. Longer exposures, necessary for the identification of fainter objects, will result in streaking of moving targets, allowing for quick identification (but lower-SNR photometry). In a 1000-second exposure, objects at 0.5, 3.0, 4.2, 14.0, 29.0, and 49.0 AU will move approximately 1085, 74, 45, 8, 3, and 1 pixels, respectively. (These distances correspond to typical observer distances for NEAs, main belt asteroids, Jupiter Trojans, Centaurs, inner-edge KBOs, and outer-edge KBOs, respectively.) Estimated diameters of minor bodies in a 1000-second exposure through the various WFI filters can be found in Figure 3. Multiple 1000-second exposures with the WFI during especially long exposures on a fixed target with the IFC or CGI would be required to confirm movement for more distant KBOs (≥50 AU). With the right analysis tools, images taken in parallel will be useful for photometry and astrometry of serendipitously detected minor bodies from the inner to the outer solar system.

### 3.4. Detector subarrays

A handful of science investigations, including targeted occultations (§5.5.1) and imaging of Jupiter and Saturn (§7.1) would make good use of imaging subarrays. Targeted occultations require faster readout times than the 2.7-second readout for an individual WFI detector and options for shorter exposure times are necessary to prevent saturating on Jupiter and Saturn. Possible subarray options could include 512×512 (for full-disk imaging of Jupiter and Saturn), 64×64 (for imaging portions of Jupiter and Saturn), and 16×16 (for fast readout of a point source). Intermediate options could also be useful depending on the readout times. Not all 18 of the WFI detectors would need to be configured with subarrays; in fact, only one would be necessary.

### 3.5. Target of opportunity (ToO) programs

ToO programs are standard at many observatories, both ground- and space-based, and we would like to see them made available for WFIRST. Quick follow-up observations (within 1 week, with faster turnaround for faster-moving objects) of newly discovered objects reported by WFIRST, LSST, and other facilities would enable orbit calculations and the initial study of physical properties. Such objects would include NEAs, comets, main belt comets, Trojan asteroids, Centaurs, and KBOs; follow-up of NEAs would require a fast turnaround time of a few days, while KBOs could be followed up a week later. A compelling use of ToO programs is for the study of the relatively new population of extrasolar asteroids on hyperbolic orbits, such as 1I/2017 U1 ('Oumuamua) [15]. The observing window for these objects is very short due to their unusual orbital inclinations, small sizes, and high velocities as they travel through the inner solar system. With the availability of ToO programs and proper planning, WFIRST, and the high-sensitivity WFI in particular, will be capable of following up on future extrasolar asteroid detections for longer periods than any other facility.

ToO programs would also be useful for triggering observations of transient events such as Centaur outbursts, fragmentation events, onset of cometary activity, volcanic eruptions on Io, and impacts on asteroids and the giant planets (a few days); and for occultations that may be uncertain at the time of proposal submission (~2-4 weeks). It is unlikely that ToO activations could occur during the microlensing survey, but we would request that they be able to interrupt any of the other four astrophysics surveys (the CGI exoplanet characterization, HLS imaging, HLS spectroscopy, and supernova surveys).

### 3.6. Large programs

Targeted survey programs would benefit from the availability of a handful of large programs during each cycle. Such programs would enable targeted surveys in the vicinity of the ecliptic for detection of asteroids, Trojans, Centaurs, KBOs, and Jupiter Family Comets (JFCs). Coverage of a few contiguous WFI fields with a cadence of a few hours and a duration of a few weeks could probe very deep in a less-crowded field than the one covered by the microlensing survey (§6.2). Such large programs are currently offered by HST and the Gemini Observatory, and are planned for JWST starting in Cycle 1.

## 4. Solar system science with astrophysics assets

### 4.1. Public interest in solar system science

WFIRST, in the tradition of other NASA astrophysics assets, is primed to deliver observations of solar system objects that will not only contribute to the advancement of the field, but also excite the public. As seen in Figure 8, solar system observations performed with HST account for a proportionally larger share of the total number of press releases compared to the time awarded to these programs. Solar system observations therefore contribute in a very significant way to the public profile of a space telescope mission, while only accounting for a small percentage of the total observing time (in HST Cycles 22-24, solar system observations accounted for only ~2.3% of the total orbit allocation and ~15.4% of the press releases).

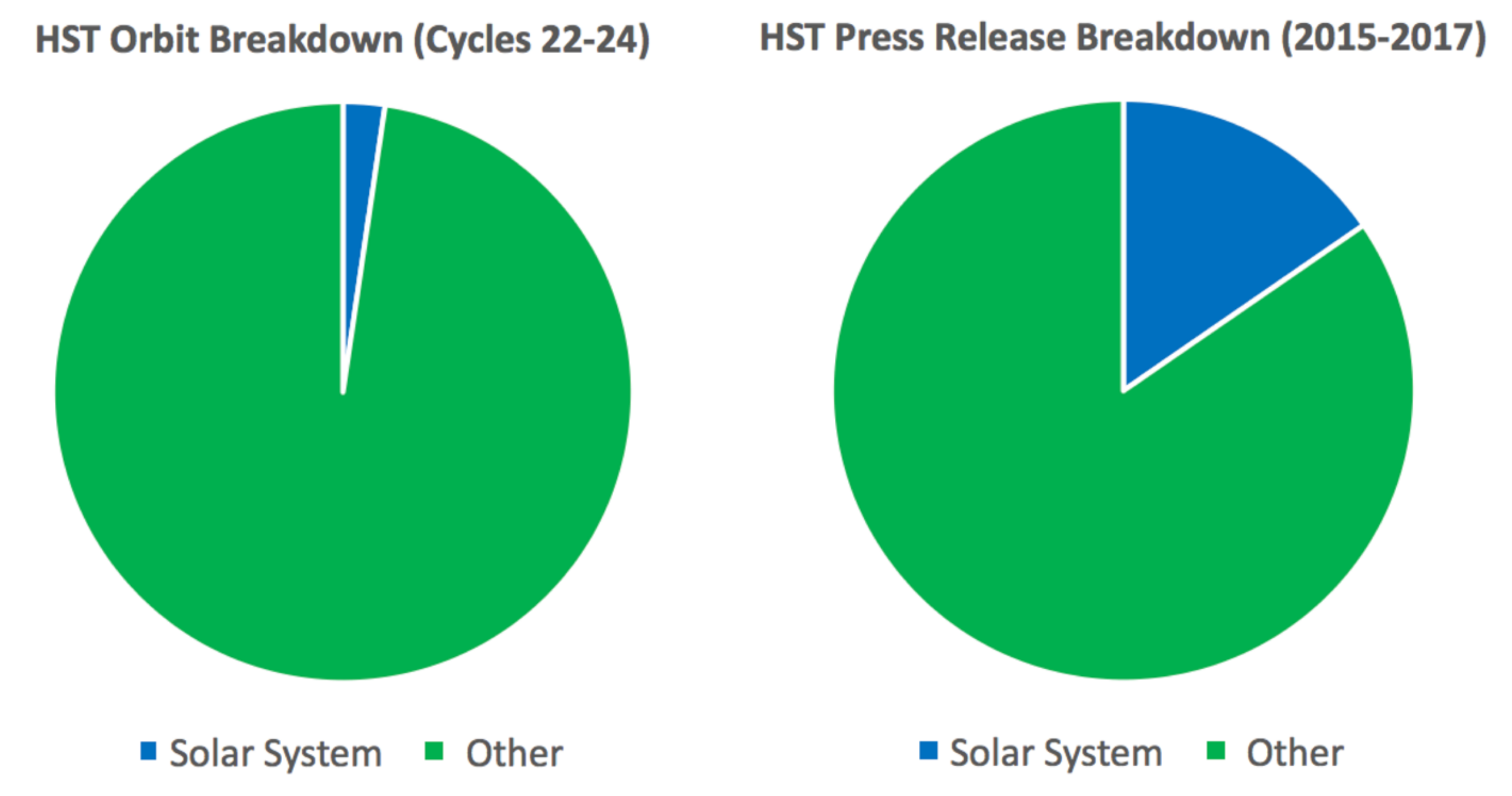

**Figure 8**: (left) Pie chart of orbit breakdown by science category for Cycles 22-24 (Oct. 1, 2014-Sept. 30, 2017). The small blue wedge represents the number of HST orbits for solar system observations, while the large green portion represents all other science categories. Solar system observations accounted for ~2.3% of the total orbits in this time span. (right) Pie chart of the press release breakdown by science category for the period covering 2015-2017, the only period for which reliable data exist. The same color scheme is used in both pie charts. This date range primarily contains press releases from programs executed in Cycles 22-24. Solar system science accounts for a disproportionately large fraction of the number of press releases (and therefore public outreach and awareness) compared to the orbits granted for solar system observing programs: ~15.4% vs. ~2.3%.

**4.2. Previous use of astrophysics assets**

The past quarter century has seen an increase in the number of astrophysics assets located in space. Furthermore, there is an ongoing shift from locating these facilities in low-Earth orbit to the Earth-Sun L2 point. This is due to the need to place these infrared facilities in a stable thermal environment away from the thermal pollution that pervades the near-Earth region and is indicative of the desire in many fields of astronomy for observations at longer wavelengths. Solar system science is one such field, and we will look to use WFIRST in the same manner as the other astrophysics assets before it.

The most successful of these previous astrophysics assets, both in terms of longevity and output, is the Hubble Space Telescope. HST has made significant contributions to our understanding of the solar system, from the atmosphere of Venus to the surface compositions of KBOs, but one area in which HST has really excelled is the detection of new satellites, especially in the outer solar system. These discoveries include two new classical Uranian satellites [16], one new classical Neptunian satellite [17], the identification and characterization of dozens of KBO binaries [18-21] a hierarchical triple system in the Kuiper Belt [22], and the recent detection of a satellite around each of the dwarf planet-class objects Makemake [23] and 2007 $OR_{10}$ [24]. In the case of the ice giants, these new, small satellites paint a more complicated picture of satellite origins and evolution: The small Uranian satellites point to a dynamic picture of moon-ring interactions [25], while the new Neptunian satellite provides evidence that the capture of Triton resulted in the destruction of the original inner satellites and accretion of the current satellites [26]. For the KBOs, calculating the orbit of a satellite provides an estimate of the system mass, which tends to be dominated by the primary. The total mass of the Kuiper Belt is currently not well-constrained because the size distribution and binary fraction of KBOs are only poorly understood. The mass of the current Kuiper Belt is important for comparison to the predicted mass of the primordial Kuiper Belt: How much material was lost from the solar system or captured into other minor body populations? WFIRST will be able to contribute to this topic, both through detection and characterization of more wide KBO binary systems and colorimetry of other minor body populations.

Some of these satellite detections were made in the Pluto system. Of Pluto's 5 satellites, only one, Charon, was discovered using a different facility [27]. The 4 minor satellites, Styx, Nix, Kerberos, and Hydra, were all discovered in deep images from HST [11,28,29]. Other important observations of the Pluto system were made with HST, including surface maps of Pluto and Charon [30] that provided valuable pre-flyby context for NASA's New Horizons mission and the discovery of potential chaotic rotation of the minor satellites [31]. The target of the New Horizons extended mission, 2014 $MU_{69}$, was also discovered using HST [32].

Astrophysics assets have provided valuable support and follow-up for other planetary missions. NASA's Spitzer Space Telescope obtained mid-infrared spectra of the Deep Impact target, comet 9P/Tempel, shortly after the impactor hit the surface [33]. These spectra provided valuable information on the composition of the excavated material, and the chemical make-up of the comet as a whole. Spitzer was also responsible for the discovery of Saturn's largest, most diffuse ring, the Phoebe ring, that went undetected by the in-situ Cassini spacecraft [34]. But not all observations in support of a planetary mission have been simultaneous. A tentative detection of water vapor around Ceres was reported from far-infrared observations made with the Herschel Space Observatory prior to the Dawn spacecraft's arrival in 2015 [35]. This detection supports the idea that some main belt asteroids formed at larger heliocentric distances and migrated inwards. HST provided long-term follow-up observations of images from Voyagers 1 and 2 of Titan and Triton, resulting in the detection of seasonal changes on both satellites [36,37]. (The Titan observations were made only a few months after the launch of HST in 1990.) Other HST observations have provided context and momentum for future missions, such as Europa Clipper, with the recent detection of water ice plumes around Europa [38,39]. WFIRST is set to operate during some of these future missions and will therefore be able to provide simultaneous support observations, including hazard mitigation, and possibly generate momentum for the next generation of planetary missions in the 2030s and 2040s.

Large observing programs with astrophysics assets have enabled the study of sizeable samples of minor body populations. Expansive, consistent data sets were produced with Spitzer, Herschel, the Wide-field Infrared Survey Explorer (WISE), and the repurposed NEOWISE mission (an ongoing, as of early 2018, extension to the WISE mission following the depletion of coolant) [40,41]. The "TNOs are Cool" Herschel program was a survey of KBOs at far-infrared wavelengths for the purpose of calculating diameters and albedos [42]. At the time of this writing, 13 papers based on these data have been published. Two of those papers combined Spitzer and Herschel data to better constrain the thermal properties of Centaurs and KBOs [43,44]. Spitzer also produced a consistent data set for the study of the size and activity of a significant fraction of the known JFC population, resulting in a new classification scheme for comets [45,46]. WISE, and the later NEOWISE extension, were dedicated survey missions in the far-infrared that observed minor bodies from a wide variety of different populations, including ~100,000 main belt asteroids, 5 active asteroids, 163 comets, and 52 Centaurs and KBOs [47-49]. According to the MPC, WISE is credited with the discovery of 3832 numbered minor bodies, including the first and only known Earth Trojan asteroid [50]. Earth Trojans are difficult objects to detect from the surface of the Earth because they typical transit ~2 hours before sunset or after sunrise (§5.1). Surveys with astrophysics assets can also operate in the time domain. An ongoing Hubble 2020 program, the Outer Planet Atmospheres Legacy (OPAL), will obtain complete maps of each of the giant planets once a year for every year that HST continues to operate in order to resolve long-term atmospheric changes, such as the evolution of features in Jupiter's atmosphere [14].

Some solar system observations made unexpected use of astrophysics assets or produced very surprising results. For instance, X-ray emission from Pluto was detected using the Chandra X-ray Observatory, but this detection at such a large distance from the Sun (~32 AU) remains a mystery [51]. Active comets are known to produce X-rays and were previously studied with Chandra [52]. Also, at a heliocentric distance of 32 AU, but at longer wavelengths, Herschel provided an albedo measurement of comet Hale-Bopp ~15 years after perihelion; the albedo was much higher than that of a typical comet, suggesting freeze-out of fresh material onto the comet nucleus [53]. Other novel investigations have been performed with the Ultraviolet/Optical Telescope (UVOT) on the Swift Gamma-Ray Burst mission, including a calculation of the rotation period of the dwarf planet Eris [54], observations of active asteroids [55], and observations of comet C/2013 A1 (Siding Spring) during its close approach to Mars [56]. Sometimes unexpected opportunities present themselves, as is the case with the repurposed Kepler mission, now known as K2. Following the loss of half of the spacecraft's reaction wheels in 2013 (necessary components for accurate pointing of the telescope) the mission was altered so that it could only observe fields along the ecliptic. This resulted in high-cadence time-domain observations of solar system objects such as the rotation light curves of Jupiter Trojans [57], the KBO 2007 $OR_{10}$ [58], the irregular satellites of Uranus [59], and Neptune's atmosphere [60]. The slow rotation period observed for 2007 $OR_{10}$ led to a search for a satellite that was ultimately discovered in archival HST images [24] and the Uranus field also yielded observations of over 600 main belt asteroids, including 86 new light curves [61]. The solar system community, drawing on these previous experiences, is prepared to develop clever investigations and take advantage of unexpected opportunities with WFIRST.

### 4.3. Synergies with JWST, LSST, and NEOCam

It is anticipated that both JWST and the Large Synoptic Survey Telescope (LSST) will be operational during at least a portion of WFIRST's 6-year nominal mission beginning in the mid-2020s. The goal for JWST is a 10-year mission beginning in Fall 2020, and therefore has the potential to operate in conjunction with WFIRST at L2 for up to 3-4 years. (JWST's nominal mission is from 2020-2025, which could result in no overlap with WFIRST.) The plan for LSST is to begin its primary 10-year survey in 2022, resulting in simultaneous operations over a large portion of WFIRST's nominal lifetime. NEOCam currently has no project timeline but could conceivably be operational during the WFIRST nominal mission.

*4.3.1 JWST*

WFIRST and JWST provide complementary, rather than overlapping, capabilities to each other. Though both will orbit around the Earth-Sun L2 point, the FOR of WFIRST is nearly 50% larger than that of JWST (54°-126° compared to 85°-135°). Assuming both will be operating at the same time, an object on the ecliptic will first be observable by WFIRST up to one month prior to being observable by JWST, depending on orbital geometry. Then the object will be observable by both for ~1.5 months, followed by 1-2 weeks when it is only observable by JWST. Then the process repeats in reverse for the trailing window. Together, WFIRST and JWST will be able to observe solar system objects along the ecliptic for nearly half of each year.

JWST has a larger primary mirror (~7 times the light gathering power), a broader wavelength range (0.6-28.8 μm), and a handful of different spectroscopy modes, while WFIRST's major strength for solar system observations is its 0.28 $deg^2$ FOV (Fig. 2) and higher imaging, spectroscopic, and coronagraphic sensitivity. The science cases for these two observatories are, again, complementary: the JWST cases [4] typically focus on a handful of objects from important populations and are heavy on spectroscopy, while the WFIRST cases are directed more at surveys, larger sample sizes, and serendipitous detections with its considerable imaging FOV (see §5 and §6).

During the potential period of simultaneous operations, the two observatories could work together to increase the science gains from both. Time-domain studies of variable objects, such as comets, can be "handed off" as the object passes through the FOR of each observatory, as discussed above. WFIRST is likely to identify more minor bodies than JWST, with the brighter objects being passed to JWST for spectral observations. With the proper orbital alignment, one could even imagine WFIRST and JWST being used simultaneously for observing a stellar occultation by a minor body. Many other interesting uses of WFIRST and JWST either in tandem or in sequence may be developed and could someday lead to extraordinary discoveries in our own solar system.

*4.3.2. LSST*

The LSST is an 8.4-meter ground-based telescope currently under construction on Cerro Pachón in Chile. Beginning in 2022, LSST will undertake a 10-year survey of the entire southern sky, and a small portion of the northern sky along the ecliptic, covering a total of approximately 25,000 $deg^2$. As currently planned, the same area of sky will be observed every 3-4 days (the exact cadence is still open to community input). Covering such a large area of sky in such a short time period requires (1) a large FOV and (2) short exposures. LSST accomplishes this with a FOV of 9.6 $deg^2$, or ~34 times larger than the WFI FOV, and two 15-second exposures through each of the 6 available SDSS filters (*ugrizy*) at each sky position. The 5-σ limiting magnitudes per visit are expected to be: $u$=23.9, $g$=25.0, $r$=24.7, $i$=24.0, $z$=23.3, and $y$=22.1 [62]. Objects brighter than V~16, including all the planets, will saturate the LSST detector; an estimate of the saturation limit for the WFI is not currently available.

The proposed cadence of the LSST survey will provide a unique opportunity for the identification and study of minor bodies in the solar system, including improved orbit determinations, colorimetry, and rotation light curves. Consistent monitoring of small body light curves will enable the early detection of transient events such as outbursts, fragmentation, and other unexpetected activity. Specific objectives include the identification of additional objects on hyperbolic orbits [15], taxonomic classification of NEAs, secular changes in color and rotation period for active asteroids and comets, identification of new Trojan asteroids of Mars and the giant planets, and detection of additional "extreme" Kuiper Belt Objects, to name a few [63]. However, the downside to the LSST's fast cadence is that the individual observations themselves will not probe very deep. Identification of new, faint objects could be quickly followed-up with deeper observations by WFIRST, assuming an appropriate target of opportunity (ToO) program is in place (§3.5). Such a ToO program should be capable of triggering within 1 week from the desired start time of the observations (depending on the object's apparent rate of motion) to prevent loss of the newly identified object.

An interesting prospect with LSST is the identification of dwarf planet-class Kuiper Belt Objects (KBOs) passing through the galactic plane. These regions are typically avoided in targeted surveys and a ~1000-km diameter KBO could conceivably be detected in 30 seconds of imaging with LSST; this assumes adequate subtraction of field stars in the LSST pipeline. Even with large orbital uncertainties shortly after detection, the 0.28 $deg^2$ FOV of the WFI would have no issues acquiring the object for the necessary follow-up observations, or in removing the background stars due to the stable PSFs provided by space-based observations. Such observations would include deeper imaging to gain a better understanding of the physical properties of the KBO.

The SDSS filters on LSST will cover the near-UV to near-infrared range from 0.32-1.05 μm, with WFIRST covering the near-infrared range from 0.6-2.0 μm. The overlap between these two wavelength ranges provides a means of calibrating the data sets for full spectral photometry between the near-UV and the near-infrared.

Another important synergy between WFIRST and LSST is complementary sky coverage. A major detractor for both WFIRST and JWST is their inability to observe solar system objects at opposition, when they are at their brightest. This is especially important for faint minor bodies that may be more difficult to observe at other solar elongation angles, and for phase angle studies. LSST potentially fills that gap by providing coverage for objects at opposition, as long as they are observable in the southern half of the sky.

*4.3.3. NEOCam*

WFIRST may also provide complementary observations to those of the Near-Earth Object Camera (NEOCam). NEOCam was not officially selected in the previous round of Discovery-class missions but was provided with an additional year of funding in January 2017 to explore possibilities for future implementation as part of an extended Phase A study. NEOCam was designed specifically to identify and characterize a larger number of currently unknown potentially hazardous asteroids (PHAs) and has the capability to measure the diameters and thermal inertias of these objects [64]. As originally designed, NEOCam images at infrared wavelengths from 4-5 μm and 6-10 μm. Simultaneous surveys by LSST, WFIRST, and NEOCam would therefore cover a very large region of the sky from the visible into the mid-infrared. NEOCam would also be able to observe objects moving at very high apparent rates that cannot be tracked by WFIRST. The combination of LSST's fast cadence, WFIRST's field of view and field of regard, and NEOCam's tracking capability would potentially result in a complete catalog of PHAs that could threaten life on Earth.

## 4.4. Planetary mission support

NASA is currently developing a handful of new planetary missions in the Discovery, New Frontiers, and Flagship categories that will be operational in the 2020s and beyond. These missions are in various stages of development, and WFIRST will be able to provide varying levels of support for most, if not all, of them. Missions planned for the 2030s would benefit from observations for the purpose of defining science goals; currently operating missions would benefit from complementary observations made by WFIRST; and for missions nearing the end of their nominal operation period, WFIRST would provide valuable information for guiding extended mission proposals. A perfect example of a NASA astrophysics asset supporting a planetary mission was the discovery of the New Horizons extended mission target, 2014 $MU_{69}$, by HST [32]. WFIRST's deep imaging capability may someday do the same for a future, as yet unplanned, flyby mission to the outer solar system.

WFIRST will be in a position to provide valuable context for the results of the Discovery-class mission Lucy, planned to launch in 2021. From 2025 to 2033, Lucy will encounter 1 main belt asteroid and 5 Jupiter Trojans [65]. The Trojan targets include 3548 Eurybates, the largest member of a collisional family of Trojan asteroids; 11351 Leucus, an object with a rotation period potentially greater than 500 hours [66]; and 617 Patroclus, a binary Trojan with components of roughly equal size [67]. Observations with WFIRST would place the flyby results in context through further study of the Trojan population as a whole (§7.4.3). WFIRST could contribute to our understanding of the origins of these minor bodies, identify more collisional families through improved orbital solutions and colorimetry, identify new binaries, and measure rotation periods for a larger sample of Trojans. WFIRST should also be able to efficiently probe the relevant regions of sky to detect Jupiter Trojans down to very small sizes in order to provide additional remote follow-up targets for Lucy.

Two New Frontiers mission concepts were down-selected in December 2017 for further study and one mission will be selected for execution in July 2019 with launch planned for 2024/2025: Dragonfly [68] and the Comet Astrobiology Exploration Sample Return (CAESAR) [69]. The Dragonfly concept is to place a science payload into a multi-rotor vehicle that can vertically take off and land on different locations of Titan's surface. The primary science objectives are to follow-up on the results of the Huygens probe by studying the composition of Titan's atmosphere, surface, and sub-surface. Dragonfly would arrive at Titan in 2034, well into any WFIRST mission extension. WFIRST would therefore be able to provide evidence of seasonal changes on regional scales during Titan's northern hemisphere Fall in the late 2020s (§7.2.3) and provide valuable information about potential landing sites for Dragonfly. The CAESAR concept is to visit comet 67P/Churyumov-Gerasimenko in 2029 and return a sample of the comet to Earth in 2038. This comet was previously visited by the European Space Agency's Rosetta spacecraft and the rationale for returning to this comet is to use that prior information to guide spacecraft and instrument design in order to increase the probability of success. By the time CAESAR arrives at 67P, the comet will have undergone 2 full revolutions around the Sun since the Rosetta mission. Comparison of ground-based observations and those expected from JWST and WFIRST (§7.3.2 and §7.3.3) with the Rosetta in-situ data will facilitate an assessment of the environment surrounding the comet, which is important for evaluating the safety of the spacecraft.

WFIRST would also provide useful observations in support of NASA Flagship missions, like the Europa Clipper [70]. Set to launch in the early 2020s and arrive at Jupiter in the mid-2020s, the purpose of this mission is to study the potential habitability of Europa's sub-surface ocean through compositional and geologic analysis. Europa Clipper will orbit Jupiter and make multiple flybys over the surface of Europa. WFIRST will be able to provide support for these science goals through time-domain, longitudinal studies of Europa's surface composition. WFIRST may be able to indirectly identify locations of recent or ongoing plume activity on Europa through spatially-resolved spectroscopy of the surface with the IFC: regions with recent activity will be rich in salts and have weaker water ice absorption (§7.2.2). Longitudinal coverage over a large time baseline would help to identify regions of interest for Europa Clipper to focus on during its flybys over Europa's surface. (The JWST NIRSpec instrument will be better suited for spectral observations of Europa, but they will also be possible with the WFIRST IFC.)

The planets Uranus and Neptune have not been visited by spacecraft since Voyager 2 flew by them in 1986 and 1989, respectively. This fact was recognized in the 2013-2022 Decadal Survey, prompting the development of a science definition team and a report on possible Flagship missions to visit one of these giant planets. Flight paths and science goals were identified for a Flagship mission to each planet. A mission to Uranus would launch between 2030-2034 and arrive between 2041-2045; a mission to Neptune would launch between 2029-2030 and arrive between 2042-2043. By the time the WFIRST mission begins, the target, Uranus or Neptune, will have been chosen for the Flagship mission, and WFIRST observations could potentially help guide the specific science objectives and flight paths of each mission. For example, WFIRST's wide field of view will provide deep imaging of the Hill spheres of these giant planets in only a handful of separate pointings, contributing to the inventory of irregular satellites (§5.3). Discovery and orbital characterization of a new irregular satellite (perhaps one that is a different color from other irregular satellites in the system) may prompt a flight path that takes the spacecraft on a close approach to the object, as was done with Cassini for Saturn's irregular satellite, Phoebe.

## 5. High-impact science investigations

### 5.1. Earth Trojan asteroids

The Trojan asteroids of Jupiter and Neptune are valuable tracers of solar system dynamical evolution, providing information specifically about the outward migration of Neptune [71,72]. The origin of the Mars Trojan population is currently up for debate, but they could also be a population of objects captured into their current orbits early in solar system history [73]. The currently identified Venus and Uranus Trojans are thought to be minor bodies from other populations captured into temporary Trojan orbits [74,75]. No Saturn Trojans have been heretofore identified but are thought to be highly unstable due to the gravitational influence of Jupiter; the lack of detections to this point is not surprising given their short residence times [76]. To date, one Earth Trojan has been discovered and its stability remains a complicated issue [50,77]. Earth Trojans are notoriously difficult to detect from the ground due to their small solar elongation angles, which restricts observability to twilight at high airmasses. This particular object, 2010 $TK_7$, was detected by WISE, a survey telescope, and future identifications will likely need to wait for the launch of WFIRST.

LSST is set to increase the number of known Mars, Jupiter, Uranus, and Neptune Trojans significantly and is likely to discover the first Saturn Trojans, if they exist. Spectroscopy and imaging of Trojans (and all minor bodies, for that matter) with well-characterized orbits will be one of the strengths of JWST. With these niches filled, WFIRST will be best suited for surveying the L3, L4, and L5 points of Mars and the giant planets. However, LSST will not operate during the day and JWST has a more restrictive field of regard (85°-135°) than WFIRST, preventing observations almost everywhere within its own orbit, including observations of Earth Trojans. HST could conceivably search for additional Earth Trojans, but the search area is large and would need to be covered by the relatively small field of view of ACS or WFC3 (Fig. 2). Only a few images would be obtainable during each orbit (<1 hr) and the targets are relatively faint. The maximum brightness of 2010 $TK_7$ is V~21.2 and, based on the fact that it is the only known Earth Trojan, is likely to be one of the larger members of the population. In other words, an HST survey would be prohibitively time-consuming. WFIRST's large FOV and its more favorable field of regard (Fig. 4) make it the ideal (and, in fact, only) facility for undertaking a search of the Earth-Sun L4 and L5 points. Placement of WFIRST at the Earth-Sun L2 point (~1.01 AU) means that an object exactly at the L4 or L5 point (60° ahead of or behind the Earth) will be at a solar elongation angle of ~59°; the minimum solar elongation that WFIRST can observe is 54°, resulting in a ~5° buffer. (In reality, Trojan asteroids of any planet would librate around the L4 or L5 point, resulting in periods when the object would not be within the WFIRST field of regard and other periods when it would be present in the field of regard for a longer duration. We present the calculation for an object exactly at the L4 or L5 point for illustrative purposes.) The detection and study (colorimetry, spectroscopy, astrometry, etc.) of Earth Trojans is therefore an investigation unique to WFIRST, with every other major facility on the ground or in space unable to observe the L4 and L5 points for various reasons. This high-impact investigation requires moving target tracking based on the apparent rate of motion of 2010 $TK_7$ of ~25-90 mas/s.

Dynamical modeling suggests that Earth Trojans are not as rare as current observations might imply [77]. Stability around the L4 and L5 points exists for orbital inclinations of 0°-20° and between 28° and 40°. The stability for more highly inclined Trojans is potentially very long. WFIRST will play a very important role in the discovery of additional Earth Trojans, both because it is the only facility that can do so and because of the WFI's large FOV. The faintness of the targets and the large distribution of orbital inclinations means that a large area of sky needs to be covered to a decent depth (V~21-24). Near-infrared colors and spectra of the Earth Trojans will also help inform us about their origins and dynamical evolution, since color and surface composition are good indicators of a minor body's formation environment.

Discovery of one or more Earth Trojans in orbits that are stable over long time periods (comparable to the age of the solar system) would be especially significant. Such a body could be from the same population as the Jupiter and Neptune Trojans and would be relatively easy for a spacecraft to reach, regardless of the object's inclination. Any Earth Trojan would be in the plane of the Earth's orbit once every six months, resulting in the possibility of multiple flybys as part of a Discovery-class mission. The in-situ study of one or more Earth Trojans thought to originate in the outer solar system would be transformative for our ideas of how the solar system formed and evolved.

### 5.2. Binary asteroids with the CGI

The CGI can be effectively used for observations of solar system objects, particularly for observations of binary asteroids and the detection of activity around asteroids (§7.4.2) and Centaurs (§7.5.1). In this section, we focus on CGI observations of main belt asteroids in order to identify or confirm the presence of satellites. The study of binary asteroids is important for understanding the formation and physical processes that operated, or continue to operate, on these inner solar system minor bodies. Characterization of the satellite's orbit can provide mass and density estimates for the primary; the primary's porosity can also be constrained when taking into account information from rotation light curves. The CGI requires that targets be brighter than V=14, rendering similar CGI observations of many Jupiter Trojans, Centaurs, and KBOs impossible.

The wavelength range of the CGI is 0.43-0.98 μm and the inner working angle (IWA) is 0.15″. (For JWST, the smallest achievable IWA is with the NIRCam short-wavelength bar coronagraph combined with the F182M filter; it has a minimum IWA of 0.23″, over 50% larger than what the CGI is capable of.) In theory, with a contrast of $10^{-9}$, the faintest detectable satellite is V=36.5 (5-σ detection), but the real limit is set by the sensitivity of the CGI, which is not yet well-constrained, but it is safe to assume that the limit is brighter than V=36.5. The IWA corresponds to 220, 325, 430, and 540 km at observer distances of 2, 3, 4, and 5 AU, respectively. Observations of faint satellites with short orbital periods are therefore possible with the CGI. As an example, the Ida/Dactyl system, with Ida at V=13.6 and Δm=6.7, is composed of a primary and a satellite with mean diameters of 31.4 and 1.4 km, respectively [78]. The semi-major axis of Dactyl is not well-constrained (between 85 and 355 km), but additional constraints could be placed on its orbit with WFIRST imaging observations [79]. At ~2.8 AU from the Sun, Dactyl could be observed with the CGI if its semi-major axis is on the larger end of the range. Discovery of similar asteroid binaries is presently only possible with spacecraft flybys, but an increase in the number of known systems is possible with WFIRST and the CGI.

### 5.3 Irregular satellites

The irregular satellites of the giant planets represent a diverse collection of minor bodies potentially captured early in solar system history. Their exact origins remain uncertain due to the large diversity of orbital parameters, the lack of well-constrained physical characteristics and surface compositions, and the small number currently known (~100). While all irregular satellites are the result of capture, the source populations are still debated. Some irregulars are thought to have originated in the Kuiper Belt, like Saturn's moon Phoebe [80] and Neptune's moon Triton [26], but other satellites do not fit this paradigm, implying multiple source populations. (Interestingly, Phoebe and Triton are the irregular satellites best studied by spacecraft, and both are thought to have originated in the Kuiper Belt.) As a way to better understand their origins, the irregular satellites of Jupiter and Saturn were categorized into families based on their orbital parameters (inclination, eccentricity, semi-major axis, direction of motion). The intent was to group irregular satellites together based on known properties that suggest similar origins either from the same source population or through collisional processes [81-84]. Additional information on the physical properties and surface compositions of these objects would contribute to a more accurate classification but is work best-suited for JWST. Probing the giant planet Hill spheres for new, faint irregular satellites would help to fill out the size distribution of these populations and would help inform origin theories. This work is best-suited for WFIRST.

Increasing the number of known irregular satellites requires a large imaging field of view and deep imaging in a wide-band filter, both of which are capabilities of the WFI. In a 1000-second exposure, the WFI can image point sources down to V~27.7 in the W149 filter. This translates to the detection of irregular satellites down to approximately 0.3, 1.0, 4.5, and 11.4 km in diameter around Jupiter, Saturn, Uranus, and Neptune, respectively. We use these diameter limits to estimate the number of irregular satellites that WFIRST would detect in combination with an estimated size distribution. We used the most up-to-date model size distribution estimates for the irregular satellites [85] which are consistent with observed size distributions [86]. Considering both prograde and retrograde satellites, and assuming that WFIRST discovers only half of the unknown irregular satellites that are above these diameters, it would find approximately 1000, 200, 100, and 5 irregular satellites around Jupiter, Saturn, Uranus, and Neptune, respectively (compare this to 60, 40, 10, and 5 currently known irregular satellites). The order-of-magnitude increase in the number of detected objects is because current surveys just barely reach the steep collisional size distribution, whereas WFIRST will go significantly deeper.

The search regions of interest around the giant planets (i.e., the solid angle of the Hill spheres as seen from Earth at opposition) are 4.7, 3.0, 1.5, and 1.5 $deg^2$, respectively [87]. With WFIRST's 0.28 $deg^2$ field of view, around 6-17 separate pointings, with two dithered images made at each pointing to cover chip gaps, are required for complete coverage of the Hill spheres surrounding each planet. However, according to [88], satellites are not truly stable over the entire volume of the Hill sphere; the maximum semi-major axis for an irregular satellite is only ~67% of the Hill radius, and only for a retrograde satellite. The size of this "stability region", as well as the semi-major axis of the furthest known irregular satellite around each giant planet are presented in Table 6. Considering these values, only 52%, 18%, 9%, and 26% of the volumes of the Hill spheres of Jupiter, Saturn, Uranus, and Neptune, respectively, are known to be occupied by irregular satellites. In other words, a very large fraction of the stable volume of the giant planet Hill spheres remains to be thoroughly searched. This is not surprising given the small FOV of ground-based telescopes and the difficulty in obtaining enough observing time to cover the entirety of the Hill spheres to an appropriate depth for detection of small irregular satellites. WFIRST will provide the necessary FOV and sensitivity for deep irregular satellite searches.

To determine irregular satellite orbits would require ~5 sets of observations covering ~2 years. Because WFIRST would only require ~40 pointings with 1000-second exposures for each of these observation sets, the total observing time required to increase the number of known irregular satellites by the numbers described previously is only ~60 hours. Either as part of these follow-up observations or as an additional campaign, colors and/or rotational curves could be investigated as well, giving significant insight into the origin and evolution of these irregular satellites. These insights, when placed in context, are important for understanding the surfaces of regular satellites and for constraining solar system formation models [89-91].

**Table 6. Giant planet Hill spheres**

| Planet | Hill radius ($10^7$ km) | Stability region ($10^7$ km) | Semi-major axis of furthest known satellite (km) | % of Hill sphere known to be occupied |
|---|---|---|---|---|
| Jupiter | 5.32 | 3.56 | 2.86 | 52 |
| Saturn | 6.53 | 4.38 | 2.45 | 18 |
| Uranus | 7.00 | 4.69 | 2.09 | 9 |
| Neptune | 11.6 | 7.77 | 4.93 | 26 |

While the WFI will have a large FOV for a space-based telescope, it will be small in comparison to the new Hyper Suprime-Cam (HSC) imager on the 8.2-meter Subaru telescope [92]. The HSC has a 1.5 deg$^2$ FOV with a plate scale of 0.17″/pixel and can achieve a 5-σ imaging depth of 27.8 in the SDSS *g* filter in 3600 seconds (1 hour), given superb conditions (New Moon, 0.5″ seeing, and transparency of 1.0). The WFI provides a comparable imaging depth (27.7, 5-σ) in only 1000 seconds. For the WFI to cover the FOV of the HSC to the same depth would require 6 pointings, for a total of 6000 seconds. However, due to the WFI's higher sensitivity, it can take shorter exposures than the HSC to reach a comparable imaging depth. While the HSC will outperform the WFI in terms of areal coverage, the WFI will outperform the HSC in terms of depth, which is important for the detection of very faint irregular satellites (and minor bodies). Additionally, the weather conditions required for the HSC to reach a comparable depth to the WFI are very restrictive, meaning less opportunities for irregular satellite searches compared to a space-based platform like WFIRST.

### 5.4. Inner Oort Cloud objects

The most distant solar system minor bodies are time capsules that preserve information from the formation of the solar system, and their orbits possibly provide clues to the presence of an as-yet-undetected giant planet [93]. These objects are distinguished from KBOs by their very large perihelion distances (>100 AU). Long-period comets are members of this population and provide useful compositional information, but we must wait until they make the long journey to the inner solar system. It is also unknown just how many of these objects are present in the outer reaches of the solar system. Identification of objects orbiting at large heliocentric distances would be the next step in understanding the formation and evolution of the solar system and would provide better context for the population of long-period comets and various KBO populations, including the Scattered Disk Objects (SDOs) and detached (large-perihelia) objects.

At the present time, no objects have been detected that are >100 AU from the Sun at the time of the observations; the farthest known object at this time is the large KBO Eris, currently near aphelion at ~96 AU from the Sun [94]. The region beyond the Kuiper Belt, but interior to the Oort Cloud proper, is sometimes referred to as the "Inner Oort Cloud" or the "Hills Cloud" [95]. Frequently identified as the source of returning long-period comets, the Inner Oort Cloud does not share the same isotropic distribution of inclinations with the Oort Cloud proper [96]. Many objects have been identified with orbits that take them beyond 100 AU, but none have been detected while in this region, likely because of the extremely large orbital distances and physical characteristics of these bodies (small size, low albedo) that combine to produce very low reflected flux. High sensitivity and large amounts of observing time are required to make a detection of a small body beyond 100 AU.

Arbitrarily long exposures would be ideal for the detection of Inner Oort Cloud objects but, in reality, there is a practical limit set by the flux of cosmic rays. According to [97], JWST at the Sun-Earth L2 point will experience ~3.3 cosmic ray events per square centimeter per second averaged over a solar cycle and assuming a shielded focal plane, which will be true for WFIRST [2]. The area of each individual WFI pixel is 100 $\mu m^2$, meaning that ~0.33% of pixels would be hit by a cosmic ray in a 1000-second exposure, and ~3.3% would be hit in a 10,000-second exposure. However, this assumes that each cosmic ray affects only one pixel. It is more likely that 5 total pixels will be affected by each cosmic ray hit: the central pixel that is directly hit as well as the 4 adjacent pixels. So, for a 1000-second exposure ~1.65% of pixels would be affected by cosmic ray strikes, while ~16.5% would be affected for a 10,000-second exposure.

The imaging depth for a 5-$\sigma$ detection in a 10,000-second exposure through the W149 broadband filter is 32.67 [2]. However, to reduce the effects of cosmic rays, we break the 10,000-second exposure into ten 1000-second exposures. The imaging depth for one 1000-second exposure through the W149 filter is 27.67, so co-addition of these 10 frames results in an imaging depth of 30.17, about 2.5 magnitudes brighter than for a single 10,000-second exposure. Scaling fluxes from Charon using the geometric albedo and radius from the New Horizons flyby [3] and the heliocentric distance, observer distance, and V magnitude on an arbitrary date from JPL Horizons, we calculated the V magnitude for Inner Oort Cloud objects of varying sizes and heliocentric distances (Fig. 9). We assumed a geometric albedo for these objects of 0.10, similar to SDOs. Based on these results, a 2000-km dwarf planet class object would be detectable out to distances of ~800 AU and a 200-km object would be detectable out to ~250 AU. The smallest detectable object at the inner edge, 100 AU, would be <50 km in diameter.

The Oort Cloud is theorized to be a spherical distribution of comets, meaning the distribution of objects across the sky is relatively uniform, and is the reservoir of dynamically new comets [96,98,99]. However, the shape of the Inner Oort Cloud is possibly more disk-like, centered on the ecliptic, with a larger spread in ecliptic latitude as heliocentric distance increases. This means a higher density of objects in a smaller area of sky, increasing the chances that any particular pointing could detect an Inner Oort Cloud object. A potential Inner Oort Cloud survey could be combined, as a GO program, with a deep imaging survey of faint astrophysical sources, such as high-redshift galaxies, that would not be detected as part of the currently planned surveys. The ideal survey region would be a few tens of degrees above or below the ecliptic plane to avoid the majority of minor bodies in the solar system, and outside of regions heavily populated with galaxies, to avoid source confusion. The general strategy would be to obtain ten 1000-second exposures of a particular area of sky, then return at a later time (no later than 1 day) and obtain another set of ten 1000-second exposures with the second field offset from the first in order to increase areal coverage while probing deeper than a single visit could provide. The overlap of the two fields should be ~25% for deeper imaging over a reasonable portion of the first field while still providing a useful increase to the survey area. This process would then continue over days, weeks, or months to build up a large, deep survey region. An object at 100 AU, on the closest edge of the Inner Oort Cloud, will move ~4 pixels in 10,000 seconds, and ~32 pixels over the course of one day and would therefore be the easiest to detect. More distant objects at 600 AU will only move ~2 pixels over the course of a day, so re-imaging different quadrants of earlier fields over the course of 3-4 days would enable detection of objects over a wide range of heliocentric distances. The imaging depth and cadence of the LSST survey is not adequate for detection of Inner Oort Cloud objects (§4.3.2) and the survey area required is prohibitively large for a JWST GO program, so this investigation is only possible with WFIRST. Additionally, the large field of view of WFIRST would contribute to a larger survey area in a shorter amount of time compared to many other ground- or space-based telescopes.

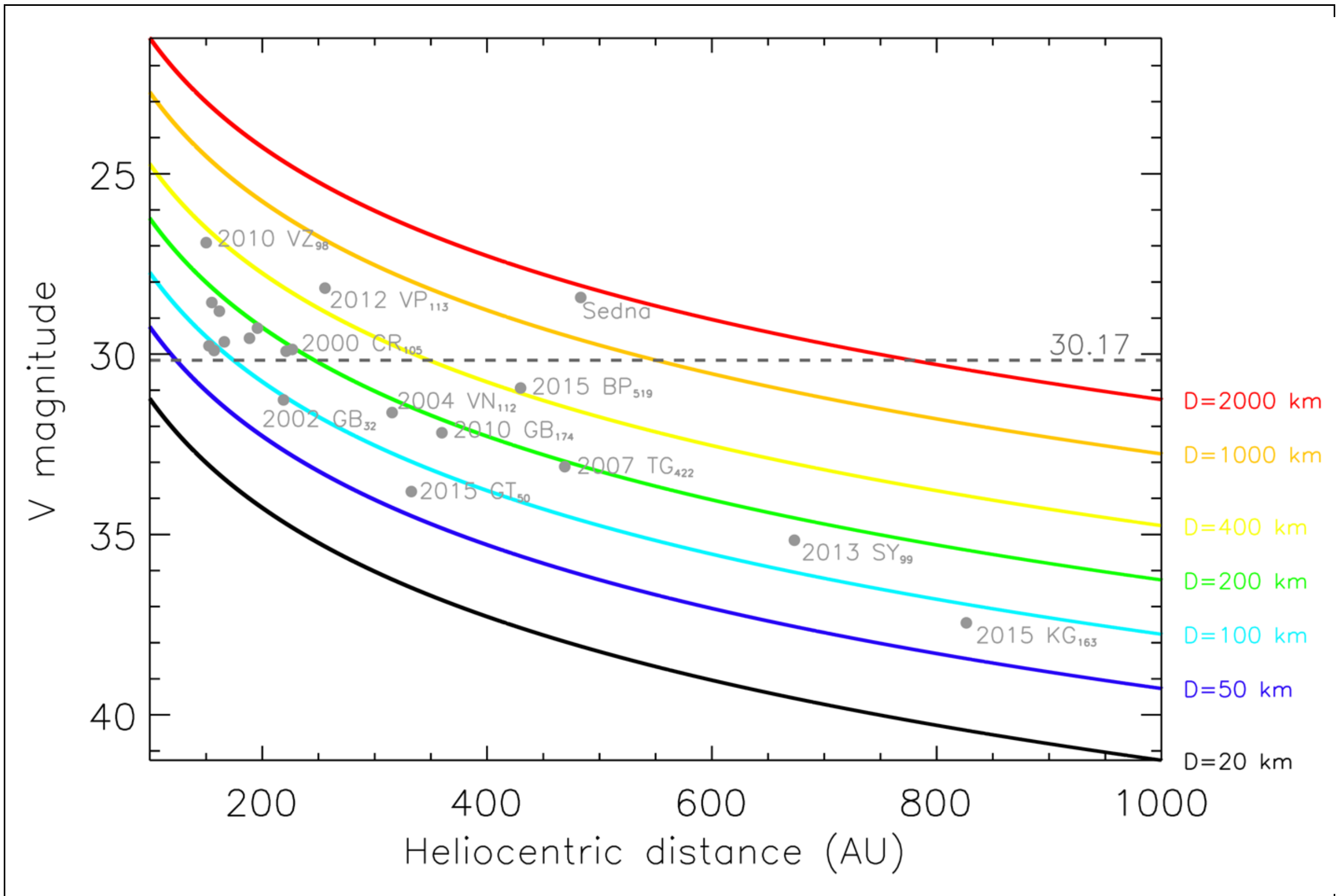

**Figure 9:** V magnitude as a function of heliocentric distance for Inner Oort Cloud objects of different diameters. The geometric albedo of these objects is assumed to be 0.10, comparable to Scattered Disk Objects (SDOs). The curves correspond to different Inner Oort Cloud object diameters, with labels on the right side of the plot. The dashed line marks the V magnitude limit (30.17) for detection in a co-added set of ten 1000-second exposures through the W149 filter. The grey labeled points show extreme KBOs ($q$>30 AU and $a$>150 AU) with decent orbital solutions if they were to be observed at the distance of their semi-major axes. Their estimated V magnitudes are scaled from present-day V magnitudes, heliocentric distances, and observer distances obtained from JPL Horizons. Note that Sedna's diameter is ~1000 km and its geometric albedo is ~0.32 [100], which places it close to the D=2000 km curve.

## 5.5. Occultation science

*5.5.1. Targeted occultations*

Occultations of a background star by a solar system object are sources of a variety of important information for the determination of sizes, albedos, and shapes of main belt asteroids, Jupiter Trojans, Centaurs, and KBOs [101-105]; the detection of ring systems around Uranus, the Centaur Chariklo, and the dwarf planet Haumea [106-108]; the detection and study of Pluto's atmosphere [109]; and the study of planetary atmospheres [110].

Occultation events are rare and sometimes difficult to predict depending on the quality of the solar system target's orbital solution. Expanding coverage beyond the surface of the Earth increases the chances that a particular event will be observable. With its orbital location around the Earth-Sun L2 point, WFIRST will be in position to observe occultations that will likely not be visible from Earth or other spacecraft. (In fact, JWST Guaranteed Time Observation (GTO) program 1271 will attempt a proof-of-concept for space-based occultations.) Due to their slower apparent rates of motion, Centaurs and KBOs will be the primary targets; under the right conditions, WFIRST will be able to observe occultations by faster-moving main belt asteroids and Jupiter Trojans.

In order to achieve the proper imaging cadence (≥2 Hz, with ~25 Hz ideal for faster-moving targets), a small subarray must be available in the larger field of view for faster readout times (§3.4). The maximum readout time required is 0.5 second, with a desired readout time of 0.04 second. The readout time for individual WFIRST detectors is 2.7 seconds, and the guide boxes are anticipated to be 16×16 pixels. The readout cadence of this guide box will be 5.8 Hz, above the minimum required cadence for occultation science.

An occultation observation will consist of an occultation imaging series at ~5.8 Hz with minimal or no dead time between images; the desired rate is determined by the sky-plane velocity of the occulting body, ideally to match or exceed the Fresnel scale (Fig. 10). Notably, a cadence of 10 Hz is required to match the Fresnel scale for bodies in the outer solar system, such as KBOs. If a body were to have a Pluto-like atmosphere with a scale height of 60 km, a cadence of 5 Hz is always sufficient to achieve 10 points per 60-km scale height, a rule-of-thumb that allows for sufficient resolution to fit for the temperature structure. The imaging series is tracked at the sidereal rate, maintaining exact positioning on the occulted star as the occulting body passes through the field. (Non-sidereal tracking is needed for the exceptional case of mutual events when, for example, a KBO satellite passes in front of its primary.) Separate photometry of the star and body is required before and/or after the main occultation event. Occultation durations depend on the size of the occulting body and the sky-plane velocity and can range from 1 second to several hours (for giant planet and ring events). KBOs as small as ~15 km can be resolved with the targeted occultation technique on WFIRST, although for bodies this size there will be considerable positional uncertainties in the predictions.

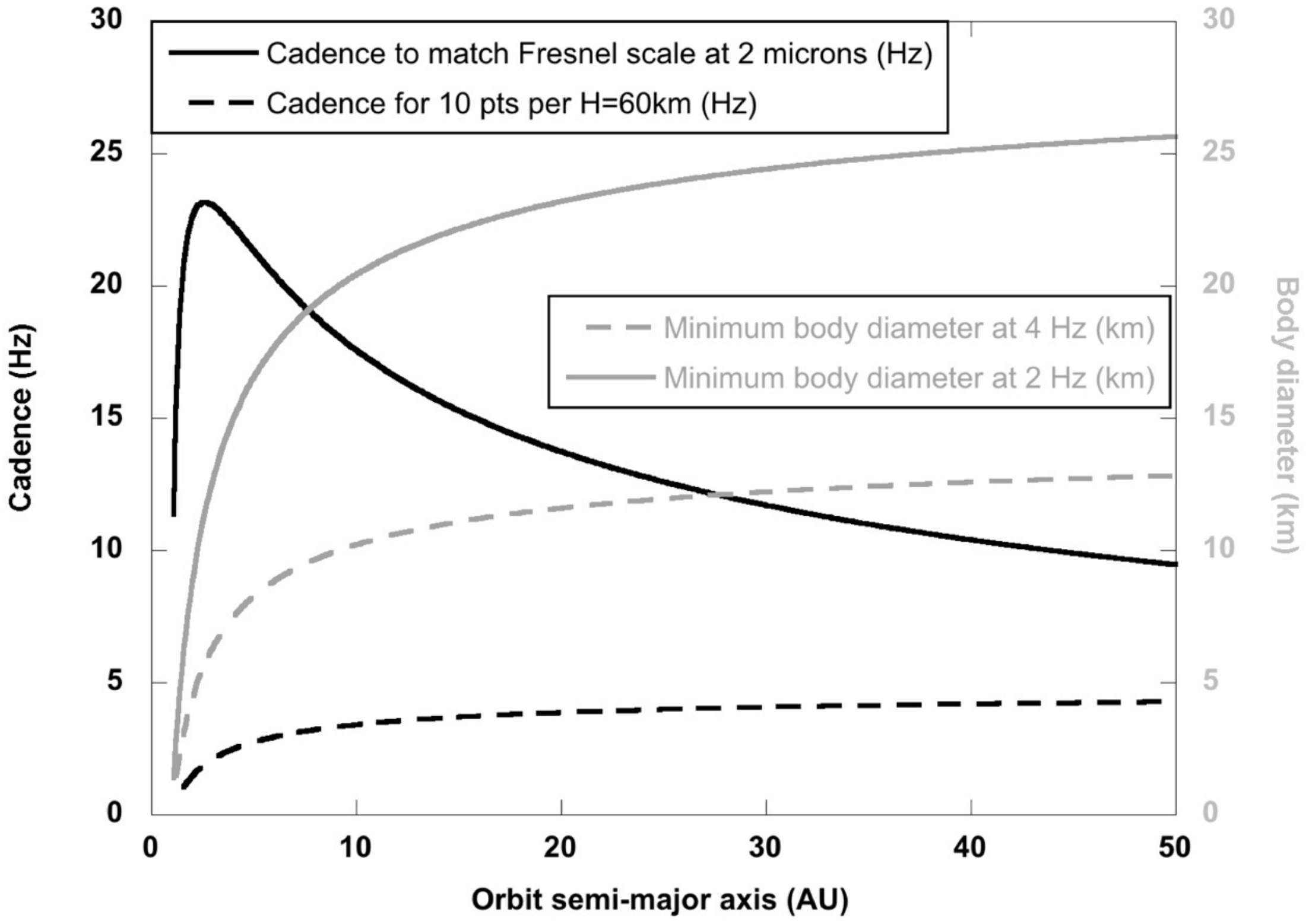

**Figure 10:** Cadence (black) and body diameter (grey) limits for occultations, as a function of body orbital semi-major axis. The curves in black present the cadence required (solid line) to match the Fresnel scale at 2 μm or to achieve 10 points per 60-km scale height in a hypothetical atmosphere (dashed line), as a function of distance. The grey curves predict the minimum body size that would be resolved with a minimum of two points across the body, for an observing cadence of 2 Hz (solid line) and 4 Hz (dashed line).

*5.5.2. Serendipitous occultations*

The method of detecting and characterizing serendipitous occultations of background stars by foreground minor bodies is presently the only way to estimate the population of KBOs in the sub-kilometer size range; however, this method does suffer from the random and unrepeatable nature of the observations. Observing serendipitous occultations with WFIRST is possible and was previously attempted with the Fine Guidance Sensors (FGS) on HST [111,112], in X-rays with the Rossi X-ray Timing Explorer [113], and with ground-based observing programs [114-116]. Detections of sub-kilometer sized KBOs were reported from ground-based programs [114,116] and two detections were reported in over 30,000 "star-hours" from the FGS [112]. However, in light of recent work on the size-frequency distribution of KBOs based on the impact crater populations of Pluto and Charon [117], these detections are now considered likely to be false positives.

Two FGS detectors are used for guiding purposes with HST, and each tracks its own guide star, so there are 2 star-hours per hour of science observations. JWST will track one star at a time using its FGS, but it is unclear if these data will be saved and transmitted to the ground. If these data are available for analysis, the cadence would be ~15.6 Hz and 1 hour of observations would correspond to 1 star-hour. LSST will not perform high-cadence imaging of stars, so serendipitous occultations cannot be mined from these data.

This same process of analyzing guide star images from HST is possible with WFIRST for imaging, integral field spectroscopy, and coronagraphy observations, though at a lower cadence than the HST FGS (5.86 Hz vs. 40 Hz) [2]. However, this cadence is above the 2 Hz threshold and is sufficient for identifying serendipitous occultations by KBOs and other distant solar system objects. In fact, WFIRST should provide an even better opportunity to identify serendipitous occultations than HST because each of the 18 detectors in the WFI focal plane can track its own guide star [118]. Thus, up to 18 hours of star time per hour of science observations will be possible with WFIRST. Guiding will occur during the majority of the surveys and GO observations, resulting in a very large data set over WFIRST's mission lifetime for the potential detection of serendipitous occultations.

## 6. Solar system science with the WFIRST astrophysics surveys

In this section, we discuss the design of the astrophysics surveys and how they could be used for solar system science, specifically the serendipitous identification of new minor bodies (i.e., near-Earth asteroids, main belt asteroids, comets, active asteroids, Trojan asteroids, Centaurs, Kuiper Belt Objects, Inner Oort Cloud objects, extrasolar asteroids) and irregular satellites. Serendipitously detected minor bodies and irregular satellites will naturally be streaked on the detector, with the length of the streak dependent on the distance and relative velocity to the object. Orbital and photometric information can be extracted from these streaks, but the detection limits will be less sensitive than those listed in Table 1 due to the flux being spread over a larger number of pixels. The sections below are by no means an exhaustive list of possible investigations. The Guest Investigator (GI) program will provide funding for analysis of data from the WFIRST astrophysics surveys. Like the GO program, funding through the GI program will be a competitive process.

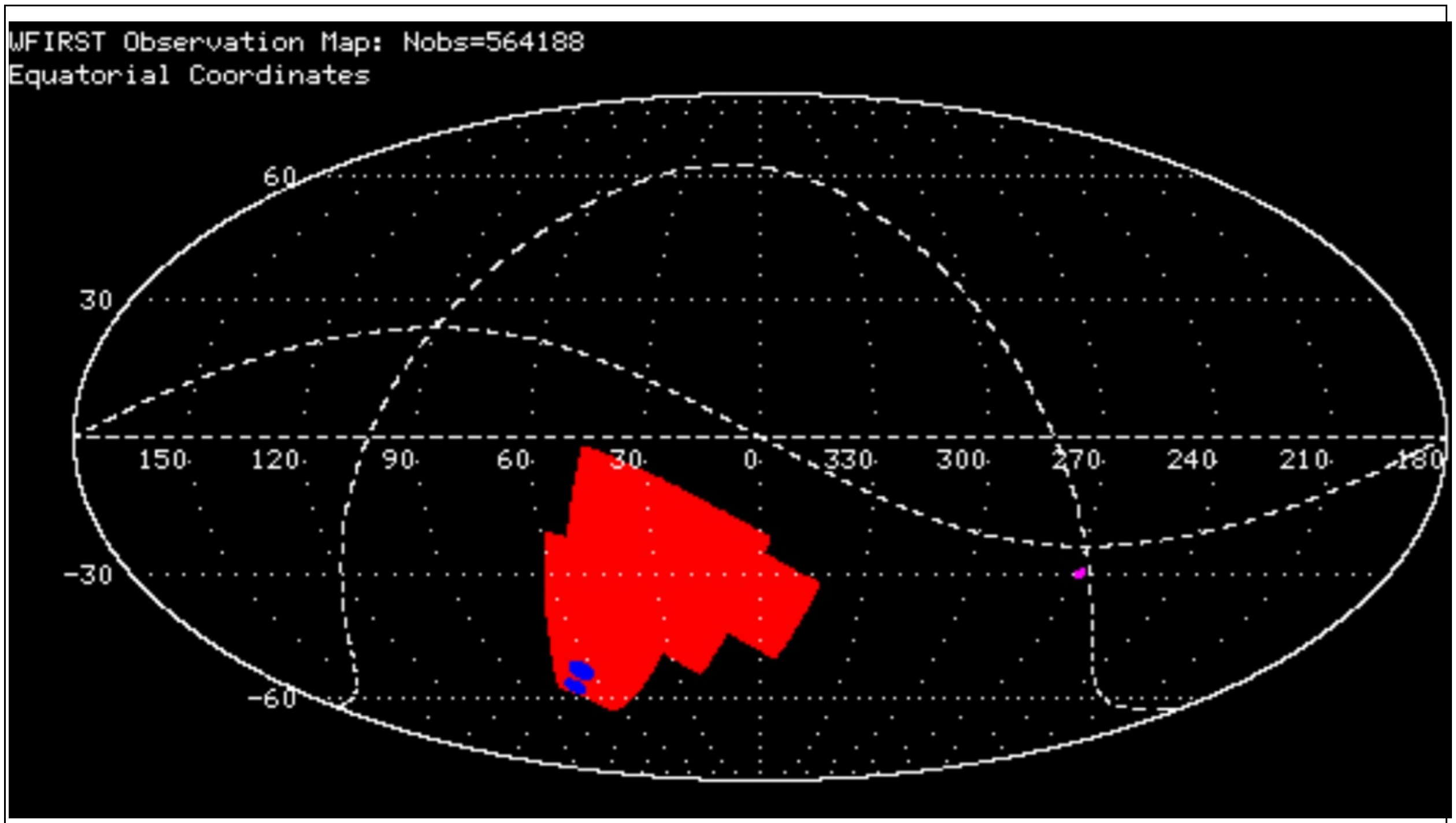

**Figure 11**: Survey coverage of the pre-defined WFIRST surveys. The large red region corresponds to the survey region of the High-Latitude Survey (HLS). The blue regions are the supernova survey fields. The small magenta region corresponds to the patrol area of the microlensing survey near where the galactic plane and the ecliptic cross. The patrol area for the CGI survey is not yet defined. The sinusoid-shaped dashed line represents the ecliptic; the other dashed line represents the galactic plane. The equatorial coordinate system is used. (Image from [2])

### 6.1. High-Latitude Survey (HLS)

WFIRST's High-Latitude Survey (HLS) is made up of WFI imaging and IFC spectroscopic components, with the purpose of observing galaxies found in a large region of the southern sky. The rest of this section focuses on using the imaging component, with its predefined strategy, for the detection of KBOs; however, imaging observations could also be possible in parallel with the spectroscopic component as a GO proposal with user-defined parameters (§3.3). The imaging component of the HLS will include observations of minor bodies that serendipitously fall within the 0.28 deg$^2$ FOV. The proposed ~2200 deg$^2$ patrol field (large red region in Fig. 11) for the four-filter (Y106, J129, H158, F184) HLS will observe KBOs on inclined, dynamically excited orbits. (This survey could also be used to search for a distant giant planet [93], if it exists and is not found prior to the launch of WFIRST in the mid-2020s.) The multi-band WFI photometry will measure the surface composition of Centaurs and KBOs, important for understanding their formation history in the early planetesimal disk and later migration under the gravitational influence of Neptune to their present orbits. The presence of KBOs in WFI imaging is strongly dependent on where WFI images on the sky. KBOs have a complex and finely structured population distribution, with the majority of the population in low-inclination, low-eccentricity orbits, placing them near the ecliptic; the sky density of KBOs decreases rapidly further from the ecliptic. These KBOs likely formed closer to the Sun and were emplaced on their present orbits by Neptune's migration [119], so their composition and dynamics are particularly useful probes of planetary formation and migration history.

By the time of WFIRST, the survey efforts of the Dark Energy Survey (using the DECam instrument at the Cerro Tololo Inter-American Observatory), followed by the LSST, will provide a comprehensive catalogue of all $r$<24.5 KBOs in this region of sky in Sloan *griz*. Assuming a heliocentric distance of 40 AU, an observer distance of 39 AU, a geometric albedo of 0.08, an average $g$-$r$ color for KBOs of 0.65 [120], and using the SDSS to Johnson filter conversions from R. Lupton*, this $r$ magnitude limit corresponds to >70 km. The serendipitous WFI imaging will therefore augment the spectral photometry for known KBOs, which will have well-determined orbits with small sky uncertainties within the FOV. Current estimates for the size of the high-latitude KBO population are exceptionally uncertain: observations of a few hundred KBOs to $r$~25 (diameter of ~55 km) are plausible [121,122]. WFIRST will reach $r$~27 (diameter of ~20 km) with each set of exposures in a particular filter during the HLS survey [2]. This limit is 2.5 mag fainter (10 times lower flux) than the DES and LSST surveys. Put a different way, this means that WFIRST will be able to observe objects ~3 times smaller at the same distance and observe objects of a particular size ~3 times further away compared to DES and LSST.

Serendipitous discovery of new KBOs is very likely based on the observing strategy of the HLS. The strategy, as currently defined, is to observe ~9 deg$^2$ through one filter, switch filters, and cover the same area again until it has been observed through all 4 filters. The exposure length for each filter at each dither position (of which there will be 4 for each pointing) will be 174 seconds. The full ~9 deg$^2$ area will take ~5.4 hours to cover in each filter (slightly longer for the J129 filter due to one additional dither), and approximately a day for that area to be covered through all 4 filters. If we conservatively define "detection" of a moving target as a shift in the centroid position by one WFI pixel (0.11″), objects out to heliocentric distances of ~17 AU could be detected between 174-second exposures and out to ~100 AU in only ~45 minutes. Assuming sufficient brightness in each filter, motion of all solar system objects will be detectable with the 1-day cadence of the HLS; these detections will also be more robust than provided by a 1-pixel shift in the centroid position. WFIRST will return to the same patch of sky every ~0.5 years and execute the same strategy at a different roll angle. The return visits could eventually allow mining of moving target discoveries from the transient, non-stationary catalogs [123-125].

The compositional classes of KBOs are only beginning to be understood. KBOs have a steep size distribution and orbit at large heliocentric distances, making most of them small and faint, even for space-based facilities. A detailed examination of the population in the proposed WFI bandpasses with multi-band photometry will provide substantial new compositional information. The WFIRST wavelength regions are barely explored for the majority of KBO populations. Ground-based observations require exorbitant amounts of time to provide sufficient SNR to investigate compositional variation. Whether these minor planets are detectable in the observed WFI bandpass will depend on their albedo, diameter, and distance from the Sun. A strong bias will apply to favoring detections at orbital perihelion. KBOs exhibit a range of colors from solar to substantially redder-than-solar [126]. Observations of KBOs through multiple filters will result in near-infrared spectral slopes that provide a first-order look at compositional differences. Thus, detection of new minor bodies with the HLS could probe an interesting subset of the color phase space of the <150 km diameter KBO population, but it is likely to be challenging to map to dynamical populations and thus composition and migration history unless time is invested in a GO/ToO program to obtain better-quality orbits for these faint KBOs.

---

*http://www.sdss3.org/dr8/algorithms/sdssUBVRITransform.php

### 6.2. Microlensing survey

The WFIRST microlensing survey is designed to detect light curve variations due to a host star with a planet passing in front of a background star. This is an effective way to detect planets at further distances from their host stars, and complements the transit method, which is useful for observing planets at smaller orbital distances. This survey will cover 7-8 contiguous WFI fields (1.96-2.24 $deg^2$) near the galactic bulge between RA 265° and 267° and declination -30° to -26° (the small magenta region in Fig. 11). These 7-8 fields will be observed during 6 non-consecutive 72-day campaigns over the 6-year WFIRST nominal mission. WFIRST will cycle through the 7-8 fields for the entirety of the 72-day campaign, with 15 minutes spent on each field. The majority of the observations will be through the wide band W149 filter (ten 52-second exposures plus overheads); every 12 hours the fields will be observed through the Z087 filter (two 290-second exposures plus overheads). This will result in ~33,000 epochs through the W149 filter and ~700 through the Z087 filter.

This large number of epochs, and even larger number of individual exposures, will enable detection and further study of slower-moving solar system objects, such as KBOs, without much additional effort. An unmodified microlensing survey could detect KBOs down to a V magnitude of 30.2 [127]; this equates to objects ~11 km in diameter, assuming a heliocentric distance of 40 AU and a geometric albedo of 0.04. KBO satellites would be detectable down to a V magnitude of 31.0 (~7 km) due to the benefit of a smaller search area. Satellites within 10 mas (~0.1 pixel) could be identified around primaries with a maximum V magnitude of 25.0. Detection of these satellites would result in a significant increase in the total number of known KBO binary systems (~80 are currently known) and thus an increase in known system masses. The cadence of this survey will allow KBOs and their satellites to be identified in multiple images through the W149 filter over a time baseline of ~days for the construction of rotation light curves of both components.

### 6.3. Supernova survey

The supernova survey, as defined by the science definition team [2], is set to take place over a 2-year timespan with three imaging depth "tiers." The purpose is to discover and characterize supernovae, with the ultimate goal of identifying new Type Ia supernovae to test cosmological theories of the expansion of the universe. The potential survey fields are shown in blue in Fig. 11; it is currently unclear if one or both of these fields will be used. The plan is for each field to be visited every 5 days, resulting in a total of 146 visits over the 2-year survey duration. Each visit is 30 hours long, with 8 hours for imaging with the WFI and 22 hours for spectroscopy with the IFC.

Three different "tiers" of imaging depth will be achieved within the 8 hours of imaging each visit. Table 7 presents the details for each tier. Total times for each tier are 3.0, 2.0, and 3.0 hours, respectively. Individual WFI exposure times range from 13 to 265 seconds. The 22 hours of IFC observations will be broken into short, medium, and long exposures for each supernova observed. The duration of the medium and long exposures will be 1.3 and 1.8 times the duration of the short exposure, with the short exposure time determined by the redshift of the supernova. Individual IFC exposure times range from 27.39 seconds to 2560.2 seconds.

**Table 7. WFIRST supernova survey imaging depths [2]**

| Tier | Area ($deg^2$) | # of pointings | Filter 1 | Depth per exp. (mag) | Filter 2 | Depth per exp. (mag) |
|---|---|---|---|---|---|---|
| 1 | 27.44 | 98 | Y106 | 22.3 | J129 | 22.4 |
| 2 | 8.96 | 32 | J129 | 24.6 | H158 | 24.5 |
| 3 | 5.04 | 18 | J129 | 26.2 | H158 | 26.1 |

The proposed supernova survey fields will be well off the ecliptic, resulting in fewer serendipitous detections of solar system minor bodies. However, images from the WFI portion of the survey and potential parallel images taken during the IFC portion could be used to constrain the number of minor bodies in high-inclination orbits, including NEAs, Centaurs, and comets. As currently defined, objects at observer distances of 0.5, 3.0, 4.2, 14.0, 29.0, and 49.0 AU can be detected down to 0.02, 0.3, 0.5, 5, 20, and 56 km in diameter, respectively, with the J129 filter in tier 3 (assuming a geometric albedo of 0.08). The proposed survey regions of the supernova survey are in the southern hemisphere, and therefore overlap with the LSST survey region. WFIRST will have an advantage over LSST (in this small part of the sky) because it will probe deeper, enabling identification of fainter objects. Statistics on the number of detections (or non-detections), even in such a small area of sky, would result in useful constraints on the high-inclination minor body population.

It is unclear at this time whether or not parallel observations with the WFI can be made during IFC observations due to data volume constraints, or if a GO proposal would be allowed to attach parallel observations to supernova survey observations since the targeted field would likely be within the survey area (§3.3). If parallel observations do not cause data volume issues, it seems likely that the designers of the survey would include parallel imaging observations to increase the chances of discovering additional supernovae. Another complication is that the currently defined strategy, as described above, was found to be non-ideal for the detection of new supernovae [128]. Suggestions to improve the survey include adding additional filters to the imaging portion, removing tier 1 and shifting the time to tiers 2 and 3, or removing the IFC component altogether. All of these potential changes must be considered when evaluating the effectiveness of supernova survey data for the detection of minor bodies from specific populations.

### 6.4. CGI survey

Due to the late addition of the CGI instrument to the WFIRST mission, the coronagraphy survey has yet to be well-defined. Information on the goals of the survey, sky coverage, and target selection is not currently available. However, these observations could be ideal for parallel observations with the WFI (see section §3.3 on parallel observations), assuming that the science policy allows parallels and does not prevent them from being attached to CGI survey observations.

## 7. Additional science investigations

### 7.1. Giant planets

Due to the brightness of Jupiter and Saturn, and the sensitivity of the WFI detectors, it is unlikely that imaging observations of these giant planets will be possible without the implementation of subarrays (§3.4). Spectra of Jupiter and Saturn, though at a very low resolution (R~100), may not result in saturation, but would likely be of limited use. Only the broadest absorption features would be identifiable in the spectra (Fig. 12). The JWST NIRSpec instrument is expected to be more adept at providing new insights into the composition of giant planet atmospheres due to the choice of a high-resolution (R~2700) grating. It is currently unclear if JWST/NIRCam observations of Jupiter or Saturn will be possible due to saturation concerns. In general, JWST will be a more appropriate tool for many of the science investigations described below, but JWST's nominal (2020-2025) and expected (2020-2030) mission lifetimes suggest that WFIRST may outlast it. In this sense, it is useful to consider what science WFIRST could carry out in a post-JWST world.

The relatively lower brightness of the ice giants Uranus and Neptune will allow WFIRST to obtain images and spectra of theses planets. The timing of WFIRST's nominal mission, the late 2020s to early 2030s, is set to occur during seasonal periods of Uranus and Neptune that have not yet been studied with modern instruments. For instance, northern hemisphere (System III) summer on Uranus will begin in 2030; due to Uranus' extreme axial tilt (~98°), the southern hemisphere will be in complete darkness and the northern hemisphere will experience constant illumination. WFIRST could therefore enable some of the first calculations of wind velocities surrounding Uranus' North pole and provide further constraints for investigating seasonal variations on Uranus. These data would also provide a useful comparison to previous results on the circulation patterns in Uranus' atmosphere [129,130]. Clouds in Neptune's southern hemisphere increased in size and albedo in the years leading up to the solstice in 2005 (northern winter/southern summer) [131]. Each season on Neptune spans ~40 years, so seasonal changes are relatively slow, with the current increase in cloud activity thought to be observable until the early 2020s. WFIRST may be well-placed to observe this expected transition, as the midway point (cross-quarter) between the 2005 solstice and 2046 equinox occurs between 2025 and 2026.

In the IFC, the angular diameters of Uranus (~3.7″) and Neptune (~2.4″) will result in approximately 74 and 48 pixels across their disks, respectively, using the 0.05″/pixel plate scale. The entire disk of Neptune will fit in the IFC FOV (3.0″ × 3.15″) but Uranus would require a 4-tile mosaic to cover its entire disk. Spectral imaging of Uranus and Neptune with the IFC will reveal $CH_4$ profiles in these planets' atmospheres as a function of latitude and altitude. Time-domain studies over the life of the WFIRST mission would add the time dimension to these maps and provide a useful baseline for future space- and ground-based observing programs that aim to study the dynamics of the ice giant atmospheres.

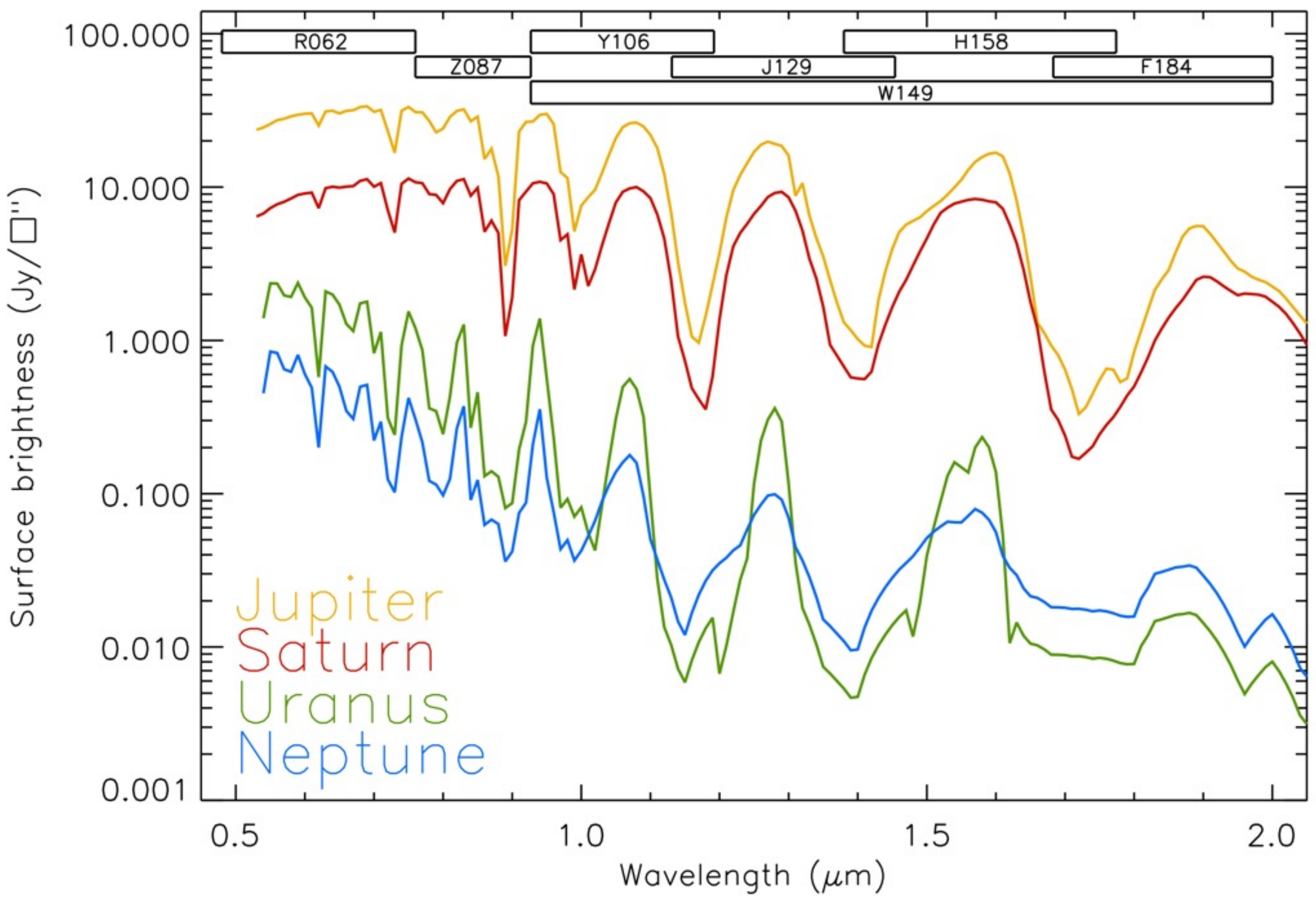

**Figure 12**: Surface brightness of the giant planets from 0.5 to 2.0 μm plotted at the resolution of the IFC [132-134]. WFI filter bandwidths are presented as boxes across the top of the plot, with the filter name labeled within each box (see Table 1 for filter bandpasses).

## 7.2. Giant planet satellites

Following the end of the Cassini mission in September 2017, there are no other missions planned to study the giant planet satellites that are likely to arrive at their destination prior to the launch of WFIRST: ESA's JUICE and NASA's Europa Clipper missions are currently slated to arrive in Jupiter orbit between 2025 and 2030, no Saturn missions are planned, and a Uranus or Neptune mission is in the early planning stages. Ground- and space-based facilities, such as WFIRST, are therefore needed to provide preparatory observations of these satellite systems prior to any future missions and to help define their science objectives. The below snapshots of possible giant planet satellite observations are presented in order to show what WFIRST is capable of, and not necessarily to describe new and ground-breaking science.

### *7.2.1. Tracking volcanoes on Io*

Identification of ongoing volcanic activity on Io is best done between 3 and 5 μm, a spectral regime outside of WFIRST's range, but within the range of JWST [135]. Near-infrared observations between 0.6 and 2.0 μm do not provide a definitive identification due to the contamination from reflected sunlight: bright regions could be regions of volcanic activity or just high-albedo areas. However, such observations would provide useful information if they were made while Io is in Jupiter's shadow. This removes the flux due to reflected sunlight and any bright regions detected are therefore the result of thermal emission and possible volcanic activity. For 2 hours every ~42 hours (Io's orbital period) Io is in Jupiter's shadow, allowing many opportunities for these observations. The spatial resolution of the IFC (~20 pixels across the disk) is sufficient for identifying large, hot regions indicative of ongoing volcanic activity.

#### *7.2.2. Spectral evidence for plume activity on Europa*

Salts and hydrated minerals, possibly brought to Europa's surface by plume activity [39,136,137], are easily detectable through identification of weakened water ice absorption features at the low spectral resolution (R~100) of the IFC [138]. Assuming that an IFC resolution element is ~0.2″, regions ~750 km (1/4 of a Europa diameter) in diameter would be resolvable at the distance of Jupiter (~5.2 AU). This equates to a total of ~18 resolution elements covering Europa's disk. Comparing the depth and shape of water ice absorption bands between 1.4-1.7 and 1.9-2.0 μm for different areas of Europa's surface will provide clues about the locations of ongoing or recent activity.

#### *7.2.3. Tracking of clouds and surface features on Titan*

Understanding Titan's methane meteorology is of critical importance for several reasons. At the simplest level, the methane humidity controls the radiative balance of the lower atmosphere, through collisional-induced opacity that is substantially due to $CH_4$-$N_2$ pairs [139]. Thus, methane stops the atmosphere from cooling too far and freezing out entirely [140]. The methane fraction at the cold trap (tropopause) is also important for controlling how much methane enters the stratosphere, where it diffuses upwards and engages in photochemistry, producing haze [141]. Third, methane precipitation is crucial for understanding the history of the surface. Titan's surface is carved by river channels at low latitudes, and the Huygens landing site is apparently a dried-up seabed [142]. At higher latitudes, the surface is pock-marked with holes that resemble collapsed karst-like terrain and may be periodically filled with liquid methane [143]. At the north polar regions, there are large methane seas the size of North America's Great Lakes [144]. The co-evolution of the surface and atmosphere is currently much debated, since Titan apparently had much greater liquid covering in the past [145], perhaps analogous to Mars in this respect. WFIRST will be capable of tracking ongoing surface and atmospheric changes on Titan shortly after equinox in the mid-2020s and into northern hemisphere winter. Tracking of seasonal methane clouds is possible at the spatial resolution provided by the IFC (~14 pixels across the disk) with a key focus area on methane windows at 0.85, 0.93, 1.08, 1.27, and 1.59 for detection of clouds [146] in the lower atmosphere. Low values of spectral ratios at 1.59/1.08 μm and 1.59/1.27 μm relative to the 1.27/1.08 μm ratio are indicative of regions dominated by surface water ice and lacking in liquid methane [147]. Time-domain studies with the IFC would be used for identifying regions that go from being dominated by water ice to liquid methane, or vice versa, signaling seasonal changes.

#### *7.2.4. Spectral characterization of smaller satellites*

The origin of many of the small regular and irregular satellites of the giant planets are currently unknown, but near-infrared spectra can provide valuable clues. As described in §5.3, the irregular satellites are grouped into families based on shared orbital and color characteristics, but these criteria are imperfect and surface composition is really needed to be more confident of family membership. Surface composition is also a useful tracer of the population of origin or collisional origin of the irregular satellites. However, very few irregular satellites have known surface compositions, complicating this determination [9,148]. The origins of some of the small regular satellites of the giant planets also remain uncertain. For example, Amalthea, the largest inner regular satellite of Jupiter, probably did not form at its current distance [149]; further spectral studies are needed to better understand the origin of the inner satellites and the evolution of the Jupiter satellite system. The composition of the rings of Jupiter and Uranus are also of unknown composition and it is believed that these rings are replenished from the nearby regular satellites [25,150,151]. Constraining the surface composition of the nearby regular satellites would therefore provide valuable clues to the composition of the rings.

The IFC can obtain spectra with an SNR of ~10 in a 1000-second exposure for targets with V magnitudes of ~24.9 [2]. For comparison, the NIRSpec Integral Field Unit (IFU) in Prism mode (R~100) can achieve an SNR of ~10 in 1000 seconds on a V~23.8 object. At this sensitivity, the IFC could obtain new or improved near-infrared spectra of most intermediate-sized giant planet satellites, both classical and irregular. At this limiting magnitude, an SNR of ~10 could be achieved in 1000 seconds for satellites of Jupiter, Saturn, Uranus, and Neptune, of 1.0, 3.5, 15.5, and 36.5 km, respectively (assuming a geometric albedo of 0.08). This means that decent SNR spectra of all currently known satellites of Jupiter and Uranus could be made with the IFC. The same is true for a majority of Saturn's satellites, including all irregular satellites, with only a handful of small interior satellites excluded. The IFC would be capable of obtaining spectra with adequate SNR for all Neptune satellites except for the recently discovered S/2004 N1 [17]. The previous statements neglect the issue of scattered light from the giant planets; in other words, spectra of satellites with smaller semi-major axes will be significantly impacted by scattered light from the very bright nearby planet. An analysis of this effect is dependent on a firm understanding of the optical design of the telescope and is outside the scope of this paper.

### 7.3. Comets

Comets are remainders from the time of the formation of the giant planets and serve as a probe of primitive composition. Comets can be broadly grouped into two categories based on their orbital characteristics: Jupiter Family Comets (JFCs; short-period comets) and Oort Cloud comets (long-period comets). JFCs typically have orbital periods <20 years and low inclinations; their orbits have been shaped by gravitational interactions with Jupiter. The long-period comets have more elliptical orbits, cover a large range of inclinations, and originate in the Oort Cloud, a spherical reservoir of minor bodies beyond the Kuiper Belt. While comets are most recognizable by their activity (the presence of a cloud of dust and gas and/or dust tails), other potentially active objects include active asteroids, main belt comets (§7.4.2), and Centaurs (§7.5.1). WFIRST will be able to observe comets beyond where water-sublimation is dominant over other species (< 2.5 AU) and activity onset and diminishment for species not previously characterized. Since dust characteristics are best studied at near-infrared wavelengths (§7.3.2), WFIRST fills a critical niche in wavelength space. Here, we will demonstrate how WFIRST will complement JWST in wavelength, characterize surface absorption features of the nucleus in each high-SNR channel of the IFU, and utilize the statistical survey sampling afforded by the WFI. We define three science cases where WFIRST can play a key role:

- How many comets are in the Oort Cloud (§5.4)?
- What are the properties of cometary ice?
- What are the sizes and shapes of cometary nuclei?

Before we can address those questions, we will need to assess how many comets are observable by WFIRST. For this, we repeated the JWST non-sidereal rate study [152] using

WFIRST's solar elongation constraints (Fig. 13). In brief, they considered all known comets with perihelion dates between January 2010 and January 2015, independent of their discovery circumstances: 221 short-period and 172 long-period comets. We find that WFIRST can observe a greater fraction of comets than JWST due to the wider field of regard (54°–126° versus 85°–135°), despite the same rate limit of 30 mas/s. Moreover, long-period comets are much better observed within the inner solar system. However, this tracking limit still misses many observing opportunities: approximately 43% of the targets at $r_h$=1.5 AU can be tracked, dropping to 20% at 1.2 AU. Ensuring a robust non-sidereal tracking limit is critical for obtaining observations of bright and active comets near the Sun and Earth (§3.1). Doubling the non-sidereal tracking rate from 30 mas/s to 60 mas/s would significantly increase the number of comets observable by WFIRST, especially long-period comets, and enable more detailed spatial studies of the inner structure of comet comae.

Present estimates of cometary populations suggest considerable numbers of cometary bodies will be accessible to WFIRST observations. In order to establish lower bounds, we neglect activity and assume bare cometary nuclei albedos are ~5%, similar to some other outer solar system populations [153]. This means that WFIRST, with magnitude limits approaching V=26 or more, will be capable of detecting objects down to sizes of ~1 km beyond Jupiter's orbit. The total population of JFCs has been estimated to be ~2000 [154]. For long-period comets within 8 AU, lower limits suggest ~1000 comets per year will be potentially observable by WFIRST through targeted observations, both in their quiescent and active phases.

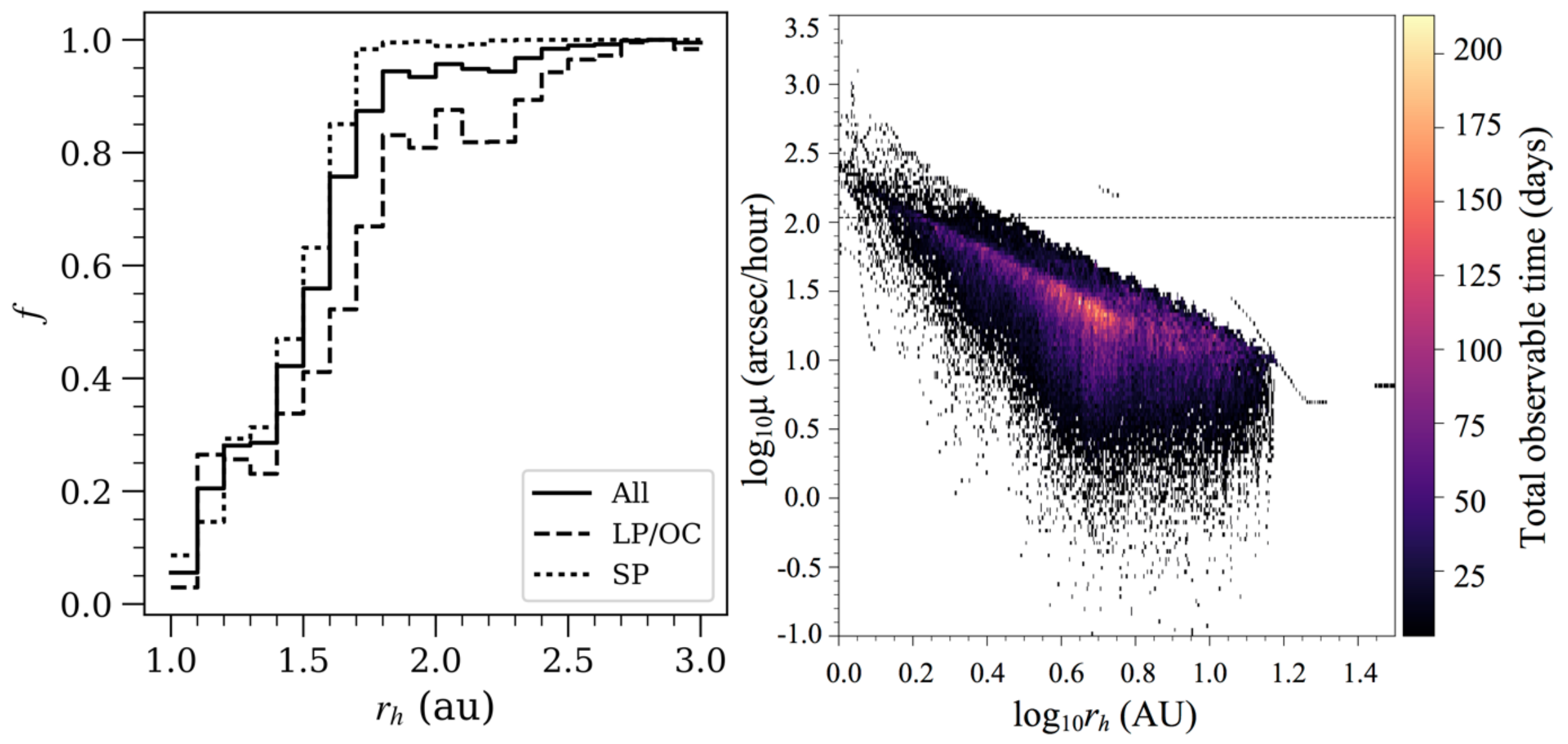

**Figure 13**: (left) Fraction of comets with perihelion dates between January 2010 and January 2015 that would be observable by WFIRST as a function of heliocentric distance. The dotted line represents short-period comets, the dashed line long-period comets, and the solid line all comets. Comets at smaller heliocentric distances will be moving at a faster non-sidereal rate than those farther away. Short-period comets are more likely to be observable due to their lower velocity compared to the long-period comets, which are on highly eccentric orbits. The nominal

30 mas/s tracking limit places a significant limit on observations of comets and their nuclei within 2.0 AU of the Sun. (right) 2D histogram of apparent rate as a function of heliocentric distance for short- and long-period comets; only times when the comets are within WFIRST's field of regard are considered. The bin value is the number of comet-days when targets are observable at that rate. The dashed line marks the 30 mas/s nominal tracking limit for WFIRST.

*7.3.1 How many comets are in the Oort Cloud?*

The densities of the populations of the primordial comet reservoirs set primary constraints on the formation and evolution of the early solar system. Linked to these quantities is the determination of the amount and nature of material presently available within the Oort Cloud [155,156]. The Oort cloud is the solar system's "deep freeze storage," where material is best preserved in its original form against the effects of insolation from the time of its emplacement. It is also the direct source of long-period comets [157]. However, the current estimates of the mass of the Oort cloud are based on comets that get relatively close to the Sun. These constraints, largely based on careful de-biasing of surveys, have set meaningful order-of-magnitude estimates [98,154,158], yet have their limitations. Specifically, the effect of giant planet perturbations in curtailing the influx of cometary bodies from the outer solar system, as manifested in the so-called "Jupiter barrier" effect, cannot be directly assessed through surveys dominated by comets in near-Earth orbits, as assumptions rely on discovery of comets within heliocentric distances where water-sublimation dominates. WFIRST's sensitivity and the planned astrophysics surveys can be used to serendipitously identify and characterize activity in much more distant comets, and thereby to assess the number of active comets beyond Jupiter's orbit relative to the presently observed fluxes within. Recent infrared observations [46,159,160,161] have shown volatile species, other than water, that are more active [162] at greater distances are ubiquitous. WFIRST can therefore place important constraints on solar system formation theories and comet evolution. Large space-based surveys, owing to their systematic sampling unencumbered by variable seeing and weather, would also be easier to de-bias [154] than ground-based surveys of comparable and even lesser depth. Large area imaging surveys with the WFI will allow us to quantify how long activity persists and at the same time to make the deepest inventory of distant comets and their reservoirs yet (see §5.4).

*7.3.2. What are the properties of cometary ice?*

Water ice is a key component of comets and was a significant planet-building material in the early solar system beyond the frost line. Therefore, assessing the physical properties of water ice in comets is essential to test comet nuclei formation scenarios. The WFI filters are too broad to determine the presence or absence of water ice absorption bands; however, the IFC has sufficient spectral coverage and resolution to map the water-ice distribution in the comae of comets through analysis of the amorphous 1.5 μm and crystalline 1.65 μm water ice absorption bands. A quantitative assessment of water ice grain sizes would only be possible by extending the long-wavelength limit of the IFC to >2.2 μm in order to disentangle the effects of grain size and abundance using the 1.5 μm and strong 2.0 μm water ice absorption features [163]. Sub-μm grains present a strong absorption feature at 2.0 μm, but not at 1.5 μm [164]; the IFC wavelength range, as currently defined, does not provide full coverage of the broad 2.0 μm absorption feature. Thus, the NIRSpec IFU on JWST would be a better choice for grain size studies, but WFIRST will be able to explore the state of water ice on the surface and in the comae of comets too faint to be observed by JWST (§7.2.4).

*7.3.3. What are the sizes and shapes of cometary nuclei?*

WFIRST will be able to constrain the sizes, shapes, and albedos of cometary nuclei by combining high angular resolution with the increased nucleus-coma brightness ratio provided by near-infrared IFC observations. Currently, only a small number of comet nuclei have been studied using coma extraction techniques but WFIRST would increase the sample size significantly [45, 154,159,165]. The removal of residual comae, possibly extant even at larger distances [159] will be enabled by the high resolution and WFIRST's stable and well-characterized PSF, as the coma extraction techniques fit the wings of the dust coma with a $(1/r)^{-n}$ profile and extrapolate to the nucleus PSF. WFIRST has an advantage over optical HST observations in that much of the near-infrared lacks any significant gas emission bands, effectively increasing the nucleus-coma brightness ratio, and facilitating the fidelity of the single-index profile model along limited azimuthal ranges to the actual coma signal. Second, the sensitivity of WFIRST allows for observations at greater distances when the object is less active. Finally, the pixel scale of the WFI is slightly higher in the near-infrared than in the WFC3 infrared channel on HST (0.11 vs. 0.13 ″/pixel). The JWST NIRCam instrument has a pixel scale of ~0.031″, smaller PSFs over the whole near-infrared range, and is slightly more sensitive than the WFI, so the use of WFIRST for the imaging investigation makes sense only in the post-JWST era. However, the IFC is more sensitive than the NIRSpec IFU at the same spectral resolution (R~100, SNR of 10 in 1000 seconds for V=24.9 vs. V=23.8). WFIRST's field of regard is also more favorable for observations of long period comets in the inner solar system than JWST, making WFIRST the go-to choice for low-resolution spectral studies short-ward of 2.0 μm. By noting surface constituents, through spectroscopic features of silicates, darkening agents like amorphous carbon, and ice species (§7.3.2), better estimates of surface albedos can be achieved.

## 7.4. Inner solar system minor bodies

*7.4.1. Asteroid families*

The study of asteroid families within the main asteroid belt, which represent disrupted parent bodies or swarms of fragments ejected from a parent asteroid's surface, informs compositional, taxonomic, and dynamical questions concerning asteroid family origin and evolution [166-168]. In particular, questions concerning actually genetic family membership, the presence of interlopers (i.e., non-genetic family members) among a family population, family space weathering trends, and the ability of asteroid taxonomies to correctly identify family members are issues that must be addressed to accurately determine the distribution and abundance of genetic asteroid family members [169-175]. Only through accurate identification of family members that originated from their respective parent asteroids can rigorous compositional and dynamical studies of individual families be conducted.

Observations utilizing the WFIRST IFC to conduct low-resolution (R~100) visible- and near-infrared reflectance spectroscopy have the potential to more strongly constrain the genetic asteroid family membership for many asteroid families as well as test asteroid family taxonomic classifications. The IFC is more sensitive than the NIRSpec IFU on JWST (R~100, SNR of 10 in 1000 seconds for V=24.9 vs. V=23.8) and can therefore obtain spectra of family members or candidates at the lower end of the size-frequency distribution. Acquisition of spectra from 0.6 to 2.0 μm outside of the obscuring telluric features of Earth's atmosphere and using the full WFIRST 2.4-meter telescope aperture will allow the detection of mineral absorption features that include olivine (~1.0 μm), pyroxene (~0.9 and ~1.9 μm), spinel (~1.4 μm), and possibly phyllosilicates (0.7 μm). Detection of these mineral absorption features will allow direct spectral and mineralogical comparisons between potential family members and the parent body, to discover non-family interlopers or diversity among a family via detections of different mineral(s), and to verify or reject a presumed mineralogy/composition based solely on a taxonomic classification.

*7.4.2. Active asteroids*

Active asteroids [176] are a newly recognized classification of solar system minor bodies that possess asteroid-like dynamical properties (typically defined as having a Tisserand parameter with respect to Jupiter of $T_J > 3$) but exhibit mass loss similar to comets. They include main-belt comets (MBCs) [177], which orbit in the main asteroid belt and whose dust emission is due to the sublimation of volatile ice, and disrupted asteroids, whose dust emission is due to physically disruptive processes such as impacts or rotational destabilization rather than sublimation (Fig. 14). Main-belt comets present an opportunity to study the nature, extent, and abundance of ice in inner solar system bodies and have implications for understanding the formation of our solar system and delivery of water to the early Earth. Disrupted asteroids present opportunities to study disruptive processes on small bodies in real time for comparison to theoretical models of those processes and also for inferring properties of their interiors. Discovery of more such objects [178] is crucial for advancing our understanding of their abundance and distribution in the inner solar system and the ranges of various physical properties associated with these objects (e.g., sizes, activity levels, active lifetimes, rotation periods, colors, etc.). Serendipitous discovery of active asteroids too faint to be identified by LSST will increase the known population of these objects and enable follow-up by other ground- and space-based telescope facilities.

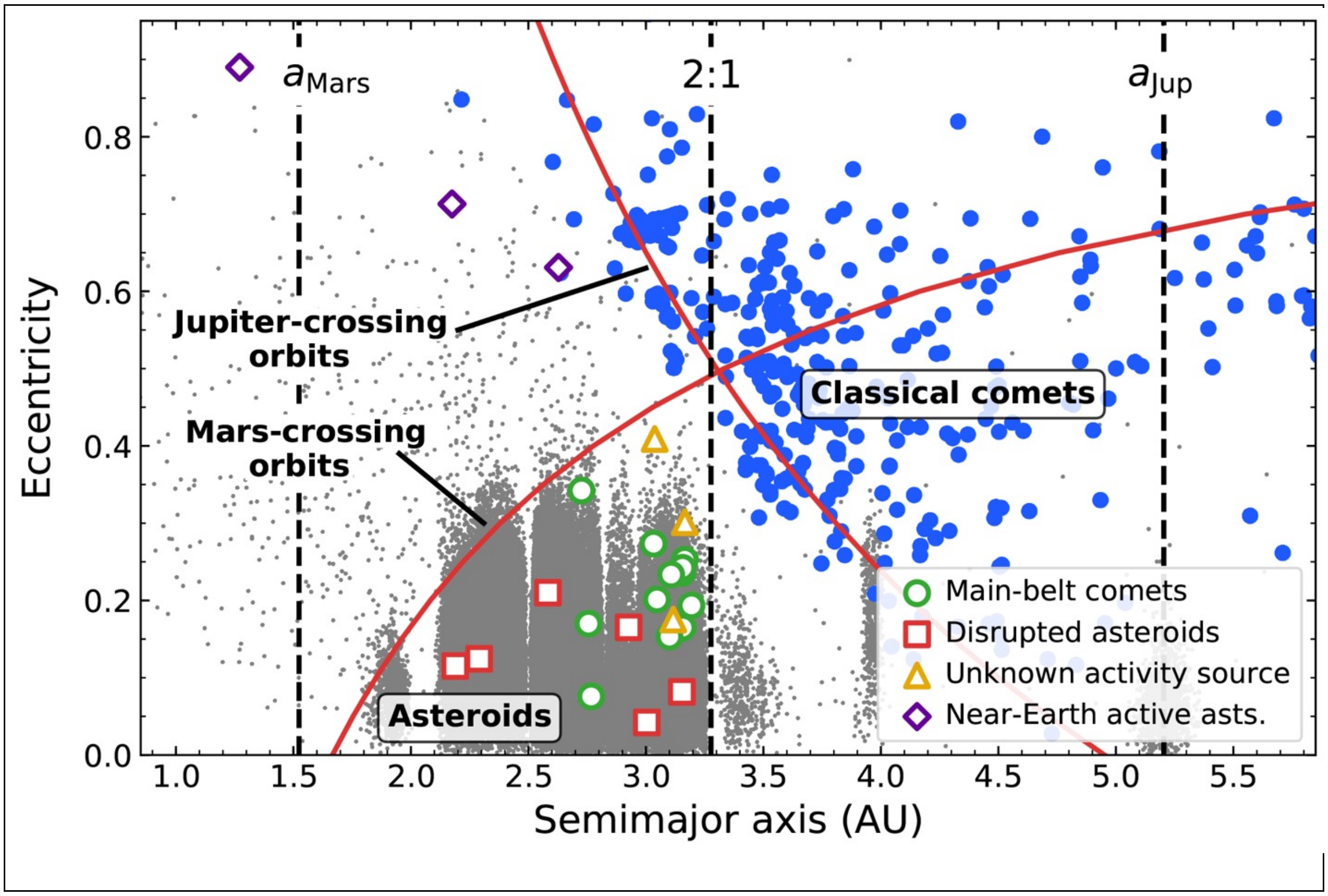

**Figure 14**: Semi-major axis vs. eccentricity for asteroids (small gray dots), classical comets (circles), and various kinds of active asteroids (open symbols). Likely main-belt comets are marked with open circles, likely disrupted asteroids are marked with open squares, active asteroids whose activity sources are currently unknown are marked with open triangles, and near-Earth active asteroids (i.e., those not found in the main asteroid belt but which have asteroid-like Tisserand's parameters with respect to Jupiter, $T_J$) are marked with open diamonds. Red curves represent Mars- and Jupiter-crossing orbits, while vertical dashed lines mark, from left to right, the semi-major axis of Mars, the 2:1 mean-motion resonance with Jupiter, and the semi-major axis of Jupiter. Active asteroids are clearly dynamically distinct from other comets, and most are dynamically indistinguishable from main-belt asteroids.

Serendipitous detection of MBCs and other active asteroids will be conducted using data obtained as part of the already planned WFIRST high-latitude (§6.1) and microlensing surveys (§6.2). The WFI wavelength range should make it more sensitive to large dust particles which should persist longer for active objects (provided that they can be ejected at speeds greater than the active object's escape velocity), widening the observability window for detection of active events. Given the field positions of these surveys (mostly far from the ecliptic), we do not expect a large number of MBCs to be discovered. There may, however, be the opportunity for discovery of non-main-belt active asteroids or other low-activity cometary bodies among the near-Earth object population which has a substantially broader inclination distribution (and therefore a broader ecliptic latitude distribution) than the main-belt asteroid population. A Guest Observer (GO) survey program to identify new active asteroids may be more successful, though more limited in observing time. Targeted observations of currently known active asteroids would extend the time-baseline of observations for the purpose of measuring potential changes in activity and could be undertaken by the CGI. Compositional studies of the ejected dust would be better handled by the NIRSpec IFU on JWST due to its broader wavelength range, but WFIRST will potentially extend the time baseline of observations into the 2030s.

*7.4.3. Jupiter Trojan asteroids*

The Jupiter Trojans either have two very distinct origins or have the same origin and are drawn from a population with a wide range of colors [6,179]. The origin of Jupiter's Trojan asteroids is still up for debate, and with the discovery of two distinct compositional classes, there may not be just one answer. These two classes have different near-infrared colors [5] and spectral properties, suggesting two different origins; the L4 and L5 clouds are home to members of both compositional classes. The IFC can provide higher-SNR near-infrared spectra than the NIRSpec IFU of more Jupiter Trojans, increasing the sample size for comparison to the irregular satellites and other minor body populations like main belt asteroids, Centaurs, Neptune Trojans, and KBOs. Previous work presents near-infrared colors for only 58 Jupiter Trojans, spread across the leading and trailing clouds [5]. Colorimetry of ~100 currently known Jupiter Trojans that lack near-infrared colors would almost double the sample size and help in determining whether or not transition objects exist and thus enable a more thorough assessment of the origins of the two distinct populations. This work would be enhanced with a K-band filter for use with the WFI (§3.2).

WFIRST is also well-placed to provide supporting observations of the Jupiter Trojans for the JAXA Solar Power Sail and NASA's Lucy Discovery-class mission (§4.4), both of which are set to launch in the early 2020s and make close flybys of Trojan asteroids later that decade.

Characterization of a larger fraction of the Trojan population by WFIRST will enhance the scientific gains from these missions by tying high-resolution imaging and spectroscopy together with population-wide trends for a more comprehensive evaluation of Trojan origins. Discovery and orbit determination of smaller Trojan asteroids could be used by the Lucy team to plan remote observations during cruise from one target to the next.

### 7.5. Outer solar system minor bodies

*7.5.1. Centaurs*

Centaurs are thought to be a transition population between Kuiper Belt Objects (KBOs) and the Jupiter Family Comets (JFCs), but their origin and evolution remain poorly understood. Two aspects of Centaurs could provide the observational evidence needed for a better understanding of this population: cometary activity and the red/gray dichotomy in the color distribution of Centaurs.

Some Centaurs have been found to be active (e.g., Chiron, P/2004 A1 (LONEOS), P/2011 S1 (Gibbs), 166P/NEAT, Echeclus, 29P/Schwassmann-Wachmann), but cometary activity is either very weak (<30 kg/s) or entirely absent [180]. A survey of cometary activity among Centaurs is fundamental for understanding if they have a common origin or if there is a difference in origin and evolution between active and non-active Centaurs. Moreover, to relate mass loss due to cometary activity to other parameters, both orbital (e.g., inclination and heliocentric distance) and optical (e.g., albedo and color), it is important to reconstruct the dynamical history of these objects. The detection of gas commonly related to cometary activity has been successful on a few Centaurs, but for many of them only an upper limit of their abundance has been calculated [181,182]. Possible constituents include ammonia ($NH_3$), methane ($CH_4$), CO, $CO_2$, methanol ($CH_3OH$), ethane ($C_2H_6$), acetylene ($C_2H_2$), hydrogen cyanide (HCN), as well as amorphous and crystalline water ice.

Centaurs are grouped in two main families based on visible color (e.g., V-R, B-R). “Red” and “gray” Centaurs are two distinct families, and no transition objects between the two families have been observed [126]. This suggests differences in origin and/or dynamical evolution. In the first scenario, red and gray Centaurs formed at different heliocentric distances in the solar nebula, with the red Centaurs retaining materials thought to result in more pronounced surface reddening (e.g., $CH_4$ and $CH_3OH$) [183]. In the second scenario, the red Centaurs are more primitive and went through fewer periods of cometary activity than the gray Centaurs, implying that the gray Centaurs underwent significantly more dynamical evolution. Previous work aimed at testing these scenarios by investigating correlations between color, albedo, and orbital parameters (e.g., semi-major axis, inclination, eccentricity, perihelion distance). It was found that gray Centaurs have higher inclination angles and lower albedos than the red Centaurs; no correlations were identified between color and the other orbital parameters [126,184]. The lack of correlation between color and a majority of the orbital parameters suggests the former scenario (differences in origin) may be the most accurate explanation, but a larger sample of near-infrared colors and spectra is needed to confirm this result.

Spectroscopy with the IFC will allow detection of most of the volatile elements that may exist on the surfaces of Centaurs. The IFC is sensitive at wavelengths where broad absorption features and overtones of crystalline water ice (1.65 μm), amorphous water ice (1.5 μm), and $CH_4$ (1.2, 1.7, 1.8 μm) are present. The IFC data will also allow for measurements of the infrared spectral slope and to relate it to band depths of volatiles. Other important measurements, particularly those of fainter Centaurs, will be possible by the imager, which covers the 0.6-2.0 μm wavelength range.

In particular, calculation of near-infrared albedos and spectral slopes will allow for discrimination between a larger number of red and gray Centaurs, as well as a comparison to the distributions of visible albedo and spectral slopes from LSST. Additionally, WFIRST's field of regard covers all declinations between the North and South ecliptic poles (with observable right ascensions dependent on the time of year), enabling observations of recently discovered high-inclination Centaurs [185,186]. These Centaurs, while only a small sub-set of the entire population, may be related to comets from the Oort Cloud and determination of their colors and surface compositions should provide further information for evaluating current theories of Centaur origins and evolution. LSST will be unable to observe a large portion of the northern sky, so detection of high-inclination Centaurs at these declinations will only be possible with WFIRST. Spectral observations of Centaurs, especially smaller, fainter members of the population, are more appropriate for WFIRST's WFI than the NIRSpec IFU on JWST due to the WFI's higher sensitivity (R~100, SNR of 10 in 1000 seconds for V=24.9 vs. V=23.8).

*7.5.2. Kuiper Belt Objects*

Based on dynamical considerations, the population of Kuiper Belt Objects (KBOs) is the likely reservoir of the Centaurs and Jupiter Family Comets (JFCs), both of which are relatively short-lived compared to the 4.5 Gyr history of the solar system. KBOs, including the dynamically-excited KBOs, likely formed in the outer solar system and were swept out to their present location and emplaced via resonances with Neptune [187]. There they have resided for most of the age of the solar system, and because they have not undergone a prolonged period of intense insolation, are primitive samples of the solar system's primordial composition. Over time, these bodies may be perturbed into planet-crossing orbits and evolve first into Scattered Disk Objects [188], and then Centaurs [189, 190], which in turn may evolve into JFCs through continued interaction with Jupiter.

The compositional link is less clear, however, between KBOs and the Centaurs and JFCs. As mentioned in §7.5.1, Centaurs fall into two different color groups, with minimal correlation between color and dynamical properties. KBOs have red and gray sub-populations [126] that are more clearly correlated with their dynamical properties [189], but the correlation between the KBO and Centaur color groups is not established, nor is the compositional source of these color differences, though different theories have been proposed [183]. Further comparison of the color distributions of the Centaur and KBO populations, particularly the Scattered Disk Objects, is needed to assess potential links between the two populations. Similar distributions in near-infrared colors, as noticed for the visible colors [191], would suggest similar surface compositions and provide a stronger link between the two populations.

In a 1000-second exposure, WFIRST can observe objects at the 5-$\sigma$ level with a V magnitude of 27.1 in each of the Z087, J129, and H158 filters, respectively. This corresponds to a 5-$\sigma$ detection of KBOs at 40 AU with diameters of ~25 km in each filter (Fig. 3). Significantly higher SNR can be achieved on brighter objects, a category that includes most, if not all, currently known KBOs and Centaurs. This investigation is more compelling for JWST/NIRCam, due to its larger suite of filters, dual-channel imaging, and higher sensitivity, but it is important to understand WFIRST's capabilities for the eventual post-JWST world.

In 1000 seconds, the IFC can obtain an SNR of 10 in each R~100 resolution element for a V~24.9 object, which is a full magnitude fainter than the NIRSpec IFU. The IFC could therefore provide high-SNR near-infrared spectra of a large majority of the currently known KBO population. Spectral studies of intermediate-sized objects from the ground is a difficult task, requiring many hours of time on-target to achieve low-SNR spectra [9,10,192-194]. In only 1000 seconds, the IFC would provide useful spectra for constraining the presence of water ice on the surfaces of intermediate-to-small KBOs. Previously identified trends in water ice abundance with absolute magnitude (which is related to diameter and geometric albedo) suggest that there is a transition from water-rich to water-poor surfaces between absolute magnitudes of 3 to 5 [193,194]. One possible explanation is that this transition corresponds to a KBO mass where differentiation can occur, with smaller objects remaining undifferentiated. Thus, the IFC could contribute in a significant way to the understanding of interior processes on KBOs early in solar system history.

Continued study of KBO binaries and multiple systems, of which over 80 are currently known, are valuable in that they provide mass and density constraints on the system as a whole. However, it is not clear how these systems formed, whether through co-formation, collisions, or capture [195]. Measurement of the rotation periods of the individual components provides a useful indicator of the tidal state of the system for evaluation of the formation mechanism. WFIRST will provide deep imaging through all filters, with objects detectable in a 1000-second exposure down to V=27.7 and V=27.2 through the W149 and Z087 filters, respectively [2]. This depth will allow observations of some of the faintest known KBOs, including the smallest minor satellites of Pluto: Styx and Kerberos. The angular resolution of the WFI is adequate for separating the components of a handful of KBO systems, including Pluto/Charon/Styx/Nix/Kerberos/Hydra, Eris/Dysnomia, Haumea/Namaka/Hi'iaka, Quaoar/Weywot, Teharonhiawako/Sawiskera, and Mors-Somnus, among others [20,196]. See §6.2 for a discussion on the prospects for serendipitous detection of new KBO satellites with WFIRST.

The sensitivity of the WFI and the timing of the mission can be leveraged to increase the time baseline of rotational light curve observations for the faint minor satellites of Pluto (Styx, Nix, Kerberos, Hydra) and the satellites of Haumea (Hi'iaka, Namaka). Observations of Pluto's minor satellites prior to the New Horizons flyby in 2015 suggested that they are chaotically tumbling, with unpredictable pole orientations from one epoch to the next [31]. A surprisingly short rotation period was calculated for Hi'iaka, the larger of Haumea's two satellites [197]. This work found a rotation period much faster than the satellite's orbital period, which at first glance points to a system that has undergone little tidal evolution. This could have significant implications for the formation of Haumea's satellites, which are usually considered to have formed after the giant impact that produced the Haumea collisional family [198]. There is currently no rotation period for the smaller of the two satellites, Namaka. Continued monitoring of these satellites' rotation periods requires time on space-based telescopes such as HST, JWST, and WFIRST (these satellites are generally too faint for usable data to be obtained from the ground). WFIRST is expected to be in operation later than either HST or JWST and can therefore continue monitoring these satellites into the 2030s, ultimately providing a time baseline of 15-20 years to search for changes in their rotation periods or any repeating trends that would hint at interesting dynamical processes.

## 8. Summary

The WFIRST Solar System Working Group identified a broad range of science investigations with all three instruments that cover a large fraction of the different solar system populations, including near-Earth asteroids, main belt asteroids, comets, Trojan asteroids, giant planets, giant planet satellites, Centaurs, and Kuiper Belt Objects. The Working Group also identified a list of the most compelling investigations that are unique to WFIRST, ground-breaking in their respective fields, and exciting to the general public, including:

- Identification of additional binary asteroid systems with similar characteristics to Ida/Dactyl using the WFIRST CGI.
- Making use of WFIRST's field of regard and field of view to search for new Earth Trojan asteroids with the WFI.
- Efficiently probe the entirety of the Hill spheres of the giant planets to potentially identify up to 1000, 200, 100, and 5 additional faint irregular satellites around Jupiter, Saturn, Uranus, and Neptune, respectively.
- Undertake a search for the first known objects orbiting in the Inner Oort Cloud (r>100 AU) with the W149 broadband filter.

Use of data from all of the planned astrophysics surveys will likely result in the identification and characterization of new minor bodies, and possibly even identification of new minor body populations. While LSST and JWST will lead the way in large-area surveys and space-based infrared observations in the 2020s, respectively, WFIRST is poised to pick up where both facilities will leave off when their primary missions end in ~2030. Incremental science with WFIRST is expected to be productive, but only if moving target tracking is commissioned from the onset of science operations.

We encourage the solar system community to continue to follow developments for the WFIRST mission. Some aspects of the mission are still uncertain, such as the final suite of instruments and the prospects for moving target tracking. Continued efforts to identify innovative solar system programs are also encouraged so that the best science can be performed starting as early as the first cycle. The opportunities presented by WFIRST are unique and exciting, and only an informed solar system science community will be able to make the most of them.

**Acknowledgements**

First and foremost, the authors would like to thank the two anonymous reviewers who provided comments that helped to significantly improve and focus this paper. We would also like to thank Jeff Kruk for his comments and clarifications, as well as Ed Nelan and Jason Kalirai for their advice. Ray Villard and Lou Strolger at the Space Telescope Science Institute provided the information necessary for evaluating the public impact of solar system observations made with the Hubble Space Telescope. WFIRST is a NASA Astrophysics Science Division joint project of the Goddard Space Flight Center, the Jet Propulsion Laboratory, the California Institute for Technology's Infrared Processing and Analysis Center, and the Space Telescope Science Institute. A portion of this research was carried out at the Jet Propulsion Laboratory, California Institute of Technology, under a contract with the National Aeronautics and Space Administration. This work made use of the Minor Planet & Comet Ephemeris Service (IAU Minor Planet Center). M. T. Bannister appreciates support from the UK STFC grant ST/L000709/1.

## References

1. Gehrels, N., Spergel, D., WFIRST SDT Project, 2015. Wide-Field InfraRed Survey Telescope (WFIRST) mission and synergies with LISA and LIGO-Virgo. Journal of Phys.: Conf. Series 610, 012007.
2. Spergel, D., et al., 2015. Wide-Field InfraRed Survey Telescope – Astrophysics Focused Telescope Assets WFIRST-AFTA 2015 Report. arXiv:1503.03757.
3. Stern, S.A., et al., 2015. The Pluto system: Initial results from its exploration by New Horizons. Science 350, aad1815.
4. Milam, S.N., et al., 2016. The James Webb Space Telescope's plan for operations and instrument capabilities for observations in the Solar System. PASP 128, 018001.
5. Emery, J.P., Burr, D.M., Cruikshank, D.P., 2011. Near-infrared spectroscopy of Trojan asteroids: Evidence for two compositional groups. AJ 141, 25.
6. Wong, I., Brown, M.E., 2017. The bimodal color distribution of small Kuiper Belt Objects. AJ 153, 145.
7. Stauffer, J., et al., 2018. The science advantage of a redder filter for WFIRST. arXiv:1806.00554.
8. Grundy, W.M., Schmitt, B., 1998. The temperature-dependent near-infrared absorption spectrum of hexagonal H2O ice. JGR 103, 25809-25822.
9. Brown, M.E., 2000. Near-infrared spectroscopy of Centaurs and irregular satellites. AJ 119, 977-983.
10. Barkume, K.M., Brown, M.E., Schaller, E.L., 2008. Near-infrared spectra of Centaurs and Kuiper Belt Objects. AJ 135, 55-67.
11. Weaver, H.A., et al., 2006. Discovery of two new satellites of Pluto. Nature 439, 943-945.
12. Hestroffer, D., et al., 2002. Asteroids observations with the Hubble Space Telescope. I. Observing strategy, and data analysis and modeling process. A&A 391, 1123-1132.
13. Grundy, W.M., et al., 2015. The mutual orbit, mass, and density of the large transneptunian binary system Varda and Ilmarë. Icarus 257, 130-138.
14. Simon, A.A., Wong, M.H., Orton, G.S., 2015. First results from the Hubble OPAL program: Jupiter in 2015. ApJ 812, 55.
15. Meech, K.J., et al., 2017. A brief visit from a red and extremely elongated interstellar asteroid. Nature 552, 378-381.
16. Showalter, M.R., Lissauer, J.J., 2006. The second ring-moon system of Uranus: Discovery and dynamics. Science 311, 973-977.
17. Showalter, M.R., et al., 2013. The Neptune system revisited: New results on moons and rings from the Hubble Space Telescope. AAS/Division for Planetary Sciences 45, #206.01.
18. Noll, K.S., et al., 2002. Detection of two binary trans-Neptunian objects, 1997 $CQ_{29}$ and 2000 $CF_{105}$, with the Hubble Space Telescope. AJ 124, 3424-3429.
19. Stephens, D.C., Noll, K.S., 2006. Detection of six trans-Neptunian binaries with NICMOS: A high fraction of binaries in the cold classical disk. AJ 131, 1142-1148.
20. Grundy, W.M., et al., 2011. Five new and three improved mutual orbits of transneptunian binaries. Icarus 213, 678-692.
21. Noll, K.S., Parker, A.H., Grundy, W.M., 2014. All bright cold classical KBOs are binary. AAS/Division for Planetary Sciences 46, #507.05.
22. Benecchi, S.D., et al., 2010. (47171) 1999 $TC_{36}$, a transneptunian triple. Icarus 207, 978-991.

23. Parker, A.H., et al., 2016. Discovery of a Makemakean moon. ApJL 825, L9.
24. Kiss, C., et al., 2017. Discovery of a satellite of the large trans-Neptunian object (225088) 2007 $OR_{10}$. ApJL 838, L1.
25. French, R.S., Showalter, M.R., 2012. Cupid is doomed: An analysis of the stability of the inner uranian satellites. Icarus 220, 911-921.
26. Agnor, C.B., Hamilton, D.P., 2006. Neptune's capture of its moon Triton in a binary-planet gravitational encounter. Nature 441, 192-194.
27. Christy, J.W., Harrington, R.S., 1978. The satellite of Pluto. AJ 83, 1005-1008.
28. Showalter, M.R., et al., 2011. New satellite of (134340) Pluto: S/2011 (134340) 1. IAU Circ. 9221, #1.
29. Showalter, M.R., et al., 2012. New satellite of (134340) Pluto: S/2012 (134340) 1. IAU Circ. 9253, #1.
30. Buie, M.W., et al., 2010. Pluto and Charon with the Hubble Space Telescope. II. Resolving changes on Pluto's surface and a map for Charon. AJ 139, 1128-1143.
31. Showalter, M.R., Hamilton, D.P., 2015. Resonant interactions and chaotic rotation of Pluto's small moons. Nature 522, 45-49.
32. Spencer, J.R., et al., 2015. The successful search for a post-Pluto KBO flyby target for New Horizons using the Hubble Space Telescope. European Planetary Science Congress 2015, EPSC2015-417.
33. Lisse, C.M., et al., 2006. Spitzer spectral observations of the Deep Impact ejecta. Science 313, 635-640.
34. Verbiscer, A.J., Skrutskie, M.F., Hamilton, D.P., 2009. Saturn's largest ring. Nature 461, 1098-1100.
35. Küppers, M., et al., 2014. Localized sources of water vapour on the dwarf planet (1) Ceres. Nature 505, 525-527.
36. Caldwell, J., et al., 1992. Titan: Evidence for seasonal change – A comparison of Hubble Space Telescope and Voyager images. Icarus 97, 1-9.
37. Bauer, J.M., et al., 2010. Direct detection of seasonal changes on Triton with Hubble Space Telescope. ApJL 723, L49-L52.
38. Roth, L., et al., 2014. Transient water vapor at Europa's South pole. Science 343, 171-174.
39. Sparks, W.B., et al., 2016. Probing for evidence of plumes on Europa with HST/STIS. ApJ 829, 121.
40. Mainzer, A., et al., 2011. Preliminary results from NEOWISE: An enhancement to the Wide-field Infrared Survey Explorer for Solar System science. ApJ 731, 53.
41. Mainzer, A., et al., 2014. Initial performance of the NEOWISE reactivation mission. ApJ 792, 30.
42. Müller, T.G., et al., 2009. TNOs are Cool: A survey of the transneptunian region. EM&P 105, 209-219.
43. Lellouch, E., et al., 2013. "TNOs are Cool": A survey of the trans-Neptunian region. IX. Thermal properties of Kuiper Belt objects and Centaurs from combined Herschel and Spitzer observations. A&A 557, A60.
44. Vilenius, E., et al., 2014. "TNOs are Cool": A survey of the trans-Neptunian region. X. Analysis of classical Kuiper Belt objects from Herschel and Spitzer observations. A&A 564, A35.
45. Fernández, Y.R., et al., 2013. Thermal properties, sizes, and size distribution of Jupiter-family cometary nuclei. Icarus 226, 1138-1170.

46. Reach, W.T., Kelley, M.S., Vaubaillon, J., 2013. Survey of cometary $CO_2$, CO, and particulate emissions using the Spitzer Space Telescope. Icarus 226, 777-797.
47. Masiero, J.R., et al., 2011. Main belt asteroids with WISE/NEOWISE. I. Preliminary albedos and diameters. ApJ 741, 68.
48. Bauer, J.M., et al., 2011. WISE/NEOWISE observations of comet 103P/Hartley 2. ApJ 738, 171.
49. Bauer, J.M., et al., 2012. WISE/NEOWISE observations of active bodies in the Main Belt. ApJ 747, 49.
50. Connors, M., Wiegert, P., Veillet, C., 2011. Earth's Trojan asteroid. Nature 475, 481-483.
51. Lisse, C.M., et al., 2017. The puzzling detection of x-rays from Pluto and Charon. Icarus 287, 103-109.
52. Bodewits, D., et al., 2007. Spectral analysis of the Chandra comet survey. A&A 469, 1183-1195.
53. Szabó, G.M., et al., 2012. Evidence for fresh frost layer on the bare nucleus of comet Hale-Bopp at 32 AU distance. ApJ 761, 8.
54. Roe, H.G., Pike, R.E., Brown, M.E., 2008. Tentative detection of the rotation of Eris. Icarus 198, 459-464.
55. Bodewits, D., et al., 2011. Collisional excavation of asteroid (596) Scheila. ApJL 733, L3.
56. Bodewits, D., et al., 2015. The pre-perihelion activity of dynamically new comet C/2013 A1 (Siding Spring) and its close encounter with Mars. ApJL 802, L6.
57. Ryan, E.L., Sharkey, B.N.L., Woodward, C.E., 2017. Trojan asteroids in the Kepler Campaign 6 field. AJ 153, 116.
58. Pál, A., et al., 2016. Large size and slow rotation of the trans-Neptunian object (225088) 2007 $OR_{10}$ discovered from Herschel and K2 observations. AJ 151, 117.
59. Farkas-Takács, A., et al., 2017. Properties of the irregular satellite system around Uranus inferred from K2, Herschel, and Spitzer observations. AJ 154, 119.
60. Simon, A.A., et al., 2016. Neptune's dynamic atmosphere from Kepler K2 observations: Implications for brown dwarf light curve analyses. ApJ 817, 162.
61. Molnár, L., et al., 2018. Main-belt asteroids in the K2 Uranus field. ApJS 234, 37.
62. Ivezić, Ž., LSST Science Collaboration, 2013. LSST Science Requirements Document. https://github.com/lsst-pst/LPM-17.
63. Schwamb, M.E., et al., 2018. Large Synoptic Survey Telescope solar system science roadmap. arXiv:1802.01783.
64. Mainzer, A.K., et al., 2017. The Near-Earth Object Camera. AAS/Division for Planetary Sciences 49, #219.01.
65. Levison, H., et al., 2017. Lucy: Surveying the diversity of Trojans. European Planetary Science Congress 2017, EPSC2017-963.
66. French, L.M., et al., 2013. A troop of Trojans: Photometry of 24 Jovian Trojan asteroids. Bulletin of the Minor Planets Section of the Association of Lunar and Planetary Observers 40, 198-203.
67. Merline, W.J., et al., 2001. S/2001 (617) 1. IAU Circ. 7741, #2.
68. Turtle, E.P., et al., 2017. Dragonfly: Exploring Titan's prebiotic organic chemistry and habitability. 48th Lunar and Planetary Science Conference, #1958.
69. Squyres, S.W., et al., 2018. The CAESAR New Frontiers mission: 1. Overview. 49th Lunar and Planetary Science Conference, #1332.

70. Pappalardo, R.T., et al., 2015. Science and reconnaissance form the Europa Clipper mission concept: Exploring Europa's habitability. 46th Lunar and Planetary Science Conference, #2673.
71. Morbidelli, A., et al., 2005. Chaotic capture of Jupiter's Trojan asteroids in the early solar system. Nature 435, 462-465.
72. Sheppard, S.S., Trujillo, C.A., 2006. A thick cloud of Neptune Trojans and their colors. Science 313, 511-514.
73. Rivkin, A.S., et al., 2007. Composition of the L5 Mars Trojans: Neighbors, not siblings. Icarus 192, 434-441.
74. Alexandersen, M., et al., 2013. A Uranian Trojan and the frequency of temporary giant-planet co-orbitals. Science 341, 994-997.
75. de la Fuente Marcos, C., de la Fuente Marcos, R., 2014a. Asteroid 2013 $ND_{15}$: Trojan companion to Venus, PHA to Earth. MNRAS 439, 2970-2977.
76. Innanen, K.A., Mikkola, S., 1989. Studies on solar system dynamics: I. – The stability of Saturnian Trojans. AJ 97, 900-908.
77. Dvorak, R., Lhotka, C., Zhou, L., 2012. The orbit of 2010 $TK_7$: Possible regions of stability for other Earth Trojan asteroids. A&A 541, A127.
78. Belton, M.J.S., et al., 1996a. Galileo's encounter with 243 Ida: An overview of the imaging experiment. Icarus 120, 1-19.
79. Belton, M.J.S., et al., 1996b. The discovery and orbit of 1993 (243) 1 Dactyl. Icarus 120, 185-199.
80. Johnson, T.V., Lunine, J.I., 2005. Saturn's moon Phoebe as a captured body from the outer solar system. Nature 435, 69-71.
81. Gladman, B., et al., 2001. Discovery of 12 satellites of Saturn exhibiting orbital clustering. Nature 412, 163-166.
82. Nesvorný, D., et al., 2003. Orbital and collisional evolution of the irregular satellites. AJ 126, 398-429.
83. Sheppard, S.S., Jewitt, D.C., 2003. An abundant population of small irregular satellites around Jupiter. Nature 423, 261-263.
84. Jewitt, D., Haghighipour, N., 2007. Irregular satellites of the planets: Products of capture in the early solar system. Ann. Rev. Astron. & Astrophys. 45, 261-295.
85. Bottke, W.F., et al., 2010. The irregular satellites: The most collisionally evolved populations in the solar system. AJ 139, 994-1014.
86. Nicholson, P.D., et al., 2008. Irregular satellites of the giant planets. In: Barucci, M.A., Boehnhardt, H., Cruikshank, D.P., Morbidelli, A. (Eds.), The Solar System Beyond Neptune. University of Arizona Press, Tucson, pp. 411-424.
87. Sheppard, S.S., 2006. Outer irregular satellites of the planets and their relationship with asteroids, comets and Kuiper Belt objects. In: Daniela, L., Sylvia Ferraz, M., Angel, F.J. (Eds.), IAU Symposium 229. Cambridge University Press, Cambridge, pp. 319-334.
88. Hamilton, D.P., Krivov, A.V., 1997. Dynamics of distant moons of asteroids. Icarus 128, 241-249.
89. Bottke, W.F., et al., 2013. Black rain: The burial of the Galilean satellites in irregular satellite debris. Icarus 223, 775-795.
90. Nesvorný, D., Vokrouhlický, D., Deienno, R., 2014. Capture of irregular satellites of Jupiter. ApJ 784, 22.

91. Li, D., Christou, A.A., 2017. Orbital modification of the Himalia family during an early solar system dynamical instability. AJ 154, 209.
92. Miyazaki, S., et al., 2018. Hyper Suprime-Cam: System design and verification of image quality. Pub. Of the Astron. Society of Japan 70, S1.
93. Batygin, K., Brown, M.E., 2016. Evidence for a distant giant planet in the solar system. AJ 151, 22.
94. Brown, M.E., Trujillo, C.A., Rabinowitz, D.L., 2005. Discovery of a planetary-sized object in the scattered Kuiper Belt. ApJ 635, L97-L100.
95. Hills, J.G., 1981. Comet showers and the steady-state infall of comets from the Oort Cloud. AJ 86, 1730-1740.
96. Dones, L., et al., 2004. Oort cloud formation and dynamics. In: Festou, M.C., Keller, H.U., Weaver, H.A. (Eds.), Comets II. University of Arizona Press, Tucson, pp. 153-174.
97. Robberto, M., 2010. A library of simulated cosmic ray events impacting JWST HgCdTe detectors. JWST-STScI-001928, SM-12.
98. Oort, J.H., 1950. The structure of the cloud of comets surrounding the Solar System and a hypothesis concerning its origin. Bulletin of the Astronomical Institutes of the Netherlands 11, 91-110.
99. Kaib, N.A., Roškar, R., Quinn, T., 2011. Sedna and the Oort cloud around a migrating Sun. Icarus 215, 491-507.
100. Paál, A., et al., 2012. "TNOs are Cool": A survey of the trans-Neptunian region. VII. Size and surface characteristics of (90377) Sedna and 2010 $EK_{139}$. A&A 541, L6.
101. Millis, R.L., et al., 1987. The size, shape, density, and albedo of Ceres from its occultation of BD+8°471. Icarus 72, 507-518.
102. Sicardy, B., et al., 2011. A Pluto-like radius and a high albedo for the dwarf planet Eris from an occultation. Nature 478, 493-496.
103. Buie, M.W., et al., 2015. Size and shape from stellar occultation observations of the double Jupiter Trojan Patroclus and Menoetius. AJ 149, 113.
104. Santos-Sanz, P., et al., 2016. James Webb Space Telescope observations of stellar occultations by Solar System bodies and rings. PASP 128, 018011.
105. Dias-Oliveira, A., et al., 2017. Study of the plutino object (208996) 2003 AZ84 from stellar occultations: size, shape and topographic features. AJ 154, 22.
106. Elliot, J.L., Dunham, E., Mink, D., 1977. The rings of Uranus. Nature 267, 328-330.
107. Braga-Ribas, F., et al., 2014. A ring system detected around the Centaur (10199) Chariklo. Nature 508, 72-75.
108. Oritz, J.L., et al., 2017. The size, shape, density and ring of the dwarf planet Haumea from a stellar occultation. Nature 550, 219-223.
109. Elliot, J.L., et al., 2007. Changes in Pluto's atmosphere: 1988-2006. AJ 134, 1-13.
110. Elliot, J.L., Olkin, C.B., 1996. Probing planetary atmospheres with stellar occultations. Ann. Rev. E&PS 24, 89-124.
111. Schlichting, H.E., et al., 2009. A single sub-kilometre Kuiper belt object from a stellar occultation in archival data. Nature 462, 895-897.
112. Schlichting, H.E., et al., 2012. Measuring the abundance of sub-kilometer-sized Kuiper Belt Objects using stellar occultations. ApJ 761, 150.

113. Chang, H.-K., Liu, C.-Y., Shang, J.-R., 2016. Upper limits to the number of Oort Cloud objects based on serendipitous occultation events search in X-rays. MNRAS 462, 1952-1960.
114. Roques, F., et al., 2006. Exploration of the Kuiper Belt by high-precision photometric stellar occultations: First results. AJ 132, 819-822.
115. Zhang, Z.-W., et al., 2013. The TAOS Project: Results from seven years of survey data. AJ 146, 14.
116. Doressoundiram, A., et al., 2016. An exploration of the trans-Neptunian region through stellar occultations and MIOSOTYS. AAS/Division for Planetary Sciences 48, #120.09.
117. Singer, K.N., et al., 2016. Craters on Pluto and Charon – Surface ages and impactor populations. 47th Lunar and Planetary Science Conference, #1903.
118. Nelan, E., et al., 2016. Guide star availability for the WFIRST auxiliary FGS using GSC2.3. WFIRST-STScI-TR1604.
119. Gomes, R.S., 2003. The origin of the Kuiper Belt high-inclination population. Icarus 161, 404-418.
120. Ofek, E.O., 2012. Sloan Digital Sky Survey observations of Kuiper Belt Objects: Colors and variability. ApJ 749, 10.
121. Sheppard, S.S., Trujillo, C., 2016. New extreme trans-Neptunian objects: Towards a super-Earth in the outer Solar System. AJ 152, 221.
122. Petit, J.-M., et al., 2017. The Canada-France Ecliptic Plane Survey (CFEPS) – High-latitude component. AJ 153, 236.
123. Larsen, J.A., et al., 2007. The search for distant objects in the Solar System using Spacewatch. AJ 133, 1247-1270.
124. Brown, M.E., et al., 2015. A serendipitous all sky survey for bright objects in the outer Solar System. AJ 149, 69.
125. Gerdes, D.W., et al., 2017. Discovery and physical characterization of a large scattered disk object at 92 au. ApJL 839, L15.
126. Tegler, S.C., Romanishin, W., Consolmagno, G.J., 2016. Two color populations of Kuiper Belt and Centaur objects and the smaller orbital inclinations of red Centaur objects. AJ 152, 210.
127. Gould, A., 2014. WFIRST ultra-precise astrometry I: Kuiper Belt Objects. Journal of the Korean Astron. Society 47, 279-291.
128. Hounsell, R., et al., 2017. Simulations of the WFIRST Supernova Survey and forecasts of cosmological constants. arXiv:1702.01747.
129. Sromovsky, L.A., et al., 2015. High S/N Keck and Gemini AO imaging of Uranus during 2012-2014: New cloud patterns, increasing activity, and improved wind measurements. Icarus 258, 192-223.
130. Karkoschka, E., 2015. Uranus' southern circulation revealed by Voyager 2: Unique characteristics. Icarus 250, 294-307.
131. Sromovsky, L.A., et al., 2003. The nature of Neptune's increasing brightness: evidence for a seasonal response. Icarus 163, 256-261.
132. Clark, R.N., McCord, T.B., 1979. Jupiter and Saturn – Near-infrared spectral albedos. Icarus 40, 180-188.
133. Fink, U., Larson, H.P., 1979. The infrared spectra of Uranus, Neptune, and Titan from 0.8-2.5 microns. ApJ 233, 1021-1040.

134. Karkoschka, E., 1994. Spectrophotometry of the jovian planets and Titan at 300- to 1000-nm wavelength: The methane spectrum. Icarus 111, 174-192.
135. de Kleer, K., de Pater, I., 2016. Time variability of Io's volcanic activity from near-IR adaptive optics observations on 100 nights in 2013-2015. Icarus 280, 378-404.
136. Fagents, S.A., et al., 2000. Cryomagmatic mechanisms for the formation of Rhadamanthys Linea, triple band margins, and other low-albedo features on Europa. Icarus 144, 54-88.
137. Sparks, W.B., et al., 2017. Active cryovolcanism on Europa? ApJL 839, L18.
138. McCord, T.B., et al., 1999. Hydrated salt minerals on Europa's surface from the Galileo near-infrared mapping spectrometer (NIMS) investigation. JGR 104, 11827-11852.
139. McKay, C.P., Pollack, J.B., Courtin, R., 1991. The greenhouse and antigreenhouse effects on Titan. Science 253, 1118-1121.
140. Lorenz, R.D., McKay, C.P., Lunine, J.I., 1997. Photochemically-induced collapse of Titan's atmosphere. Science 275, 642-644.
141. Yung, Y.L., Allen, M., Pinto, J.P., 1984. Photochemistry of the atmosphere of Titan – Comparison between model and observations. ApJS 55, 465-506.
142. Tomasko, M.G., et al., 2005. Rain, winds and haze during the Huygens probe's descent to Titan's surface. Nature 438, 765-778.
143. Hayes, A., et al., 2008. Hydrocarbon lakes on Titan: Distribution and interaction with a porous regolith. GRL 35, L09204.
144. Stofan, E.R., et al., 2007. The lakes of Titan. Nature 445, 61-64.
145. Larsson, R., McKay, C.P., 2013. Timescale for oceans in the past of Titan. P&SS 78, 22-24.
146. Griffith, C.A., et al., 2009. Characterization of clouds in Titan's tropical atmosphere. ApJL 702, L105-L109.
147. Rodriguez, S., et al., 2006. Cassini/VIMS hyperspectral observations of the HUYGENS landing site on Titan. P&SS 54, 1510-1523.
148. Romon, J., et al., 2001. Photometric and spectroscopic observations of Sycorax, satellite of Uranus. A&A 376, 310-315.
149. Takato, N., et al., 2004. Detection of a deep 3-μm absorption feature in the spectrum of Amalthea (JV). Science 306, 2224-2227.
150. Burns, J.A., et al., 1999. The formation of Jupiter's faint rings. Science 284, 1146.
151. Throop, H.B., et al., 2004. The jovian rings: New results derived from Cassini, Galileo, Voyager, and Earth-based observations. Icarus 172, 59-77.
152. Kelley, M.S.P., et al., 2016. Cometary science with the James Webb Space Telescope. PASP 128, 018009.
153. Grav, T., et al., 2012. WISE/NEOWISE observations of the Hilda population: Preliminary results. ApJ 744, 197.
154. Bauer, J.M., et al., 2017. Debiasing the NEOWISE cryogenic mission comet population. AJ 154, 53.
155. Kaib, N.A., Quinn, T., 2009. Reassessing the source of long-period comets. Science 325, 1234-1236.
156. Levison, H.F., et al., 2010. Capture of the Sun's Oort Cloud from stars in its birth cluster. Science 329, 187-190.

157. Trujillo, C.A., Brown, M.E., 2002. A correlation between inclination and color in the Classical Kuiper Belt. ApJ 566, L125-L128.
158. Francis, P.J., 2005. The demographics of long-period comets. ApJ 635, 1348-1361.
159. Bauer, J.M., et al., 2015. The NEOWISE-discovered comet population and the CO+$CO_2$ production rates. ApJ 814, 85.
160. Ootsubo, T., et al., 2010. Detection of parent $H_2O$ and $CO_2$ molecules in the 2.5-5 μm spectrum of comet C/2007 N3 (Lulin) observed with AKARI. ApJL 717, L66-L70.
161. Kelley, M.S.P., et al., 2017. Mid-infrared spectra of comet nuclei. Icarus 284, 344-358.
162. Meech, K.J., Svoreň, J., 2004. Using cometary activity to trace the physical and chemical evolution of cometary nuclei. In: Festou, M.C., Keller, H.U., Weaver, H.A. (Eds.), Comets II. University of Arizona Press, Tucson, pp. 317-335.
163. Protopapa, S., et al., 2014. Water ice and dust in the innermost coma of comet 103P/Hartley 2. Icarus 238, 191-204.
164. Yang, B., et al., 2014. Multi-wavelength observations of comet C/2011 L4 (Pan-STARRS). ApJL 784, L23.
165. Lamy, P.L., et al., 2004. The sizes, shapes, albedos, ad colors of cometary nuclei. In: Festou, M.C., Keller, H.U., Weaver, H.A. (Eds.), Comets II. University of Arizona Press, Tucson, pp. 223-264.
166. Milani, A., et al., 2014. Asteroid families classification: Exploiting very large datasets. Icarus 239, 46-73.
167. Nesvorný, D., Brož, M., Carruba, V., 2015. Identification and dynamical properties of asteroid families. In: Michel, P., DeMeo, F.E., Bottke, W.F. (Eds.), Asteroids IV. University of Arizona Press, Tucson, pp. 297-321.
168. Spoto, F., Milani, A., Knežević, Z., 2015. Asteroid family ages. Icarus 257, 275-289.
169. Fieber-Beyer, S.K., et al., 2011. The Maria asteroid family: Genetic relationships and a plausible source of mesosiderites near the 3:1 Kirkwood Gap. Icarus 213, 524-537.
170. Thomas, C.A., et al., 2011. Space weathering of small Koronis family members. Icarus 212, 158-166.
171. Hardersen, P.S., et al., 2014. More chips off of asteroid (4) Vesta: Characterization of eight Vestoids and their HED meteorite analogs. Icarus 242, 269-282.
172. Reddy, V., et al., 2014. Chelyabinsk meteorite explains unusual spectral properties of Baptistina asteroid family. Icarus 237, 116-130.
173. Alí-Lagoa, V., et al., 2016. Differences between the Pallas collisional family and similarly sized B-type asteroids. A&A 591, A14.
174. Paolicchi, P., Knežević, Z., 2016. Footprints of the YORP effect in asteroid families. Icarus 274, 314-326.
175. Migliorini, A., et al., 2017. Spectral characterization of V-type asteroids outside the Vesta family. MNRAS 464, 1718-1726.
176. Jewitt, D., Hsieh, H., Agarwal, J., 2015. The active asteroids. In: Michel, P., DeMeo, F.E., Bottke, W.F. (Eds.), Asteroids IV. University of Arizona Press, Tucson, pp. 221-241.
177. Hsieh, H.H., Jewitt, D., 2006. A population of comets in the Main Asteroid Belt. Science 312, 561-563.

178. Hsieh, H.H., et al., 2015. The main-belt comets: The Pan-STARRS1 perspective. Icarus 248, 289-312.
179. Wong, I., Brown, M.E., 2016. A hypothesis for the color bimodality of Jupiter Trojans. AJ 152, 90.
180. Jewitt, D., 2009. The active Centaurs. AJ 137, 4296-4312.
181. Rauer, H., et al., 1997. Millimetric and optical observations of Chiron. P&SS 45, 799-805.
182. Paganini, L., et al., 2013. Ground-based infrared detections of CO in the Centaur-comet 29P/Schwassmann-Wachmann 1 at 6.26 AU from the Sun. ApJ 766, 100.
183. Brown, M.E., Schaller, E.L., Fraser, W.C., 2011. A hypothesis for the color diversity of the Kuiper Belt. ApJL 739, L60.
184. Bauer, J.M., et al., 2013. Centaurs and Scattered Disk Objects in the thermal infrared: Analysis of *WISE*/*NEOWISE* observations. ApJ 773, 22.
185. Brasser, R., et al., 2012. An Oort Cloud origin for the high-inclination, high-perihelion Centaurs. MNRAS 420, 3396-3402.
186. de la Fuente Marcos, C., de la Fuente Marcos, R., 2014b. Large retrograde Centaurs: Visitors from the Oort Cloud? Ap&SS 352, 409-419.
187. Dones, L., et al., 2015. Origin and evolution of the cometary reservoirs. Space Science Rev. 197, 191-269.
188. Gladman, B., et al., 2002. Evidence for an extended scattered disk. Icarus 157, 269-279.
189. Trujillo, C.A., Brown, M.E., 2002. A correlation between inclination and color in the Classical Kuiper Belt. ApJ 566, L125-L128.
190. Horner, J., Evans, N.W., Bailey, M.E., 2004. Simulations of the population of Centaurs – I. The bulk statistics. MNRAS 354, 798-810.
191. Pike, R.E., et al., 2017. Col-OSSOS: z-band photometry reveals three distinct TNO surface types. AJ 154, 101.
192. Guilbert, A., et al., 2009. ESO-Large Program on TNOs: Near-infrared spectroscopy with SINFONI. Icarus 201, 272-283.
193. Barucci, M.A., et al., 2011. New insights on ices in Centaur and transneptunian populations. Icarus 214, 297-307.
194. Brown, M.E., Schaller, E.L., Fraser, W.C., 2012. Water ice in the Kuiper Belt. AJ 143, 146.
195. Noll, K.S., et al., 2008. Binaries in the Kuiper Belt. In: Barucci, M.A., Boehnhardt, H., Cruikshank, D.P., Morbidelli, A. (Eds.), The Solar System Beyond Neptune. University of Arizona Press, Tucson, pp. 345-363.
196. Parker, A.H., et al., 2011. Characterization of seven ultra-wide trans-Neptunian binaries. ApJ 743, 1.
197. Hastings, D.M., et al., 2016. The short rotation period of Hi'iaka, Haumea's largest satellite. AJ 152, 195.
198. Brown, M.E., et al., 2007. A collisional family of icy objects in the Kuiper Belt. Nature 446, 294-296.

**Bryan J. Holler** is a Support Scientist and member of the MIRI branch at the Space Telescope Science Institute in Baltimore, Maryland, USA. He received BS degrees in astronomy and physics from the University of Maryland and his PhD in astronomy from the University of Colorado. His current research focusses on spectroscopy of Kuiper Belt Objects (KBOs) to better constrain their surface composition and evolution.

**Stefanie N. Milam** is currently the James Webb Space Telescope Deputy Project Scientist for Planetary Science at NASA's Goddard Space Flight Center, Maryland, USA. She is also the co-chair of the Solar System Working Group for the WFIRST mission. Her current research involves spectroscopic studies of comets, evolved stars, and star-forming regions with both ground and space-based facilities.